\documentclass[preprint,review]{elsarticle}
\usepackage[margin=1in]{geometry}

\journal{Computer Physics Communications}

\usepackage{graphicx}
\usepackage[export]{adjustbox}

\usepackage{amsmath,amsfonts,amssymb,amsthm,bm}
\usepackage{epsfig,multirow,float,enumerate}
\usepackage{subcaption}
\usepackage[dvipsnames]{xcolor}
\usepackage{algorithm}
\usepackage{setspace}
\usepackage[noend]{algpseudocode}
\usepackage[section]{placeins}    
\usepackage{verbatim}
\usepackage{url}

\usepackage{outlines}

\newcommand{\bx}{ {\bf x} }
\newcommand{\by}{ {\bf y} }
\newcommand{\bk}{ {\bf k} }
\newcommand{\sk}{ {\bf s}_\bk }
\newcommand{\bl}{ {\bm{\ell}} }

\newcommand{\tell}{ {\bf t}_\bl }

\begin{document} 

\begin{frontmatter}

\title{A GPU-Accelerated Fast Summation Method Based on Barycentric Lagrange Interpolation and Dual Tree Traversal}

\author[label1]{Leighton Wilson\fnref{label2}\fnref{label3}}
\ead{lwwilson@umich.edu}

\author[label1]{Nathan Vaughn\fnref{label3}}
\ead{njvaughn@umich.edu}

\author[label1]{Robert Krasny}
\ead{krasny@umich.edu}

\address[label1]{Department of Mathematics, University of Michigan, Ann Arbor, MI 48109, USA}
\fntext[label2]{Corresponding author at: Department of Mathematics, University of Michigan, 530 Church St, Ann Arbor, MI 48109, USA.}
\fntext[label3]{Co-first authors.}

\begin{abstract}
We present the barycentric Lagrange dual tree traversal (BLDTT)
fast summation method for particle interactions.
The scheme replaces well-separated particle-particle interactions
by adaptively chosen
particle-cluster, cluster-particle, and cluster-cluster approximations
given by barycentric Lagrange interpolation at proxy particles
on a Chebyshev grid in each cluster.
The BLDTT is kernel-independent
and
the approximations can be efficiently mapped onto GPUs,
where target particles provide an outer level of parallelism
and
source particles provide an inner level of parallelism.
We present an OpenACC GPU implementation of the BLDTT
with MPI remote memory access for distributed memory parallelization.
The performance of the GPU-accelerated BLDTT is demonstrated for 
calculations with different 
problem sizes, particle distributions, geometric domains, 
and interaction kernels, 
as well as for unequal target and source particles.
Comparison with our earlier particle-cluster 
barycentric Lagrange treecode (BLTC) 
demonstrates the superior performance of the BLDTT.
In particular,
on a single GPU for problem sizes ranging from $N$=1E5 to 1E8,
the BLTC has $O(N\log N)$ scaling,
while the BLDTT has $O(N)$ scaling.
In addition,
MPI strong scaling results are presented for the BLTC and BLDTT 
using $N$=64E6 particles on up to 32~GPUs.
\end{abstract}


\end{frontmatter}



\clearpage
\section{Introduction}


Long-range particle interactions are essential in 
many areas of computational physics, including 
calculation of electrostatic or gravitational potentials,
as well as discrete convolution sums in boundary element methods.
In this context consider the potential due to a set of $N$~particles,
\begin{equation}
\phi(\bx_i) = \sum_{j = 1}^N G(\bx_i,\bx_j)q_j, \quad i = 1:N,
\label{eqn:direct_sum}
\end{equation}
where $\bx_i$ is a target particle,
$\bx_j$ is a source particle with strength $q_j$,
and
$G(\bx,\by)$ is an interaction kernel;
for example,
in electrostatics $q_j$ is a charge
and
$G(\bx,\by)$ is the Coulomb kernel.
The cost of evaluating the potentials $\phi(\bx_i)$ by direct summation scales 
like $O(N^2)$, which is prohibitively expensive for large systems,
but several methods are available to reduce the cost.
Among these are mesh-based methods 
in which the particles are projected onto a regular mesh,
such as
particle-particle/particle-mesh (P3M)~\cite{Hockney:1988aa} 
and 
particle-mesh Ewald (PME)~\cite{Essmann:1995aa},
and
tree-based methods 
in which the particles are partitioned into clusters with a 
hierarchical tree structure,
such as
Appel's dual tree traversal (DTT) method~\cite{Appel:1985aa},
the Barnes-Hut treecode (TC)~\cite{Barnes:1986aa},
and
the Greengard-Rokhlin fast multipole method 
(FMM)~\cite{Greengard:1987aa,Cheng:1999aa}.
Other related methods for fast summation of particle interactions include
panel clustering~\cite{Hackbusch:1989aa},
hierarchical matrices~\cite{Hackbusch:2015aa},
and
multilevel summation~\cite{Hardy:2015aa}.
The present approach employs features of the DTT, TC, and FMM,
and
next we outline key details of these methods.



\subsection{Tree-based fast summation methods}
Tree-based fast summation methods such as the DTT, TC, and FMM
can be described as having two phases.
In the precompute phase,
the particles are partitioned into a hierarchical tree of clusters,
and
assuming the particle distribution is homogeneous,
this phase generally scales like $O(N\log N)$.
In the compute phase,
well-separated particle-particle interactions are approximated
using for example 
monopole approximations~\cite{Appel:1985aa, Barnes:1986aa}
or
higher order multipole expansions~\cite{Greengard:1987aa, Lindsay:2001aa, Dehnen:2002aa}.
Some differences in the compute phase of these methods are noted as follows.
The TC traverses the tree for each target particle to 
identify well-separated particle-cluster pairs,
while the DTT traverses two copies of the tree simultaneously to
identify well-separated cluster-cluster pairs;
both methods use a multipole acceptance criterion (MAC) for this purpose.
The FMM passes information from one level to the next,
with a uniform definition of the interaction list at each level;
an upward pass computes cluster moments,
and 
a downward pass computes potentials using
multipole-to-local and local-to-local translations.
The compute phase of the TC scales like $O(N\log N)$,
while the compute phase of the FMM~\cite{Greengard:1987aa,Cheng:1999aa}
and
DTT~\cite{Esselink:1992aa,Dehnen:2002aa}
scale like $O(N)$.
Next we review developments relevant to the present work in more detail,
including
dual tree traversal methods, kernel-independent methods,
and GPU implementations.

\subsection{Dual tree traversal methods}

The DDT methods employ cluster-cluster approximations. Early work by Appel developed a DTT-style method using monopole approximations~\cite{Appel:1985aa}.
A later algorithm independently developed by Dehnen for stellar dynamics used Taylor approximations~\cite{Dehnen:2002aa}.
Subsequent large-scale astrophysical simulations 
implemented parallel DTTs on many-core architectures~\cite{Taura:2012aa, Yokota:2013aa,Dehnen:2014aa,Lange:2014aa},
and
the DTT with Taylor approximations has been applied in molecular dynamics simulations of periodic condensed phase systems~\cite{Lorenzen:2012aa} 
and 
for polarizable force fields~\cite{Coles:2015aa}. 


\subsection{Kernel-independent methods}

Tree-based fast summation methods originally relied on 
analytic series expansions specific to a given kernel $G(\bx,\by)$.
For example,
in the case of the Coulomb and Yukawa kernels,
multipole expansions were used in the FMM~\cite{Cheng:1999aa,Greengard:2002aa},
and
Cartesian Taylor expansions were used in the TC~\cite{Duan:2001aa,Li:2009aa}.
Alternative approximations for the Coulomb kernel were introduced involving
numerical discretization of the
Poisson integral formula~\cite{Anderson:1992aa}
and
multipole expansions at pseudoparticles~\cite{Makino:1999aa}.
Eventually kernel-independent methods were developed that
require only kernel evaluations
and
are suitable for a large class of kernels.
Among these,
the kernel-independent FMM (KIFMM) uses equivalent densities
defined on proxy surfaces~\cite{Ying:2004aa},
the black-box FMM (bbFMM)
uses polynomial interpolation
and SVD compression~\cite{Fong:2009aa},
and
the barycentric Lagrange treecode (BLTC)
uses barycentric Lagrange interpolation~\cite{Berrut:2004aa,Wang:2020aa}.
A number of related proxy point methods have recently been 
developed using skeletonized interpolation~\cite{Cambier:2019aa}
and
interpolative decomposition~\cite{Xing:2020aa, Huang:2020aa}.
Several of the kernel-independent fast summation methods have been parallelized 
for multi-core CPU systems~\cite{Ying:2003aa,Lashuk:2012aa,March:2015aa,
Malhotra:2015aa,Malhotra:2016aa,Takahashi:2012aa,Huang:2020aa}.

\subsection{GPU implementations}

Many modern HPC systems rely heavily on
graphics processing units (GPUs) for high-throughput arithmetic. 
The direct sum in Eq.~\eqref{eqn:direct_sum} 
is well suited for GPU computing;
this is because the kernel evaluations $G({\bf x}_i,{\bf x}_j)$
can be computed concurrently,
where the targets ${\bf x}_i$ provide an outer level of parallelism,
while the sources ${\bf x}_j$ provide an inner level of parallelism.
Early GPU implementations of direct summation
achieved a 25$\times$ speedup over an optimized CPU implementation~\cite{Elsen:2006aa}
and 
a 250$\times$ speedup over a portable C implementation~\cite{Nyland:2009aa}.
However,
these codes still scale like $O(N^2)$
and 
there is great interest in implementing the 
sub-quadratic scaling tree-based methods on GPUs,
although
this is challenging due to the complexity of these methods
in comparison with direct summation.
In this direction there have been several GPU implementations 
of the TC and FMM~\cite{Hamada:2009aa,Burtscher:2011aa,Yokota:2011aa,Yokota:2011ac,Takahashi:2012aa,Bedorf:2012aa,Bedorf:2014aa,Boukaram:2019aa,Boukaram:2019ab,Vaughn:2019aa},
but we know of only a few GPU implementations
of the DTT~\cite{Yokota:2012aa,Fortin:2019aa}.



\subsection{Present work}

The present work contributes a 
GPU-accelerated tree-based fast summation method called 
barycentric Lagrange dual tree traversal (BLDTT).
The BLDTT employs several techniques from previous tree-based methods,
including 
(1) the DTT algorithmic structure~\cite{Appel:1985aa,Dehnen:2002aa},
(2) barycentric Lagrange interpolation~\cite{Berrut:2004aa,Wang:2020aa,Vaughn:2019aa},
and 
(3) upward and downward passes similar to those in the FMM,
but adapted to the context of barycentric Lagrange interpolation.
The BLDTT has several additional features that should be noted.
The algorithm replaces well-separated particle-particle interactions
by adaptively chosen particle-cluster, cluster-particle, and cluster-cluster approximations,
where the clusters are represented by proxy particles at
Chebyshev grid points.
The approximations are done with barycentric Lagrange interpolation
and
require only kernel evaluations,
hence the BLDTT is kernel-independent.
Similar to other tree-based fast summation methods, 
the BLDTT has a precompute phase and a compute phase;
the precompute phase scales like $O(N\log N)$ with a small prefactor,
while the compute phase scales like $O(N)$,
so the observed scaling of the BLDTT is essentially $O(N)$.

As will be shown,
the barycentric Lagrange approximations resemble the 
direct sum in Eq.~\eqref{eqn:direct_sum} 
and 
they can be efficiently mapped onto GPUs.
Based on this observation we present an OpenACC GPU implementation of the BLDTT
with MPI remote memory access for distributed memory parallelization.
The performance of the BLDTT is documented for calculations with different
problem sizes,
particle distributions,
geometric domains,
and interaction kernels,
unequal target and source particles. 
Comparison with our earlier particle-cluster 
barycentric Lagrange treecode (BLTC) shows the superior performance of the BLDTT.
In particular,
on a single GPU for problem sizes ranging from $N$=1E5 to 1E8,
the BLTC has $O(N\log N)$ scaling,
while the BLDTT has $O(N)$ scaling.
In addition,
MPI strong scaling results are presented for the BLTC and BLDTT 
using $N$=64E6 particles on up to 32~GPUs.
The BLDTT code is part of the BaryTree library 
for fast summation of particle interactions available on 
GitHub~\cite{barytree-website}.

The remainder of the paper is organized as follows. 
Section~\ref{section:bldtt-v2} describes the 
barycentric Lagrange dual tree traversal (BLDTT) fast summation method.
Section~\ref{section:implementation} 
describes our implementation of the BLDTT 
using MPI remote memory access for distributed memory parallelization 
and 
OpenACC for GPU acceleration.
Section~\ref{section:results} presents numerical results
for several test cases. 
Section~\ref{section:conclusion} gives the conclusions.


\section{Description of BLDTT fast summation method}
\label{section:bldtt-v2}

\subsection{Barycentric Lagrange interpolation}
\label{section:bary-lagrange-interp}

We briefly review the barycentric Lagrange form of polynomial interpolation in 1d~\cite{Berrut:2004aa}.  
Given a function $f(x)$ 
and 
$n+1$ points $s_k, k=0:n$, 
the Lagrange form of the interpolating polynomial is
\begin{equation}
    p_n(x) = \sum_{k=0}^n f(s_k) L_k(x),
    \label{eqn:polynomial-interpolation}
\end{equation}
where the Lagrange polynomial $L_k(x)$ 
in barycentric form is
\begin{equation}
    L_k(x) = 
    \frac{\displaystyle
    \frac{w_k}{x-s_k}}{\displaystyle\sum_{k'=0}^n \frac{w_k}{x-s_k'}}, 
    \quad w_k = \frac{1}{\prod_{j=0,j\ne k}^n(s_k-s_j)}, \quad k = 0:n.
    \label{eqn:1d-barycentric}
\end{equation} 
This work employs Chebyshev points of the second kind,
\begin{equation}
s_k = \cos\theta_k, \quad \theta_k = \pi k/n, 
\quad k = 0:n,
\end{equation}
in which case the interpolation weights are given by
\begin{equation}
    w_k = (-1)^k\delta_k, \quad k = 0:n,
\end{equation}
where $\delta_k=1/2$ if $k=0$ or $n$, and $\delta_k=1$ otherwise~\cite{Salzer:1972aa,Berrut:2004aa}. 
The algorithms described below use 
barycentric Lagrange interpolation 
in 3D rectangular boxes with interpolation points 
$\sk = (s_{k_1}, s_{k_2}, s_{k_3})$ 
given by a Cartesian tensor product grid of 
Chebyshev points 
adapted to the box.
Depending on the context,
the interpolation points $\sk$ may also be referred to as proxy particles.

\subsection{Algorithm overview}

The BLDTT fast summation method computes the 
potential at a set of $M$ target particles due to interactions with a set of $N$ source particles; 
Eq.~\eqref{eqn:direct_sum} is a special case in which the targets and sources refer to the same set of particles.
First, 
two hierarchical trees of particle clusters are built,
one for the target particles
and 
one for the source particles,
where each cluster is a rectangular box;
clusters in the target tree are denoted $C_t$ 
and 
clusters in the source tree are denoted $C_s$.
The computed potential at a target particle $\bx_i$ 
has contributions from four types of interactions
as determined by the dual tree traversal algorithm described below.
The four types are
direct particle-particle (PP) interactions of nearby particles,
and
particle-cluster (PC),
cluster-particle (CP),
and
cluster-cluster (CC) approximations of well-separated clusters.

Algorithm~\ref{alg:BLDTT} is a high-level overview of the 
BLDTT.
Lines 1-4 describe the input consisting of 
target and source particle data, 
interpolation degree $n$,
MAC parameter $\theta$, 
and the maximum number of particles in the leaves of each tree, $M_0, N_0$.
Line 5 describes the output potentials.
Line 6 builds the target tree and source tree containing
target clusters $C_t$ and source clusters $C_s$.
Line 7 is the upward pass to compute proxy charges $q_\bk$ 
at proxy particles $\sk$ in source clusters $C_s$.
Line 8 is the dual tree traversal to compute $PP, PC, CP, CC$ interactions.
Line 9 is the downward pass to interpolate potentials from
proxy particles $\tell$ to target particles $\bx_i$ in target clusters $C_t$.
The steps will be described in detail below.
We summarize the notation used in presenting the BLDTT in Table~\ref{table:notation}.

\begin{algorithm}[htb]
\caption{Barycentric Lagrange Dual Tree Traversal (BLDTT) Fast Summation Method}
\label{alg:BLDTT}
\begin{algorithmic}[1]
\State \textbf{input} target particles ${\bf x}_i, i=1:M$
\State \textbf{input} source particles and charges ${\bf y}_j, q_j, j =1:N$
\State \textbf{input} interpolation degree $n$, MAC parameter $\theta$
\State \textbf{input} max particles per target leaf $M_0$, 
max particles per source leaf $N_0$
\State \textbf{output} potentials $\phi(\bx_i), i=1:M$
\State build target tree and source tree 
\State upward pass to compute proxy charges $q_\bk$
at proxy particles $\sk$ in source clusters 
\State dual tree traversal to compute PP, PC, CP, CC interactions
\State downward pass to interpolate potentials from proxy particles $\tell$ to target particles ${\bf x}_i$
\end{algorithmic}
\end{algorithm}

\begin{table}[htb]
\centering
\renewcommand{\arraystretch}{1.2}
\begin{tabular}{ll c ll}
Symbol & Description && Symbol & Description \\\cline{1-2}\cline{4-5}

$M, N$ & 
number of target, source particles &

& $G(x,y)$ & interaction kernel \\

$\bx_i, x_i$ & 
target particle in 3D, 1D &

& $\phi(x_i)$ & 
potential at target $x_i$ \\

$\by_j, y_j$ & 
source particle in 3D, 1D &

& $\phi_{PP}(x_i, C_t, C_s)$ & 
PP potential at target $x_i$ in $C_t$ \\

$ q_j$ & source charge of $y_j$ &

& & due to sources in $C_s$ \\

$C_t, C_s$ & 
target cluster, source cluster &

& $\phi_{PC}(x_i, C_t, \widehat{C}_s)$ & 
PC potential at target $x_i$ in $C_t$ \\

$\tell, t_\ell$ & 
target proxy particle in 3D, 1D &

& & due to proxy sources in $\widehat{C}_s$ \\

$\sk, s_k$ & 
source proxy particle in 3D, 1D &

& $\phi_{CP}(t_\ell, \widehat{C}_t, C_s)$ & 
CP proxy potential at proxy target \\

$\widehat{q}_k$ & 
proxy charge of $s_k$ &

& & $t_\ell$ in $\widehat{C}_t$ due to sources in $C_s$  \\

$\widehat{C}_t, \widehat{C}_s$ & 
proxy particles in target cluster, &

& $\phi_{CC}(t_\ell, \widehat{C}_t, \widehat{C}_s)$ & 
CC proxy potential at proxy target \\

 & source cluster &

& & $t_\ell$ in $\widehat{C}_t$ due to proxy sources in $\widehat{C}_s$  \\

\end{tabular}
\caption{Summary of notation used in the BLDTT fast summation method,
PP = particle-particle,
PC = particle-cluster,
CP = cluster-particle,
CC = cluster-cluster.}
\label{table:notation}
\end{table}


\subsection{Tree building}

The target and source trees are constructed by the same routines,
described here for the target tree.
The maximum number of particles per leaf is a user-specified parameter,
$M_0$ for the target tree and $N_0$ for the source tree.
The root cluster is the minimal bounding box containing all target particles.
The root is recursively divided into child clusters, 
terminating when a cluster contains fewer than $M_0$ particles.
Division occurs at the midpoint of the cluster; 
in general the cluster is bisected in all three coordinate directions, 
resulting in eight child clusters, with two exceptions.
First, a cluster is divided into only two or four children in order to maintain a good aspect ratio, that is, a ratio of longest to shortest side lengths no greater than $\sqrt{2}$. 
Second, a cluster is divided into only two or four children to avoid creating leaf clusters with fewer than $M_0/2$ particles on average;
in particular, if a cluster contains between $M_0$ and $2M_0$ particles, 
it is divided into two children,
and
if it contains between $2M_0$ and $4M_0$ particles, 
it is divided into four children.
Upon creation, each cluster is shrunk to the minimal bounding box containing its particles,
and
a tensor product grid of Chebyshev points
adapted to the box is created;
these are also referred to as proxy particles.
After building the trees, 
the BLDTT performs the upward pass, 
dual tree traversal, and downward pass,
but before discussing these steps, 
the next subsection describes the four types of interactions which are eventually
combined to compute potentials.


\subsection{Four types of interactions}

\begin{figure}[htb]
\centering
\begin{tabular}{cc}
\hspace{-0.1in}
\includegraphics[width=0.48\linewidth] {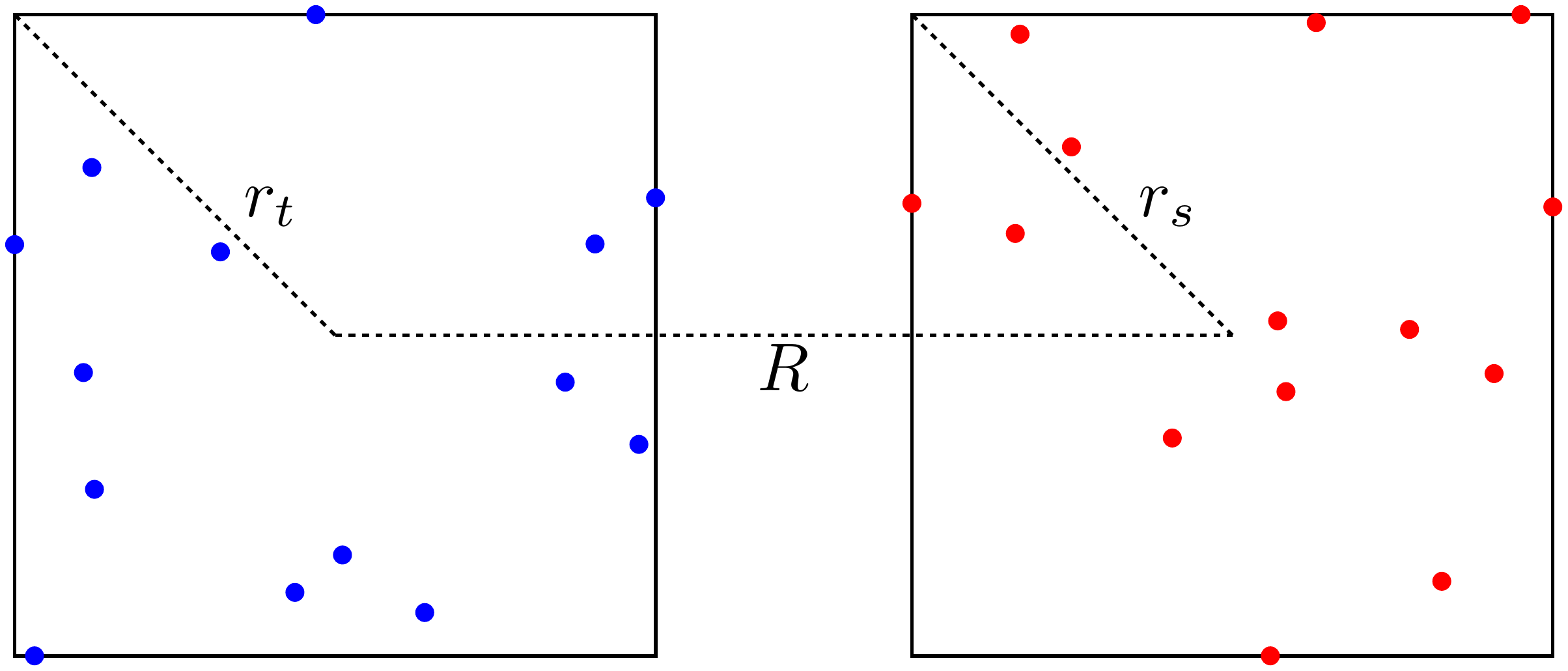} &
\hspace{0in}
\includegraphics[width=0.48\linewidth] {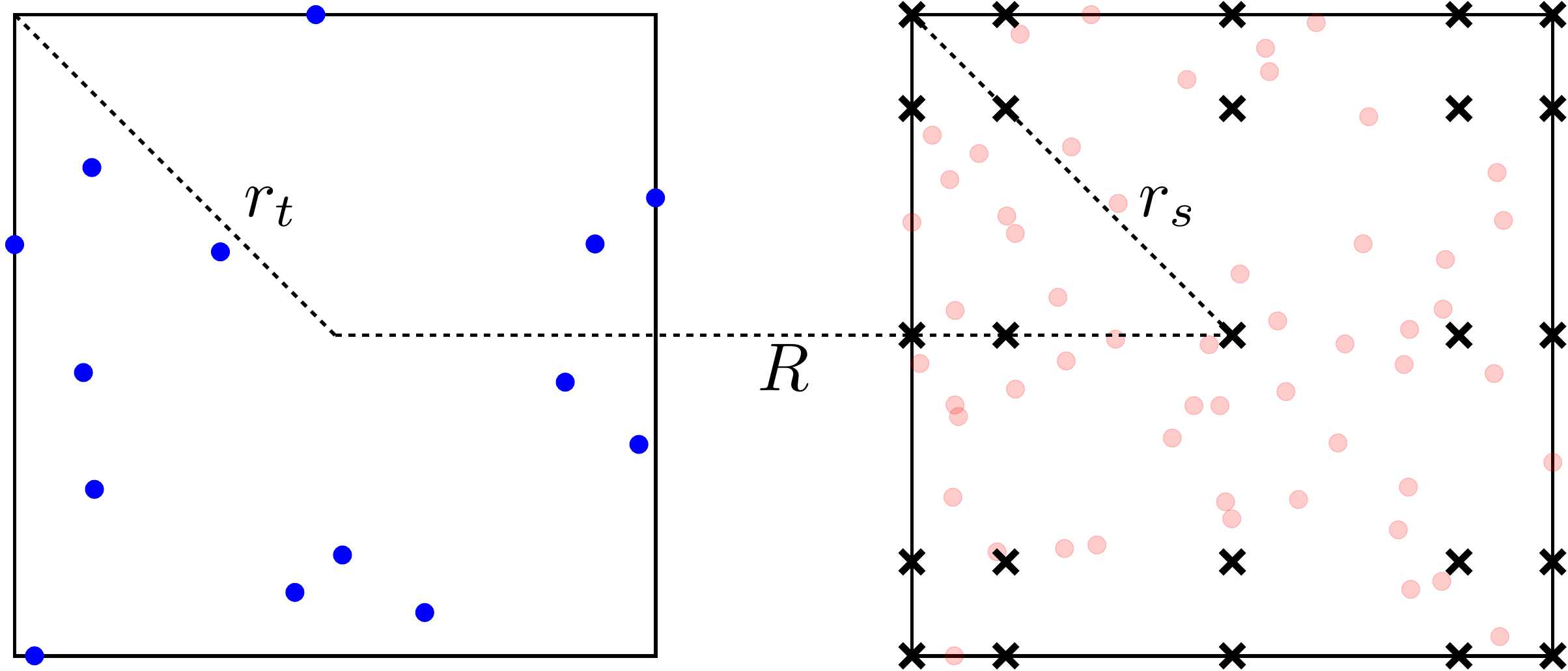} \\
\hspace{-0.1in} 
(a) particle-particle (PP) & 
(b) particle-cluster (PC) \\
~ & ~ \\
\hspace{-0.1in} 
\includegraphics[width=0.48\linewidth] {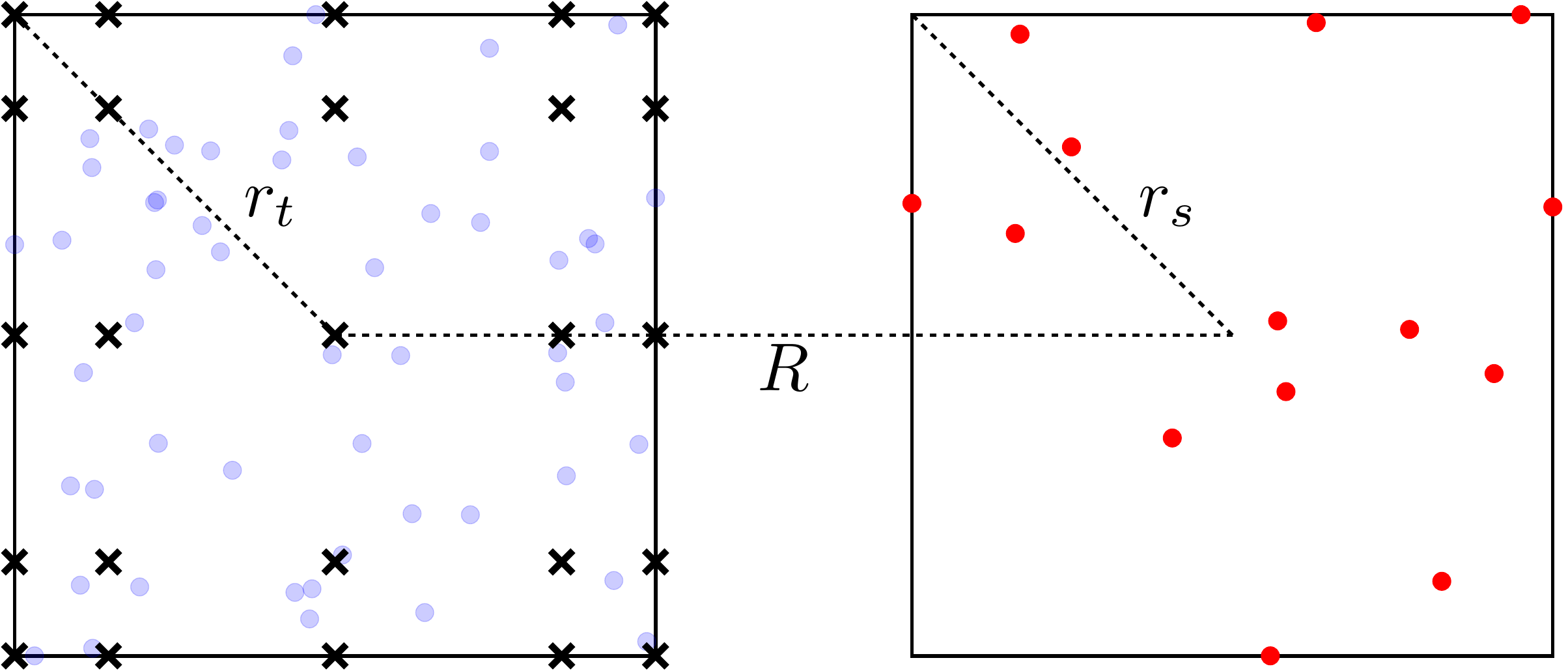} &
\hspace{0in}
\includegraphics[width=0.48\linewidth] {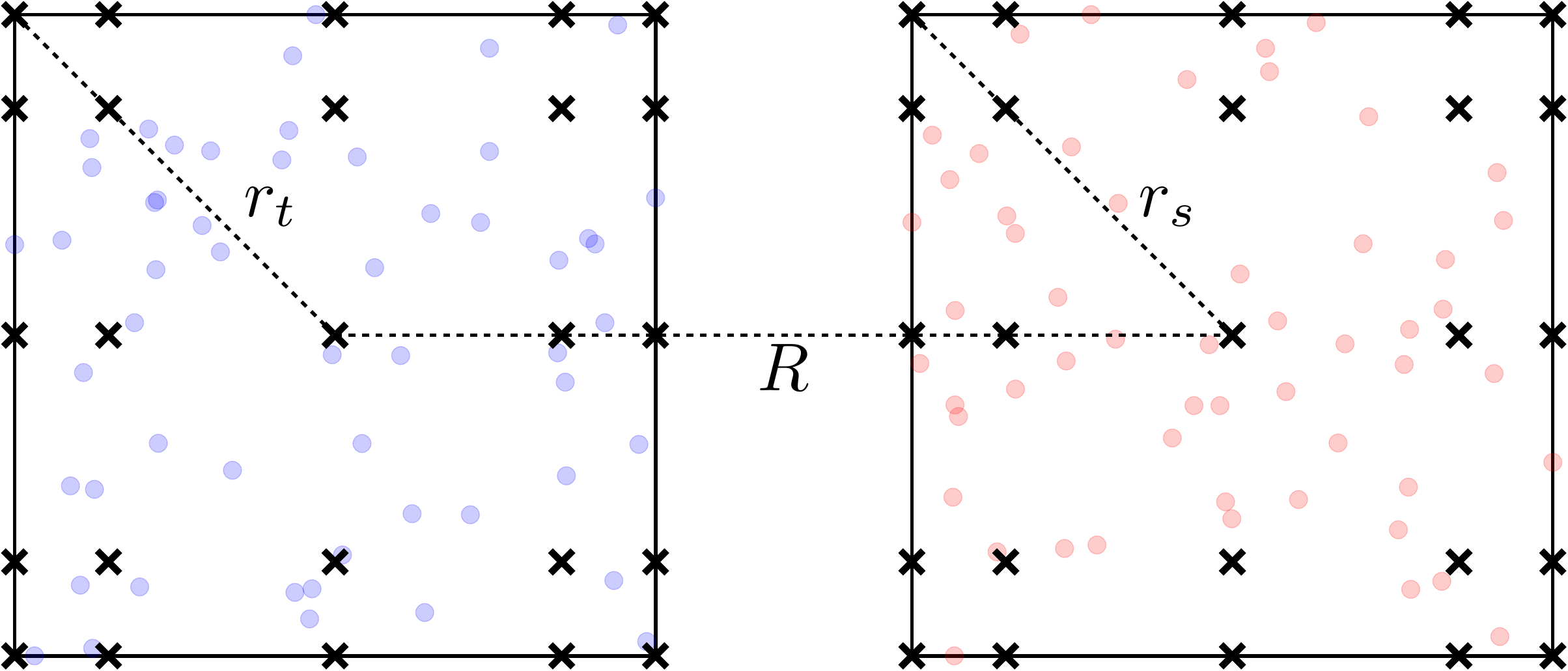} \\
\hspace{-0.1in} 
(c) cluster-particle (CP) & 
(d) cluster-cluster (CC) \\
\end{tabular}
\caption{Four types of interactions are used, 
in each case the target cluster $C_t$ on the left interacts
with the source cluster $C_s$ on the right,
(a) direct particle-particle interaction (PP),
(b) particle-cluster approximation (PC),
(c) cluster-particle approximation (CP),
(d) cluster-cluster approximation (CC),
dots are target/source particles $\bx_i, \by_j$,
crosses are target/source proxy particles $\tell, \sk$,
target/source cluster radii $r_t, r_s$,
target-source cluster distance $R$.
}
\label{fig:interactions-v2}
\end{figure}

Figure~\ref{fig:interactions-v2} 
depicts the four types of interactions 
between a target cluster $C_t$ (left, blue) 
and a source cluster $C_s$ (right, red),
where dots are target/source particles $\bx_i, \by_j$,
and
crosses are target/source proxy particles $\tell, \sk$.
Also shown are the target/source cluster radii $r_t, r_s$,
and
the target-source cluster distance $R$.
These diagrams depict 2D versions of the interactions; 
in practice the particles are distributed in 3D and the clusters are rectangular boxes.
In general, clusters can interact via their particles (dots) or their proxy particles (crosses).
Figure~\ref{fig:interactions-v2} shows the four cases:
(a) $C_t$ and $C_s$ use particles (PP),
(b) $C_t$ uses particles and $C_s$ uses proxy particles (PC),
(c) $C_t$ uses proxy particles and $C_s$ uses particles (CP),
(d) $C_t$ and $C_s$ both use proxy particles (CC).
The interactions are described in detail below;
to simplify notation we present the interactions in 1D
and
note that the extension to 3D is straightforward 
using tensor products.

\vskip 5pt
{\bf Particle-particle interaction.}
Figure~\ref{fig:interactions-v2}a depicts a PP interaction.
In this case the PP potential at a target particle $x_i \in C_t$ 
due to direct interaction with the source particles $y_j \in C_s$ 
is denoted by
\begin{equation}
    \phi_{PP}(x_i,C_t,C_s) = \sum_{y_j \in C_s}G(x_i,y_j)q_j,
    \quad x_i \in C_t.
    \label{eqn:PP-potential}
\end{equation}

{\bf Particle-cluster approximation.}
Figure~\ref{fig:interactions-v2}b depicts a PC interaction.
The kernel is approximated by 
holding $x_i$ fixed
and
interpolating with respect to $y_j$ at the 
proxy particles $s_k$ in $C_s$,
\begin{equation}
    G(x_i,y_j) \approx \sum_{k=0}^n G(x_i,s_k)L_{k}(y_j),
    \quad x_i \in C_t, \quad y_j \in C_s.
    \label{eqn:kernel-pc} 
\end{equation}
Substituting into the particle-particle interaction
and
rearranging terms yields the PC potential,
\begin{equation}
\phi_{PC}(x_i,C_t,\widehat{C}_s)  
    = \sum_{k=0}^n  G(x_i,s_k) \widehat{q}_k,
    \quad x_i \in C_t,
\label{eqn:PC-potential}
\end{equation}
where the proxy charges $\widehat{q}_k$ of the proxy particles $s_k$ are
\begin{equation}
    \widehat{q}_k =  \sum_{y_j\in C_s} L_{k}(y_{j})q_j.
\label{eqn:proxy-charges}
\end{equation}
Equation~\eqref{eqn:PC-potential} uses 
$C_t,\widehat{C}_s$ to indicate that the 
target particles $x_i$ interact with the 
proxy source particles $s_k$.

\textbf{Cluster-particle approximation.} 
Figure~\ref{fig:interactions-v2}c depicts a CP interaction.
The kernel is approximated by 
interpolating with respect to $x_i$ at the
proxy particles $t_\ell$ in $C_t$
and
holding $y_j$ fixed,
\begin{equation}
    G(x_i,y_j) \approx \sum_{\ell=0}^n 
    L_{\ell}(x_i)G(t_\ell,y_j), 
    \quad x_i \in C_t, \quad y_j \in C_s.
    \label{eqn:kernel-cp-v2}
\end{equation}
Substituting into the particle-particle interaction
and
rearranging terms yields the CP potential,
\begin{equation}
\phi_{CP}(x_i,\widehat{C}_t,C_s) = 
    \sum_{\ell=0}^n \phi(t_\ell,\widehat{C}_t,C_s) L_{\ell}(x_{i}), \quad x_i \in C_t,
    \label{eqn:cluster-particle-b}
\end{equation}
where the CP proxy potential $\phi(t_\ell,\widehat{C}_t,C_s)$
is
\begin{equation}
    \phi(t_\ell,\widehat{C}_t,C_s) = 
    \sum_{y_j \in C_s}G(t_\ell,y_j)q_j, \quad t_\ell \in \widehat{C}_t.
    \label{eqn:CP-proxy-potential}
\end{equation}
Equation~\eqref{eqn:cluster-particle-b} 
and
Eq.~\eqref{eqn:CP-proxy-potential}
use $\widehat{C}_t,C_s$ to indicate that 
the proxy target particles of $C_t$ interact with
the source particles $y_j$.
Equation~\eqref{eqn:cluster-particle-b} interpolates from the
proxy target particles $t_\ell$ 
to the target particles $x_i$.

\textbf{Cluster-cluster interaction.} 
Figure~\ref{fig:interactions-v2}d depicts a CC interaction.
The kernel is interpolated 
with respect to $x_i$ at the proxy particles $t_\ell$ in $C_t$
and 
with respect to $y_j$ at the proxy particles $s_k$ in $C_s$,
\begin{equation}
    G(x_i,y_j) \approx \sum_{k=0}^n \sum_{\ell=0}^n   
    L_{\ell}({x_i})
    G(t_\ell,s_k)
    L_{k}(y_j),
    \quad x_i \in C_t, \quad y_j \in C_s.
    \label{eqn:kernel-cc-v2}
\end{equation}
Substituting into the particle-particle interaction and rearranging terms
yields the CC potential,
\begin{equation}
    \phi_{CC}(x_i,\widehat{C}_t,\widehat{C}_s) =
    \sum_{\ell=0}^b \phi(t_\ell,\widehat{C}_t,\widehat{C}_s) 
    L_{\ell}(x_{i}), \quad x_i \in C_t,
\label{eqn:cluster-cluster-d}
\end{equation}
where the CC proxy potential $\phi(t_\ell,\widehat{C}_t,\widehat{C}_s)$ is
\begin{equation}
    \phi(t_\ell,\widehat{C}_t,\widehat{C}_s) = \sum_{k=0}^n G(t_\ell,s_k) \widehat{q}_k, \quad t_\ell \in \widehat{C}_t,
    \label{eqn:CC-proxy-potential}
\end{equation}
and
the proxy charges $\widehat{q}_k$ were defined in Eq.~\eqref{eqn:proxy-charges}.
Equation~\eqref{eqn:cluster-cluster-d} 
and
Eq.~\eqref{eqn:CC-proxy-potential}
use $\widehat{C}_t,\widehat{C}_s$ to indicate that the 
proxy target particles $t_\ell$ interact with the
proxy source particles $s_k$.
Equation~\eqref{eqn:cluster-cluster-d} interpolates
from the proxy target particles $t_\ell$ 
to the target particles $x_i$.

The following three subsections describe the rest of Algorithm~\ref{alg:BLDTT},
comprising the upward pass, dual tree traversal, and downward pass.


\subsection{Upward pass}
The upward pass computes the proxy charges $\widehat{q}_k$ 
defined in Eq.~\eqref{eqn:proxy-charges}
for the proxy particles $s_k$ 
in each source cluster $\widehat{C}_s$ in the source tree,
as required in the PC and CC approximations in Eqs.~\eqref{eqn:PC-potential} and~\eqref{eqn:CC-proxy-potential}.
Note that each source particle $y_j$ contributes to the proxy charges $\widehat{q}_k$
of exactly one cluster at a given level of the tree.
Hence with $N$ source particles and tree depth $O(\log N)$, 
computing the proxy charges directly by Eq.~\eqref{eqn:proxy-charges}
requires $O(N\log N)$ operations.

The BLDTT algorithm uses an alternative approach as follows.
Let $C_s$ denote a parent cluster with 
Lagrange polynomials $L_{k}(y)$ 
and
interpolation points $s_k$,
and
let $C_s^{i}, i=1:8$ denote the eight child clusters
with Lagrange polynomials $L^i_{k_i}(y)$
and interpolation points $s_{k_i}$.
As an alternative to Eq.~\eqref{eqn:proxy-charges},
the proxy charges of the parent $\widehat{q}_k$
are computed from the proxy charges of the children $\widehat{q}_{k_i}$
by the expression
\begin{equation}
    \widehat{q}_k = \sum_{i=1}^8\sum_{k_i=0}^n L_{k}(s_{k_i})\widehat{q}_{k_i},
    \label{eqn:1d-proxy-charges-child-to-parent}
\end{equation}
which is derived in~\ref{section:appendix-upward pass}.
The derivation uses the definitions of the 
parent and child proxy charges and the relation
\begin{equation}
L_k(y) = \sum_{k_i=0}^n L_k(s_{k_i})L_{k_i}(y),  
\label{eqn:special-polynomial-interpolation}
\end{equation}
which is a special case of Eq.~\eqref{eqn:polynomial-interpolation}.

The upward pass starts by computing the proxy charges 
of the leaves of the source tree using
Eq.~\eqref{eqn:proxy-charges}
and
then ascending to the root by Eq.~\eqref{eqn:1d-proxy-charges-child-to-parent}.
This is analogous to the upward pass in the FMM~\cite{Greengard:1987aa}
where the multipole moments of a parent cluster are obtained
from the moments of its children.
As with the FMM, 
computing the proxy charges this way requires $O(N)$ operations,
which can be seen as follows.
Computing the proxy charges for the leaves requires $O(n^3N)$ operations;
each of the $N$ source particles contributes to one leaf, which contains $O(n^3)$ proxy particles (in 3D).  
Then, evaluating the child-to-parent relation in Eq.~\eqref{eqn:1d-proxy-charges-child-to-parent} for each parent proxy charge
requires $O(n^6)$ operations
and
is independent of the number of source particles.
Since there are $O(N)$ clusters in the tree,
ascending the tree requires an additional $O(n^6N)$ operations.
Hence the operation count for the BLDTT upward pass is $O(n^3N) + O(n^6N) = O(N)$,
as it is for the FMM.


\subsection{Dual tree traversal}

The dual tree traversal determines which pairs of clusters 
in the target and source trees interact
by one of the four options described above (PP, PC, CP, CC).
Before the traversal starts,
two sets of potentials are initialized to zero,
potentials $\phi(x_i)$ at the target particles
and
potentials $\phi(t_\ell)$ at the proxy target particles.
In the course of the traversal,
the potentials $\phi(x_i)$ are incremented due to PP and PC interactions,
and
the potentials $\phi(t_\ell)$ are incremented due to CP and CC interactions.
Following the dual tree traversal,
the $\phi(t_\ell)$ are interpolated to the target particles $x_i$ 
and
combined with the $\phi(x_i)$ in the downward pass.

The dual tree traversal uses the recursive procedure
DTT($C_t, C_s$) described in Algorithm~\ref{alg:dtt},
which takes a target cluster $C_t$ and a source cluster $C_s$ as input.
Initially the procedure is called for the root clusters of the target 
and source trees. 
In what follows,
the clusters are considered to be well-separated if
$(r_t + r_s)/R < \theta$,
where $r_t, r_s$ are the target and source cluster radii 
and 
$R$ is the center-center distance between the clusters.

If $C_t$ and $C_s$ are well-separated (line 2),
then they interact in one of four ways
depending on the number of particles they contain relative to
the number of proxy particles in a cluster,
which is denoted by $n_p=(n+1)^3$ in 3D.
If $C_t$ and $C_s$ are both large (lines 3-4), 
then the CC proxy potentials are incremented using Eq.~\eqref{eqn:CC-proxy-potential};
else if $C_t$ is large and $C_s$ is small (lines 5-6),
then the CP proxy potentials are incremented using Eq.~\eqref{eqn:CP-proxy-potential};
else if $C_t$ is small and $C_s$ is large (lines 7-8),
then the PC potentials are incremented using Eq.~\eqref{eqn:PC-potential};
else $C_t$ and $C_s$ are both small (line 9)
and
the PP potentials are incremented using Eq.~\eqref{eqn:PP-potential}.
 
If $C_t$ and $C_s$ are not well-separated,
then the traversal continues as follows.
If $C_t$ and $C_s$ are leaves (lines 11-12), 
then the PP potentials are incremented using Eq.~\eqref{eqn:PP-potential}. 
Otherwise if $C_s$ is a leaf, 
then it interacts recursively with the children of $C_t$ (line 13),
while if $C_t$ is a leaf, 
then it interacts recursively with the children of $C_s$ (line 14).
Finally if $C_t$ and $C_s$ are both not leaves,
then the smaller cluster interacts recursively with the children of the larger cluster (lines 15-17).  
\begin{algorithm}
\caption{Dual Tree Traversal}
\label{alg:dtt}
\begin{algorithmic}[1]
\State \textbf{procedure} DTT(target cluster $C_t$, source cluster $C_s$)
\State \textbf{if} $(r_t + r_s)/R < \theta$ \textbf{then}
\State \quad
\textbf{if} $|C_t| > n_p$ \textbf{and} $|C_s| > n_p$ \textbf{then} 
\State \quad\quad increment CC proxy potentials $\phi(t_\ell) \mathrel{+}= \phi(t_\ell,\widehat{C}_t,\widehat{C}_s)$ for all $t_\ell \in \widehat{C}_t$ by Eq.~\eqref{eqn:CC-proxy-potential}
\State \quad \textbf{else if}
$|C_t| > n_p$ \textbf{and} $|C_s| \leq n_p$
\textbf{then} 
\State \quad\quad increment CP proxy potentials $\phi(t_\ell) \mathrel{+}= \phi(t_\ell,\widehat{C}_t,C_s)$ for all $t_\ell \in \widehat{C}_t$ by Eq.~\eqref{eqn:CP-proxy-potential}
\State \quad \textbf{else if}
$|C_t| \leq n_p$ \textbf{and} $|C_s| > n_p$
\textbf{then} 
\State \quad\quad increment PC potentials $\phi(x_i) \mathrel{+}=\phi(x_i,C_t,\widehat{C}_s)$ for all $x_i \in C_t$ by Eq.~\eqref{eqn:PC-potential}
\State \quad \textbf{else} 
increment PP potentials $\phi(x_i) \mathrel{+}= \phi(x_i,C_t,C_s)$ for all $x_i \in C_t$ 
by Eq.~\eqref{eqn:PP-potential}
\State \textbf{else}
\State \quad \textbf{if} $C_t$ \textbf{and} $C_s$ are leaves \textbf{then}
\State \quad\quad increment PP potentials $\phi(x_i) \mathrel{+}= \phi(x_i,C_t,C_s)$ for all $x_i \in C_t$ by Eq.~\eqref{eqn:PP-potential}
\State \quad \textbf{else if} $C_s$ is a leaf \textbf{then}
\textbf{for} each child $C_t^\prime$ of $C_t$ \textbf{do}
\textsc{DTT}($C_t^\prime$, $C_s$)
\State \quad \textbf{else if} $C_t$ is a leaf \textbf{then}
\textbf{for} each child $C_s^\prime$ of $C_s$ \textbf{do} \textsc{DTT}($C_t$, $C_s^\prime$)
\State \quad \textbf{else} 
\State \quad \quad \textbf{if} $|C_s| < |C_t|$ \textbf{then for}
each child $C_t^\prime$ of $C_t$ \textbf{do} \textsc{DTT}($C_t^\prime$, $C_s$)
\State \quad \quad \textbf{else for} each child $C_s^\prime$ of $C_s$ \textbf{do}
\textsc{DTT}($C_t$, $C_s^\prime$)				
\end{algorithmic}
\end{algorithm}

The DTT yields potentials $\phi(x_i)$ due to PP and PC interactions 
and proxy potentials $\phi(t_\ell)$ due to CP and CC interactions.
In the case of $N$ homogeneously distributed source and target particles, 
the operation count of the dual tree traversal has been shown to be $O(N)$~\cite{Esselink:1992aa,Dehnen:2002aa}.


\subsection{Downward pass}

The downward pass interpolates the proxy potentials $\phi(t_\ell)$ 
to the target particles $x_i$
and
increments the potentials $\phi(x_i)$.
This can be done in two ways as described below.

First note that each
target particle $x_i$ is contained in a chain of target clusters,
\begin{equation}
    x_i \in C_t^{1} \subset C_t^{2} \subset \cdots \subset C_t^{L},
\end{equation}
where the superscript denotes the level in the target tree.
The target cluster $C_t^m$ at level $m$ in the chain has
Lagrange polynomials $L^m_{k_m}(x_i)$ 
and proxy potentials $\phi(t_{k_m}^m)$ that contribute to $\phi(x_i)$,
\begin{equation}
    \phi(x_i) \mathrel{+}= 
    \sum_{m = 1}^L \sum_{k_m=0}^n L^m_{k_m}(x_{i})\phi(t^m_{k_m}),
    \label{eqn:1d-direct-downpass}
\end{equation}
where $t^m_{k_m}$ are the proxy particles of $C_t^m$,
and
the $\mathrel{+}=$ indicates that the results are aggregated 
with the potentials $\phi(x_i)$  
due to PP and PC interactions previously computed in the DTT.
In Eq.~\eqref{eqn:1d-direct-downpass}
the inner sum interpolates potential values from the 
proxy particles $t^m_{k_m}$ to the target particle $x_i$, 
and
the outer sum accumulates the results from each level in the tree.
Computing $\phi(x_i)$ 
as indicated in Eq.~\eqref{eqn:1d-direct-downpass} 
requires $O(M\log M)$ operations;
the factor $M$ is the number of target particles $x_i$,
the factor $\log M$ is the number of levels in the target tree,
and
the inner sum requires $O(n^3)$ operations independent of $M$.

Rather than interpolating from the proxy particles $t_m^k$
directly to the target particle $x_i$ as in Eq.~\eqref{eqn:1d-direct-downpass},
we utilize a recursive alternative.
In what follows,
$C_t^m$ is a parent cluster at level $m$
and
$C^{m-1}_t$ is a child cluster at level $m-1$.
The procedure interpolates the 
parent proxy potentials $\phi(t_{k_m}^m)$
to 
child proxy potentials $\phi(t_{k_{m-1}}^{m-1})$,
\begin{equation}
    \phi(t_{k_{m-1}}^{m-1}) \mathrel{{+}{=}} \sum_{k_m=0}^n L_{k_m}^m(t_{k_{m-1}}^{m-1}) \phi(t_{k_m}^m),
    \label{eqn:1d-parent-child-downpass}
\end{equation}
where the $\mathrel{{+}{=}}$ indicates that the results 
of the parent-to-child interpolation on the right
are aggregated with the child proxy potentials $\phi(t_{k_{m-1}}^{m-1})$
due to CP and CC interactions previously computed in the DTT.
This procedure starts with the root cluster of the target tree (level $m=L$) 
and 
descends to the leaves (level $m=1$).
Upon reaching the leaves, 
the proxy potentials $\phi(t_{k_1}^{1})$ 
are interpolated to the target particles $x_i$
and
aggregated with the PP and PC potentials previously computed in the DTT,
\begin{equation}
    \phi(x_i) \mathrel{+}= \sum_{k_1=0}^n L^1_{k_1}(x_{i})\phi(t^1_{k_1}).
    \label{eqn:1d-downpass-leaves}
\end{equation}
It should be noted that the expressions for $\phi(x_i)$ in
Eq.~\eqref{eqn:1d-direct-downpass} and
Eq.~\eqref{eqn:1d-downpass-leaves} are equivalent
and
this is shown in \ref{section:appendix-downward pass}.

The recursive form of the downward pass described here is similar to the local-to-local 
step in the FMM,
where the local coefficients are shifted and accumulated,
and
as in that case the operation count is reduced to $O(M)$.
In particular, the parent-to-child interpolation in Eq.~\eqref{eqn:1d-parent-child-downpass} requires $O(n^6)$ operations (in 3D), 
independent of the number of target particles $M$,
and
the tree contains $O(M)$ clusters, 
so interpolation 
from the root down to the leaves by Eq.~\eqref{eqn:1d-parent-child-downpass} 
requires $O(n^6M)$ operations.
Then the final interpolation from the leaf proxy particles
to the target particles by Eq.~\eqref{eqn:1d-downpass-leaves}
requires $O(n^3)$ operations for each target, 
yielding complexity $O(n^6M) + O(n^3M) = O(M)$ for the downward pass.

\subsection{Description of BLTC}
We briefly describe our previous barycentric Lagrange treecode (BLTC)~\cite{Wang:2020aa, Vaughn:2019aa}
which has an algorithmic structure resembling the Barnes-Hut treecode~\cite{Barnes:1986aa}.
Unlike the BLDTT which builds a tree on both the source and target particles, 
the BLTC builds a tree of clusters on the source particles
and a set of batches on the target particles,
where the batches correspond to the leaves of a target tree.
Once the source tree is built,
the BLTC computes the proxy charges for each source cluster 
directly from the source particles by Eq.~\eqref{eqn:proxy-charges}.
The source tree is then traversed for each target batch, 
starting at the root of the tree
and
checking whether a given source cluster and target batch are well-separated.
If they are well-separated  
and 
the source cluster contains more particles than interpolation points, 
then the cluster and batch interact by the PC approximation in Eq.~\eqref{eqn:PC-potential}.
If they are not well-separated, 
then the batch interacts with the children of the source cluster.
Leaves in the source tree that are not well-separated from a given target batch, 
and
source clusters that are well-separated but contain more interpolation points than particles
interact directly with the target batch by the PP interaction in Eq.~\eqref{eqn:PP-potential}.
For $M$ target particles and $N$ source particles,
the 
BLTC operation count is $O(N\log N)$ + $O(M\log N)$,
where the first term arises from the computation of the proxy charges
and
the second term arises from the source tree traversal.


\section{BLDTT implementation}	
\label{section:implementation}
The implementation of the BLDTT for multiple GPUs 
is largely similar to that 
of our previous BLTC implementation~\cite{Vaughn:2019aa},
and is available as part of the BaryTree library 
for fast summation of particle interactions 
available on GitHub~\cite{barytree-website}.
The code uses OpenACC directives for GPU acceleration 
and MPI remote memory access 
for distributed memory parallelization.
Tree building, tree traversal, 
and MPI communication of particles and clusters 
occur on the CPU, 
while the upward pass, 
particle and cluster interaction computations, 
and downward pass 
occur on the GPU.
We review here several important details 
of this implementation.

\subsection{Computing interaction lists}
We decouple dual tree traversal 
from the computation of the particle interactions.
The dual tree traversal is performed on the CPU, 
creating four interaction lists for each cluster of the target tree,
one for each type of interaction.
Each list contains the indices of the source clusters that interact with the target cluster by the given interaction type.
The interactions are then computed by directly iterating over the interaction lists;
this improves the efficiency of GPU calculations because these lists can be iterated over rapidly and GPU compute kernels (described below) can be queued asynchronously.

\subsection{MPI distributed memory parallelization}
To implement distributed memory parallelization, we use locally essential trees (LET)~\cite{Warren:1992aa}.
Particles are partitioned by recursive coordinate bisection to create compact sub-domains on each MPI rank using Trilinos Zoltan~\cite{zoltan-website,Boman:2012aa}.
Each MPI rank constructs the local source tree and local target tree for its particles.
The LET of a rank is the union of the rank's local source tree
and
all source clusters from remote ranks interacting with its local target tree.
Even though constructing the LETs requires an all-to-all communication,
the amount of data acquired by each rank grows only logarithmically 
with the problem size~\cite{Warren:1992aa}.
The construction and communication required by the LETs
are performed using MPI passive target synchronization remote memory access (RMA).
RMA is a one-sided communication model within MPI in which
an origin process can \textit{put} data onto a target process 
or \textit{get} data from a target process 
through specially declared memory \textit{windows}, 
with no active involvement from the target process. 
This approach enables each rank to construct its LET asynchronously from other ranks.

\subsection{GPU details} 
The GPU implementation uses eight compute kernels, 
two for the upward pass, 
four for computing interactions determined by the dual tree traversal, 
and two for the downward pass. 
The compute kernels are generated with OpenACC directives, 
compiled with the PGI C compiler. 
We utilize asynchronous launch of kernels in multiple GPU streams 
to hide as much latency as possible.
The approach described here generalizes to multiple GPUs in a straightforward manner, in which each GPU corresponds to one MPI rank.

The first upward pass kernel computes the proxy charges for a given leaf 
in the source tree. 
For each leaf, this kernel is launched asynchronously, 
and any further computation is blocked until all leaf proxy charges are computed. 
The second upward pass kernel computes the proxy charges of a parent cluster 
using the proxy charges of its children. 
For a given level of the tree, 
this kernel is launched asynchronously for each cluster at that level, 
and any further computation is blocked until all proxy charges at that level are computed.
The proxy charges at a given level are computed
after the computation of the proxy charges at the previous level is complete.

The four DTT kernels compute the interaction of a target cluster with a source cluster.
Each PP, PC, CP, and CC interaction launches one compute kernel. 
All such kernels are launched asynchronously
and further computation is blocked until these kernels complete.
The four interaction kernels have a similar structure 
in which the outer loop is over the target particles or proxy particles in the target cluster, 
and the inner loop is over the source particles or proxy particles in the source cluster.
Importantly, 
due to the Lagrange form of barycentric interpolation,
the inner loop iterations are independent,
unlike alternative approximation methods which are sequential.
The outer loop is mapped to the \texttt{gang} construct in OpenACC 
and the inner loop is mapped to the \texttt{vector} construct. 
Conceptually, a member of a \texttt{gang} corresponds to a thread block, 
while a member of a \texttt{vector} corresponds to an individual thread. 

The two downward pass compute kernels are similar in structure to the upward pass kernels. 
At each level of the target tree above the leaves, beginning with the root, 
the first downward pass compute kernel is launched asynchronously 
for each target cluster at that level 
to interpolate the proxy potentials of the cluster to its children. 
Further computation is blocked until all compute kernels at a given level complete. 
Finally,
the second downward pass compute kernel is launched asynchronously 
for each target leaf to interpolate the proxy potentials to the target particles.

\section{Results}
\label{section:results}

We demonstrate the BLDTT
on a series of test cases and compare its performance to that of the BLTC. 
First, we compare the scaling of the BLTC and BLDTT on problem sizes from 1E5 to 1E8 particles.
Second, we briefly demonstrate the GPU acceleration of the BLDTT over a CPU implementation.
Third, we display the performance of the BLDTT on various particle distributions: random uniform, Gaussian, and Plummer~\cite{Plummer:1911aa, Dejonghe:1987aa}. 
We also demonstrate the benefit of including CP and PC interactions in the BLDTT algorithm.
Fourth, we display the performance of the BLDTT on various geometries: a $1 \times 10 \times 10$ slab,
a $1 \times 1 \times 10$ slab,
and a spherical shell with all particles at radius 1.
Fifth, we investigate BLDTT performance in cases where the number of sources and targets are unequal.
Sixth, we show performance results for various interaction kernels,
including an oscillatory kernel, the Yukawa kernel, and the regularized Coulomb kernel.
Last, we show MPI strong scaling performance on 1 to 32 GPUs.

Except for the Plummer distribution runs, the source particle charges are random uniformly distributed on $[-1,1]$.
Except for the runs involving unequal numbers of sources and targets, 
the source and target particle sets are identical.
All computations use the Coulomb interaction kernel
except for the examples in section~\ref{section:other-kernels}.
All runs used a maximum leaf or batch size of 2000.
To facilitate comparison of the BLDTT and BLTC across this wide variety of problems, 
Figures~\ref{fig:distribution_parameter_sweep},
\ref{fig:distribution_justcc_parameter_sweep},
\ref{fig:geometry_parameter_sweep}, 
and \ref{fig:uniform_sinoverr_parameter_sweep}
all use the same axes to display time vs. error.

The calculations are done in double precision arithmetic
and the reported errors are the relative $\ell_2$ error,
\begin{equation}
    E = \left( \sum_{i=1}^M (\phi_i^{ds} - \phi_i^{fs})^2 \bigg/ \sum_{i=1}^M (\phi_i^{ds})^2\right)^{1/2}, 
\end{equation}
where $\phi_i^{ds}$ are the target potentials computed by direct summation 
and $\phi_i^{fs}$ are computed by the BLDTT fast summation method.
The error was sampled at a random subset of 0.1\% of the target particles in all cases.

The computations were performed on the NVIDIA P100 GPU nodes 
at the San Diego Supercomputer Center Comet machine, 
where each node contains four GPUs, and each GPU has 16GB of memory. 
Except for the MPI strong scaling results in section~\ref{sec:mpi}, 
each computation was run on a single GPU.
The code was compiled with the PGI C compiler using the \texttt{-O3} optimization flag.
For parallel scaling results, the Zoltan library of Trilinos \cite{zoltan-website,Boman:2012aa} was used to perform recursive coordinate bisection to load balance the particles.
As described in the previous section, tree building and interaction list building are performed on the CPU, while the upward pass, particle interactions, and downward pass are performed on the GPU.


\subsection{Problem size scaling}

\begin{figure}[htb]
\vskip 0pt
\centering
\begin{tabular}{cc}
\hspace*{-0.1in}
\includegraphics[trim=0 0 20 0, clip, width=0.51\linewidth] {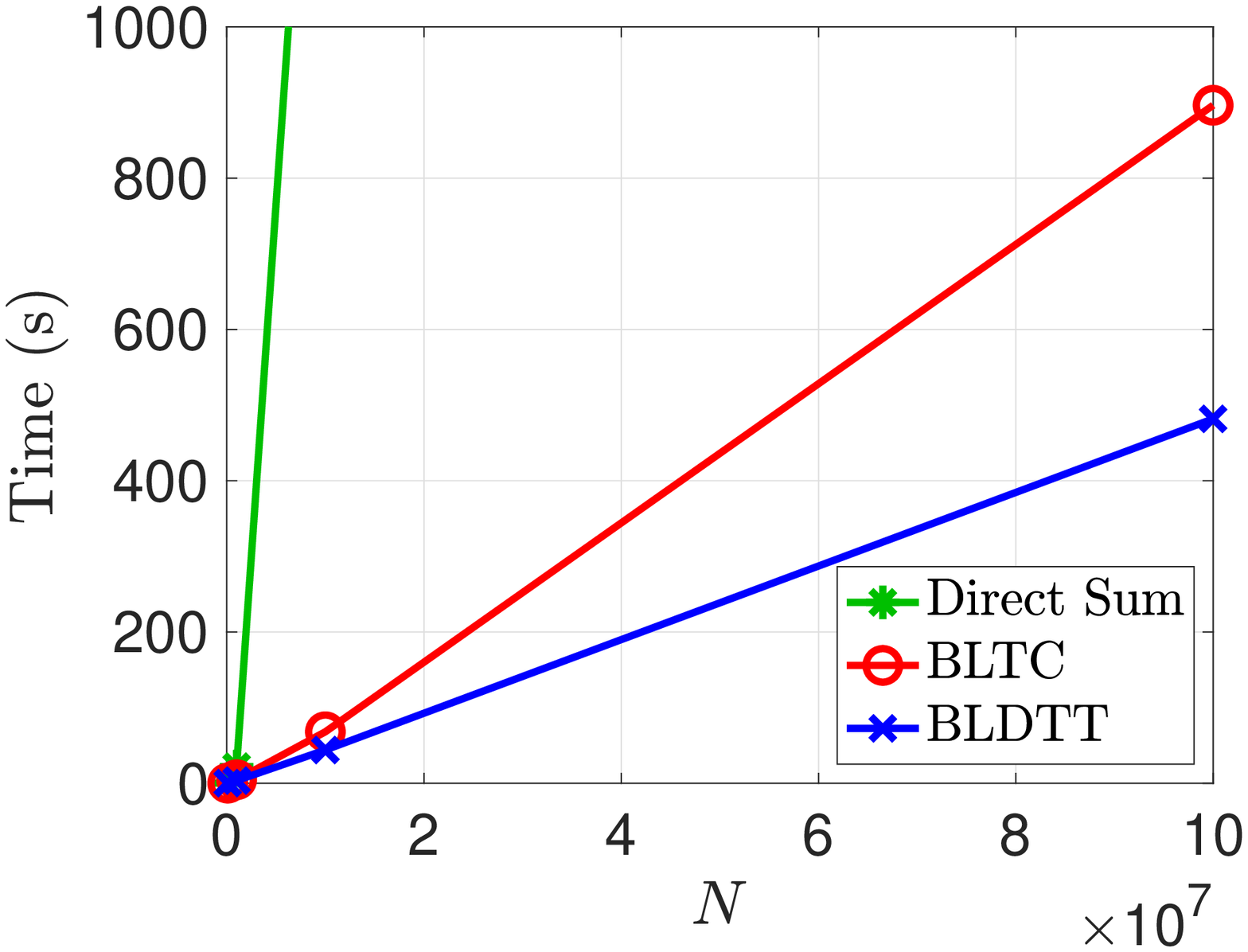} &
\hspace*{-0.2in}
\includegraphics[trim=30 0 0 0, clip,width=0.5\linewidth] {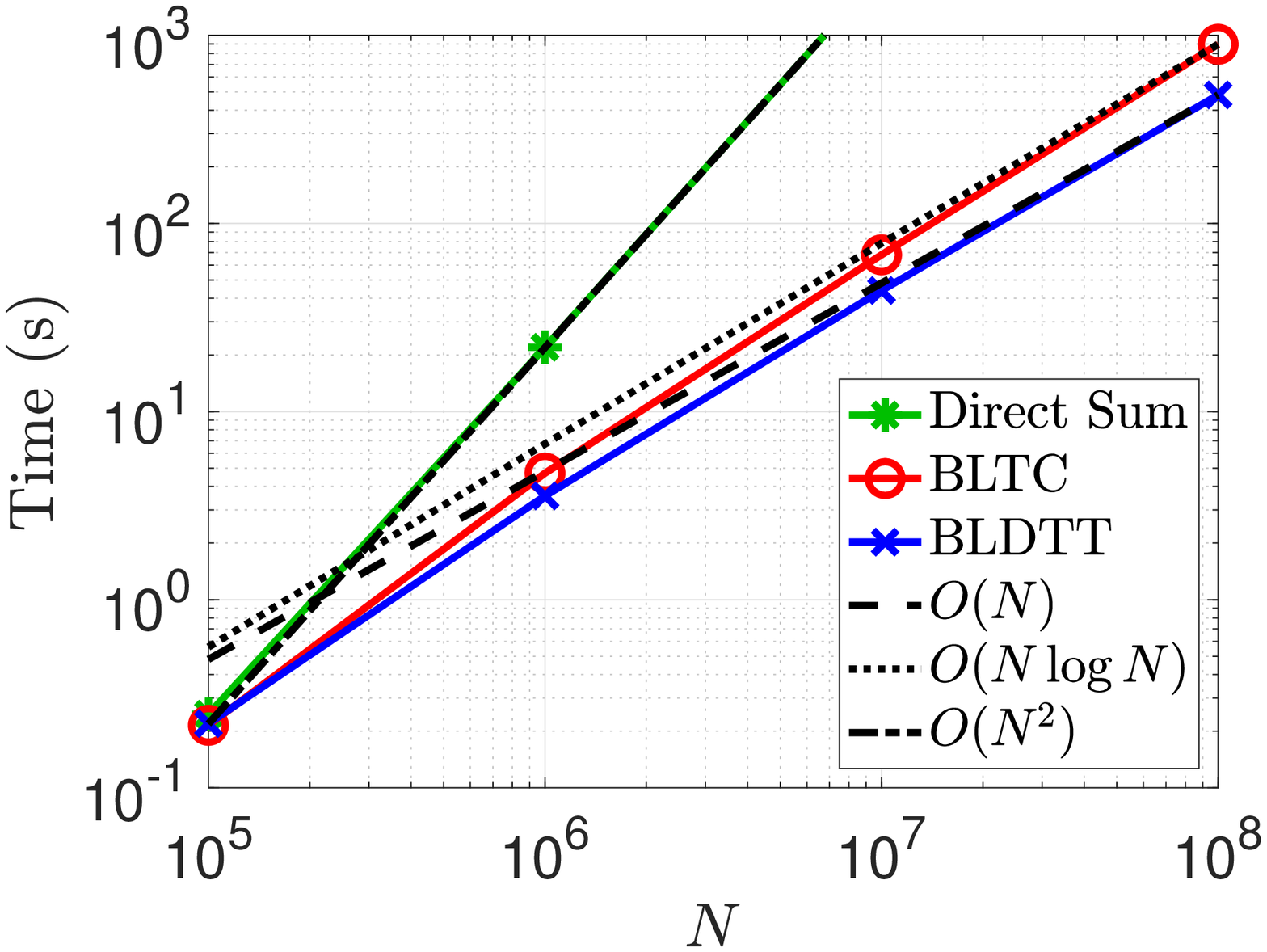} \\
\hspace*{-0.1in}
(a) linear scale &
\hspace*{-0.2in}
(b) logarithmic scale 
\end{tabular}
\caption{
Comparison of BLTC and BLDTT,
compute time (s) versus number of particles $N$=1E5, 1E6, 1E7, 1E8, 
random uniformly distributed particles in $[-1,1]^3$
interacting by the Coulomb kernel,
MAC parameter $\theta = 0.7$, degree $n = 8$ yielding 7-8 digit accuracy,
direct sum ({\color{green}green}),
BLTC ({\color{red}red}), BLDTT ({\color{blue}blue}),  
(a) linear scale,
(b) logarithmic scale,
scaling lines
$O(N^2)$ (dash-dotted),
$O(N\log N)$ (dotted),
$O(N)$ (dashed),
simulations ran on one NVIDIA P100 GPU.}
\label{fig:pc_cc_scaling}
\end{figure}

Figure~\ref{fig:pc_cc_scaling} shows the compute time (s) 
for direct summation (green), BLTC (red), and BLDTT (blue)
with $N$=1E5, 1E6, 1E7, 1E8 random uniformly distributed 
source and target particles 
in the $[-1,1]^3$ cube interacting by the Coulomb kernel.
The BLTC and BLDTT use MAC parameter $\theta=0.7$ and interpolation degree $n=8$,
yielding 7-8 digit accuracy.
Figure~\ref{fig:pc_cc_scaling}(a) is a linear plot,
showing that the BLDTT is about twice as fast as the BLTC,
and both are much faster than direct summation.
Figure~\ref{fig:pc_cc_scaling}(b) is a logarithmic plot
with reference lines scaling as
$O(N)$ (dashed), $O(N\log N)$ (dotted), and $O(N^2)$ (dash-dotted),
showing that as the problem size increases, 
the BLTC has asymptotic $O(N \log N)$ scaling,
while the BLDTT has asymptotic $O(N)$ scaling, as expected.
Table~\ref{table:pc_cc_scaling} 
records the compute time and error;
the asymptotic scaling of the compute time can be quantitatively confirmed,
and
while there is a slight increase in the error with problem size,
the BLDTT error is consistently smaller than the BLTC error.

\begin{table}[htb]
\centering
\begin{tabular}{c c c c c} 
 \hline
 $N$ & BLTC time (s) & BLTC error & BLDTT time (s) & BLDTT error \\ 
 \hline
 1E5 & 2.15E$-$1 & 1.75E$-$8 & 2.19E$-$1 & 1.58E$-$8 \\ 
 1E6 & 4.71E+0 & 1.42E$-$7 & 3.56E+0 & 3.67E$-$8 \\ 
 1E7 & 6.81E+1 & 4.68E$-$7 & 4.40E+1 & 4.12E$-$8 \\ 
 1E8 & 8.96E+2 & 9.23E$-$7 & 4.82E+2 & 4.17E$-$8 \\ 
 \hline
\end{tabular}
\caption{Comparison of BLTC and BLDTT,
number of particles $N=$ 1E5, 1E6, 1E7, 1E8,
random uniformly distributed particles in $[-1,1]^3$
interacting by the Coulomb kernel,
MAC parameter $\theta = 0.7$, degree $n = 8$,
compute time (s) from Figure~\ref{fig:pc_cc_scaling}, $\ell_2$ error,
simulations ran on one NVIDIA P100 GPU.}
\label{table:pc_cc_scaling}
\end{table}


\subsection{GPU acceleration of BLDTT}

In this subsection we compare the BLDTT running on a single NVIDIA P100 GPU to 
running on 8~CPU cores of an Intel Xeon E5-2680v3 processor at 2.50 GHz with MPI parallelization.
We perform the same four calculations above, 
so the errors are the same as those in Table~\ref{table:pc_cc_scaling}.  
Table~\ref{table:gpu-acceleration} gives the compute times,
showing that the BLDTT achieves 30-40$\times$ speedup on the GPU
compared the 8 CPU cores.
The efficiency of the BLDTT running on the GPU
is due to the independent nature of the terms in the barycentric Lagrange approximation,
which allows them to be computed concurrently.
\begin{table}[htb]
\centering
\begin{tabular}{c c c c} 
 \hline
 $N$ & CPU time (s) & GPU time (s) & speedup \\ 
 \hline
 1E5 & 7.84E+0 & 2.19E$-$1 & 35.8 \\ 
 1E6 & 1.45E+2 & 3.56E+0 & 40.7 \\ 
 1E7 & 1.40E+3 & 4.40E+1 & 31.8 \\ 
 1E8 & 1.70E+4 & 4.82E+2 & 35.3 \\ 
 \hline
\end{tabular}
\caption{Comparison of BLDTT running on 8 CPU cores and on one NVIDIA P100 GPU,
number of particles $N=$ 1E5, 1E6, 1E7, 1E8,
random uniformly distributed particles in $[-1,1]^3$
interacting by the Coulomb kernel,
MAC parameter $\theta = 0.7$, degree $n = 8$,
compute time (s),
speedup,
same errors as in Table~\ref{table:pc_cc_scaling}.}
\label{table:gpu-acceleration}
\end{table}

\subsection{Non-uniform particle distributions}
\begin{figure}[htb]
\centering
\begin{tabular}{ccc}
\hspace{-0.2in}
\includegraphics[trim=100 10 100 20, clip, width=0.28\linewidth] {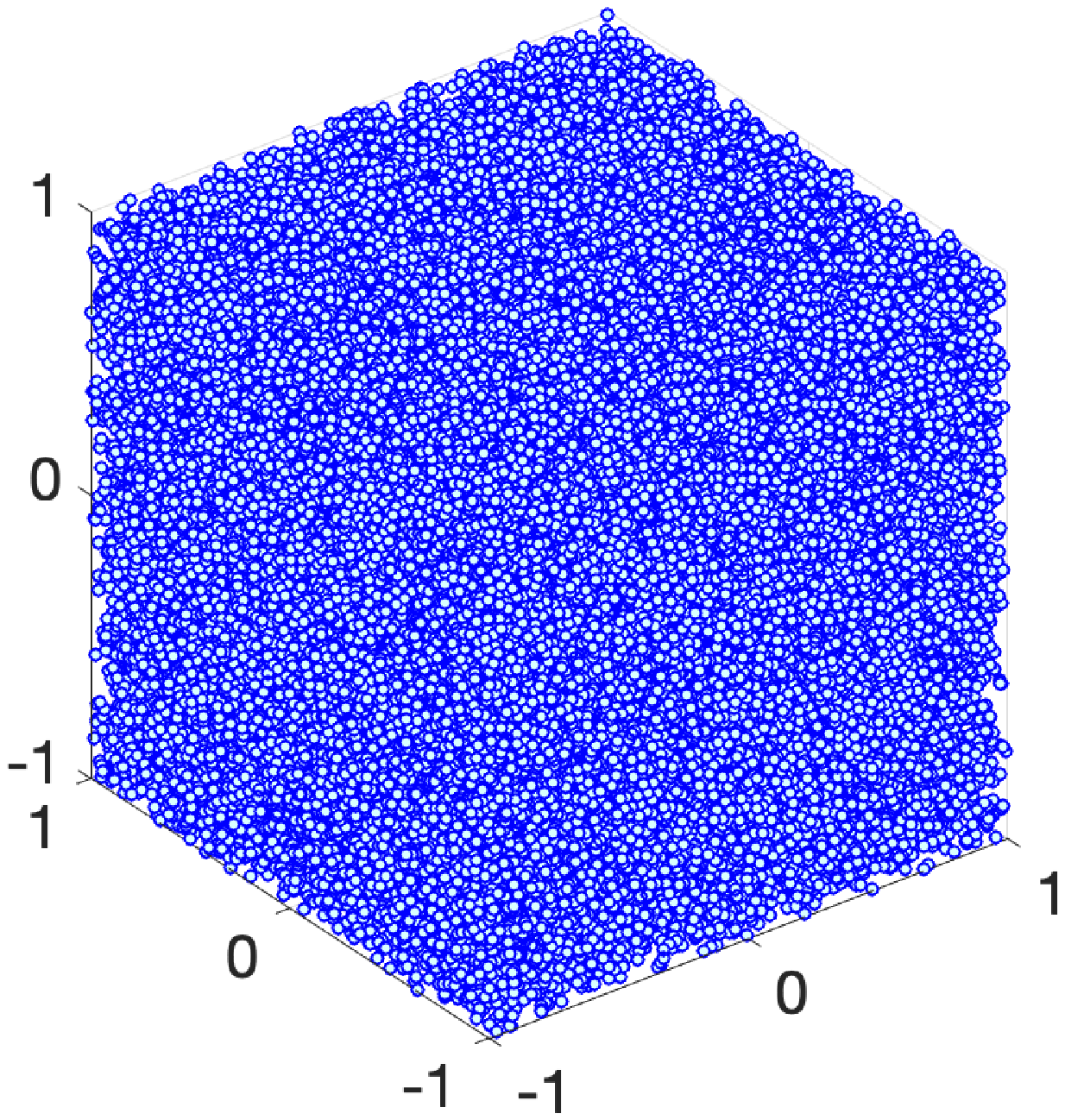}
&
\includegraphics[trim=100 10 95 20, clip,width=0.30\linewidth] {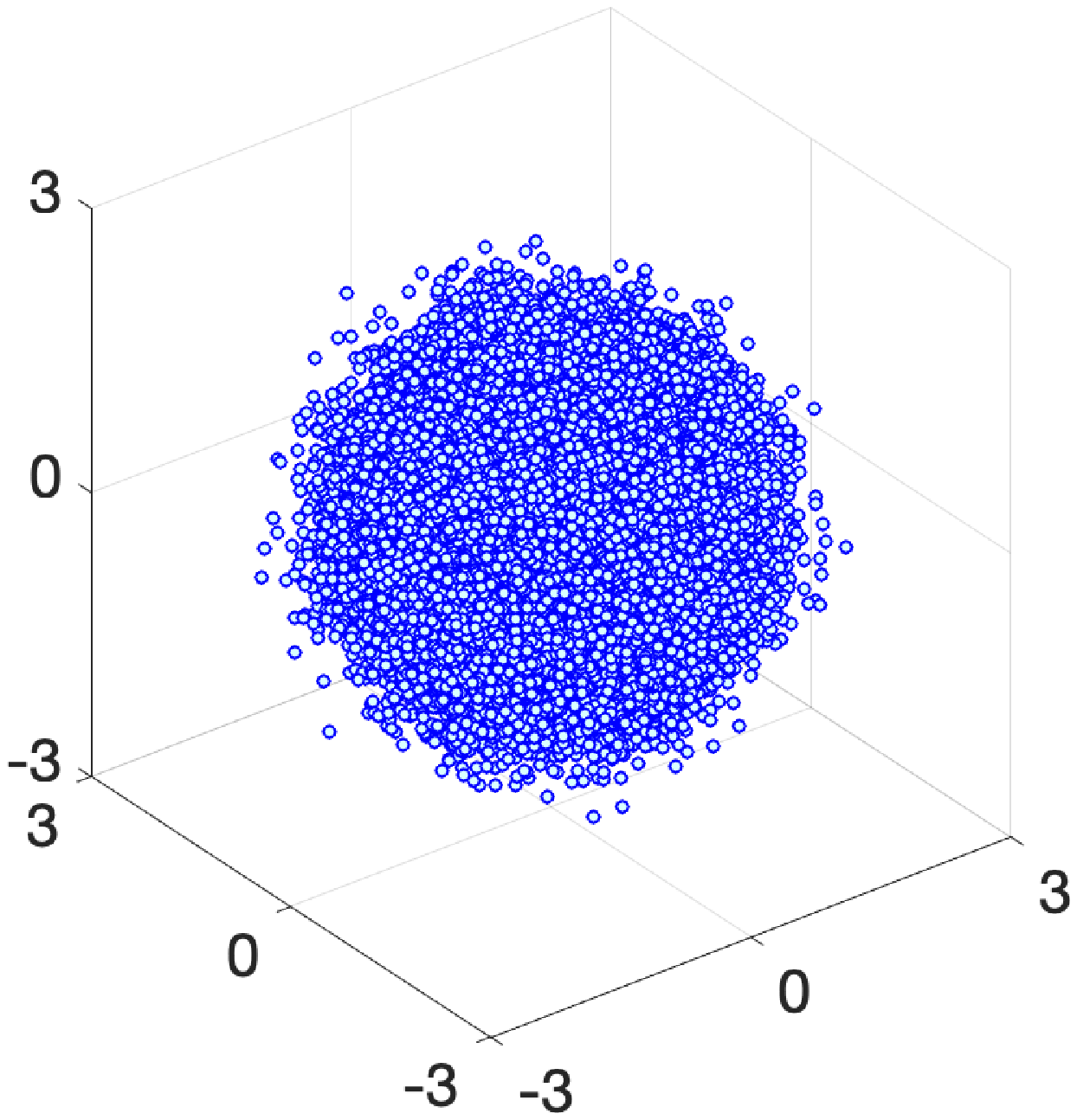} 
&
\includegraphics[trim=80 10 60 20, clip,width=0.34\linewidth] {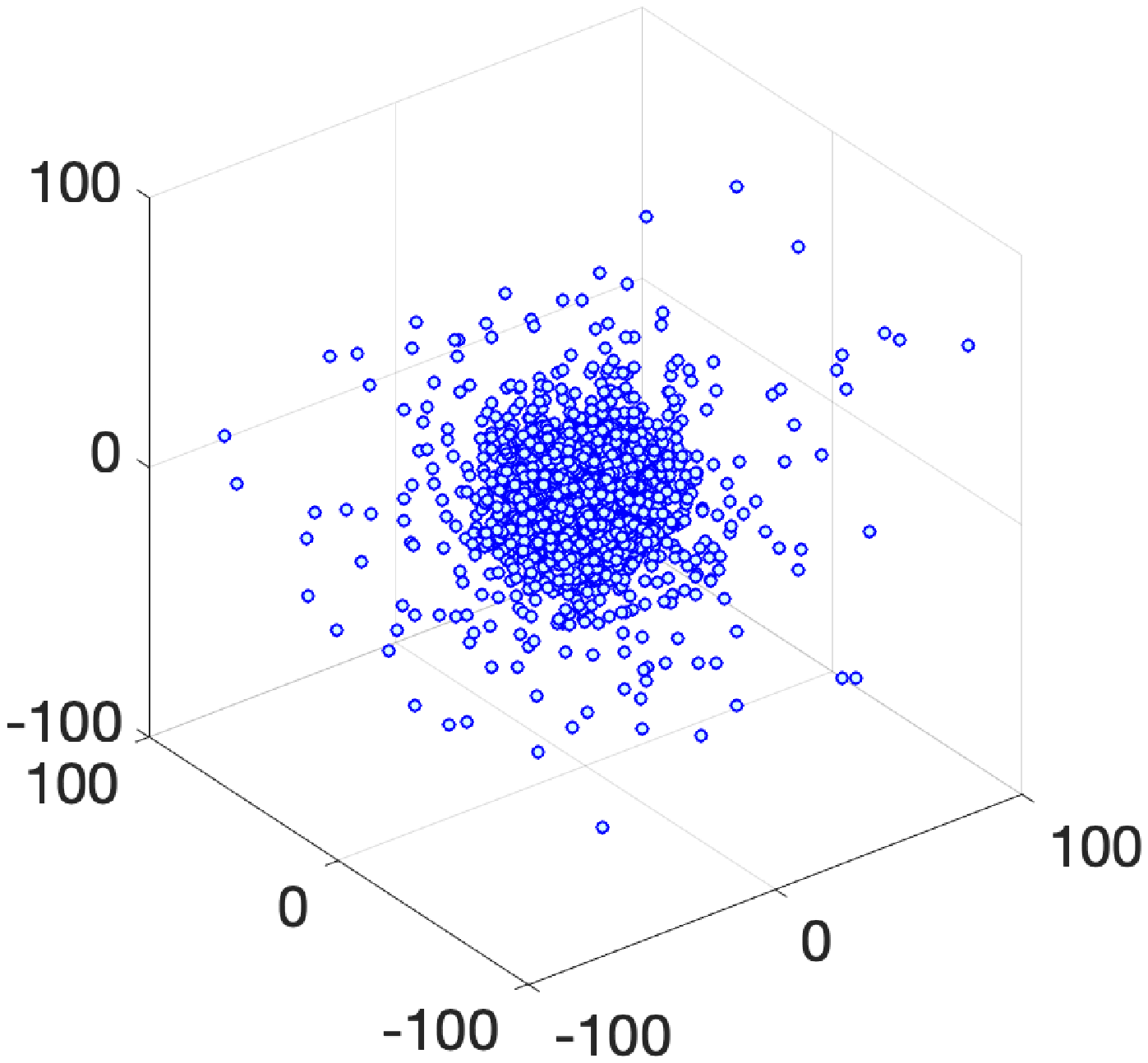}  \\
(a) Random Uniform & (b) Gaussian  & (c) Plummer \\
\end{tabular}
\caption{
Sample random distributions with $N$=4E5 particles, 
(a) uniform, (b) Gaussian, (c) Plummer.}
\label{fig:distribution}
\end{figure}

We investigate the performance of the BLDTT
for three different random particle distributions:
(a) uniform particles in $[-1,1]^3$,
(b) Gaussian particles with radial pdf
$\frac{1}{\sqrt{6\pi}}\exp\left(-r^2/6\right)$,
(c) Plummer particles~\cite{Plummer:1911aa, Dejonghe:1987aa} with radial pdf
$\frac{3}{4\pi}\left(1+r^2\right)^{-5/2}$
and
cutoff at $\pm 100$ in all three Cartesian coordinates.
The charges of the uniform and Gaussian particles are 
uniformly distributed in $[-1,1]$,
while the Plummer particle charges are set to $1/N$, 
where $N$ is the total number of particles.
To give a sense of the structure of the distributions, 
Figure~\ref{fig:distribution} depicts the three distributions
with $N$=4E5 particles.
Compared to the uniform case (a),
the Gaussian and Plummer distributions (b,c)
are more highly concentrated near the origin,
with the Gaussian decaying more rapidly away from the origin
and
the Plummer decaying less rapidly.

\begin{figure}[htb]
\centering
\begin{tabular}{ccc}
\hspace*{-0.1in}
\includegraphics[trim=4 0 15 10, clip, width=0.34\linewidth] {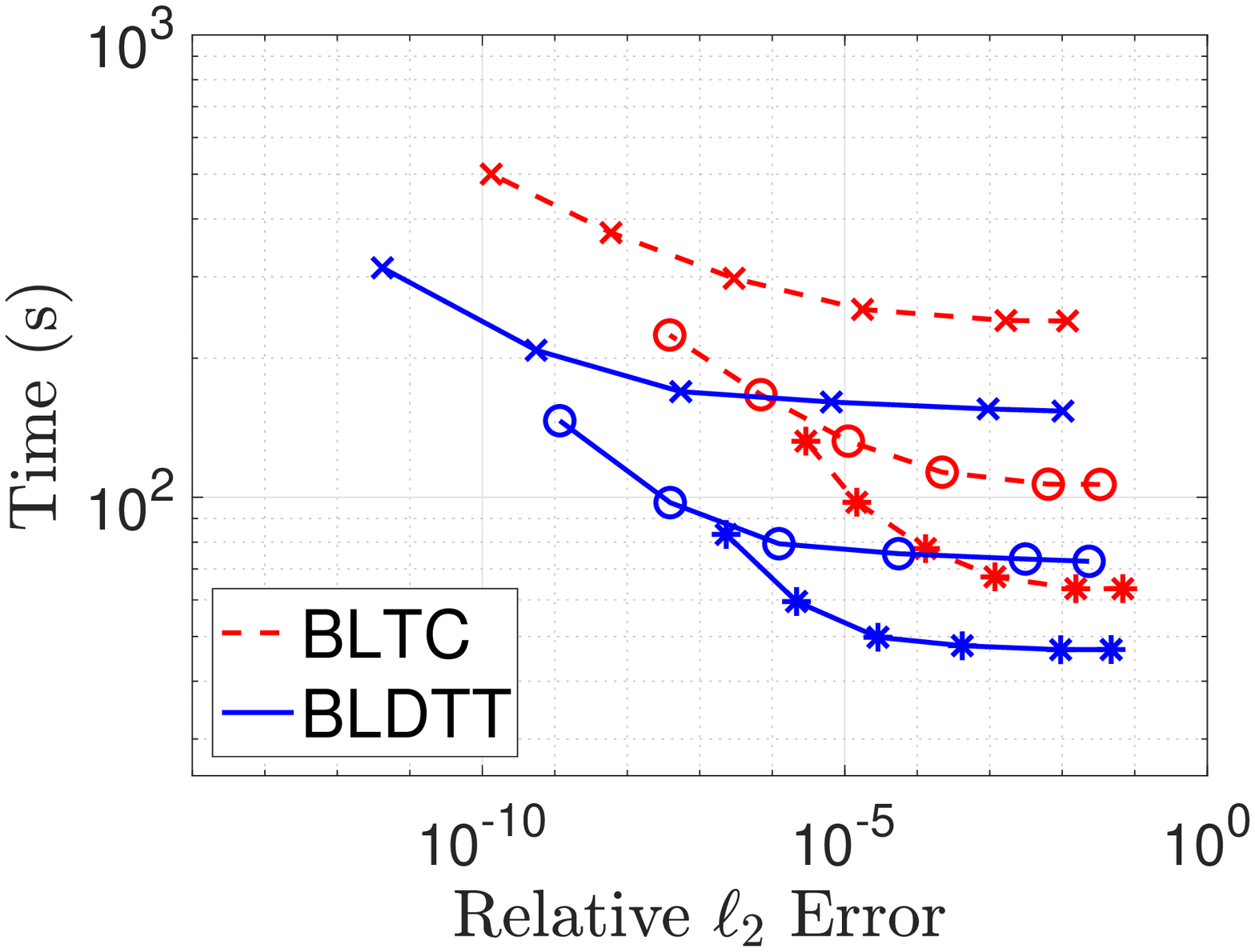} &
\hspace*{-0.2in}
\includegraphics[trim=38 0 15 10, clip, width=0.32\linewidth] {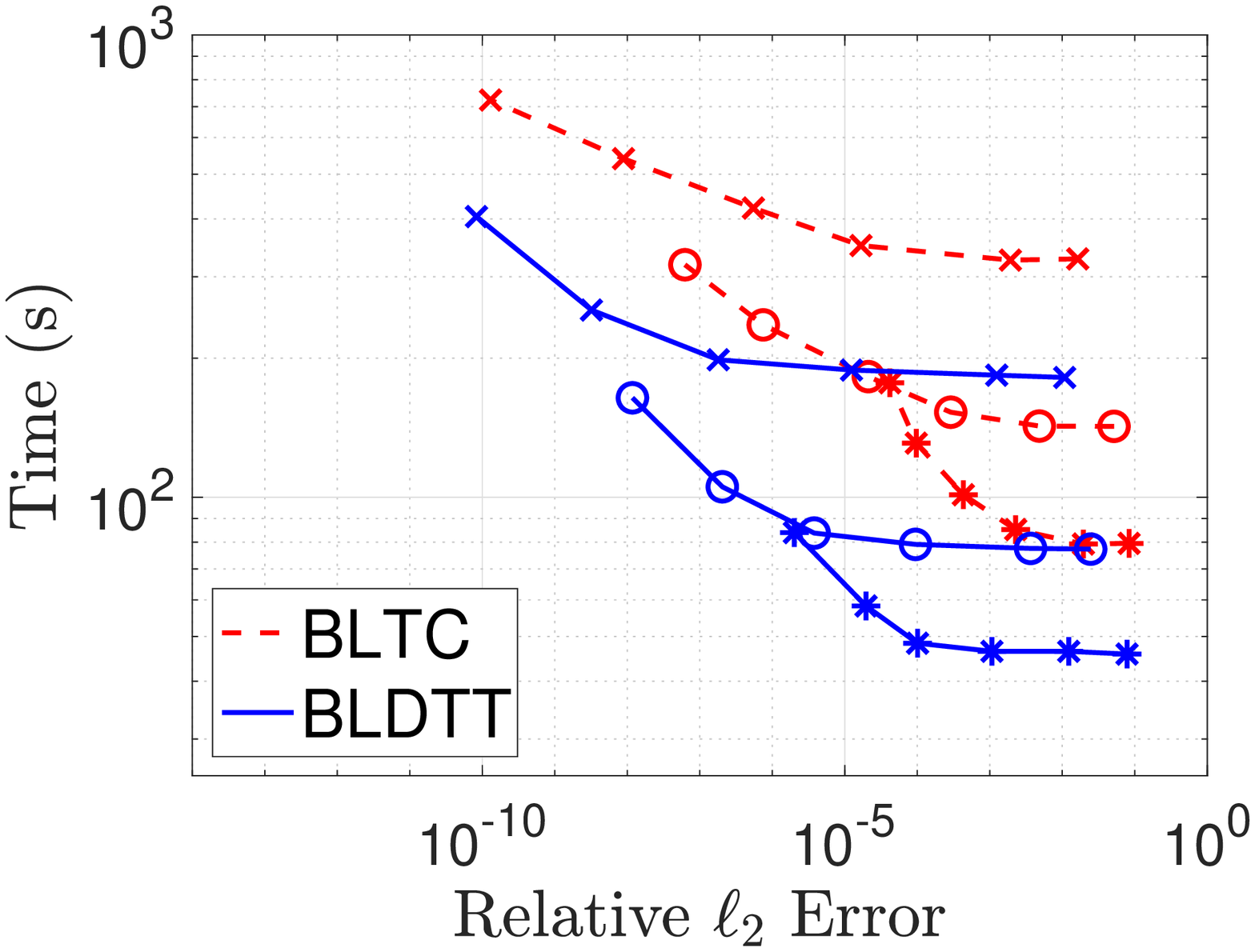} &
\hspace*{-0.2in}
\includegraphics[trim=38 0 15 10, clip, width=0.32\linewidth] {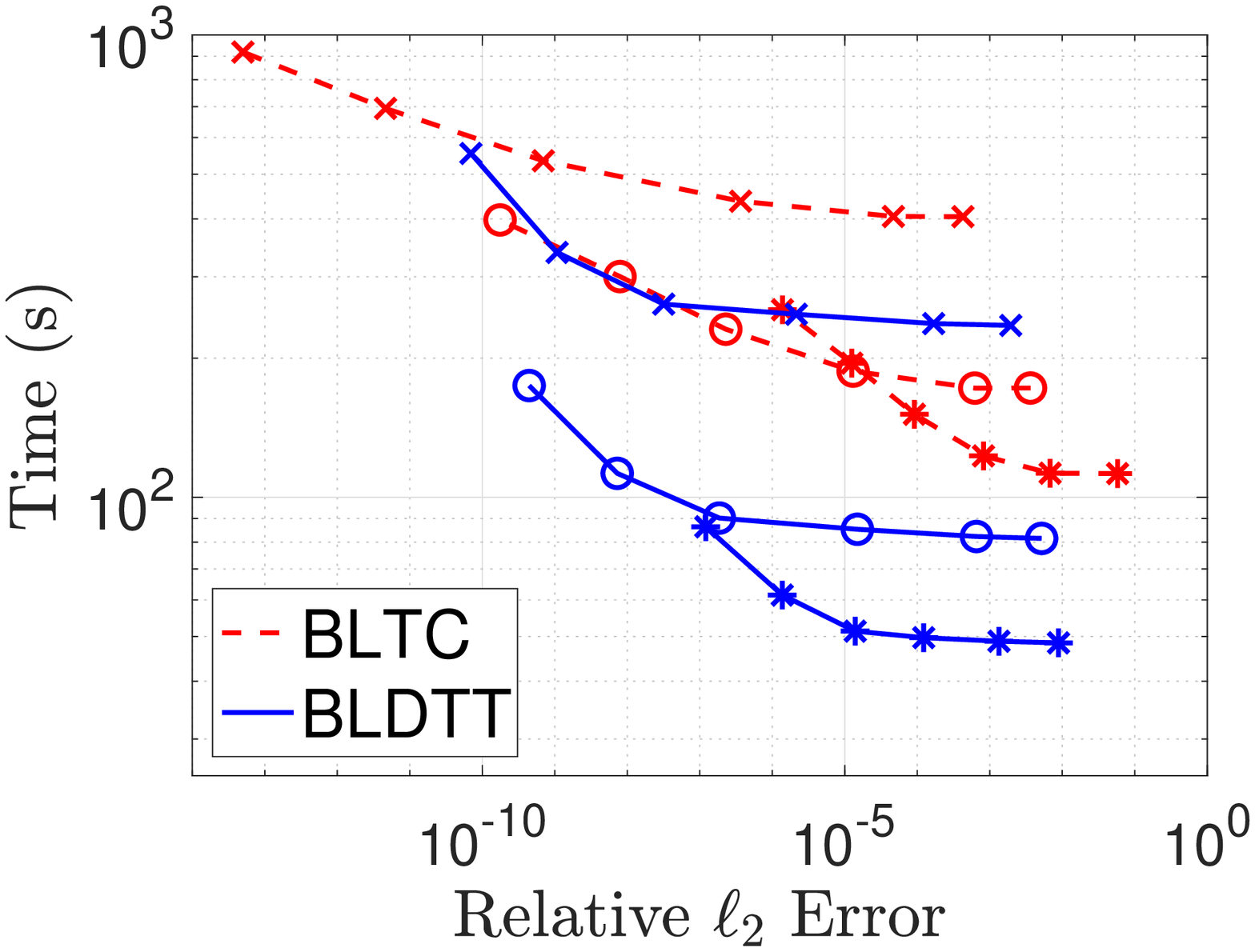} \\
\hspace*{-0.1in}
(a) Random Uniform & 
\hspace*{-0.2in}
(b) Gaussian  & 
\hspace*{-0.2in}
(c) Plummer \\
\end{tabular}
\caption{
Different particle distributions,
compute time (s) versus relative $\ell_2$ error,
$N$=2E7 random particles,
(a) uniform, (b) Gaussian, (c) Plummer, 
BLTC ({\color{red}red}, dashed),
BLDTT ({\color{blue}blue}, solid),
connected curves represent constant MAC 
$\theta$ (0.5 $\times$; 0.7 $\circ$; 0.9 $*$),
interpolation degree $n = 1, 2, 4, 6, 8, 10$ 
increases from right to left on each curve,
simulations ran on one NVIDIA P100 GPU.}
\label{fig:distribution_parameter_sweep}
\end{figure}

Figure~\ref{fig:distribution_parameter_sweep} 
shows the compute time (s) versus relative $\ell_2$ error  
for the BLDTT (blue, solid) and BLTC (red, dashed) 
on these three distributions with $N$=2E7 particles. 
Each connected curve represents constant MAC with
$\theta = 0.5$ ($\times$),
$\theta = 0.7$ ($\circ$), 
$\theta = 0.9$ ($*$),
and
the interpolation degree $n= 1, 2, 4, 6, 8, 10$
increases from right to left on each curve.
For these parameter choices
the errors span the range from 1 digit to 10 digit accuracy. 
Large $\theta$ is more efficient for low accuracy
and
small $\theta$ is more efficient for high accuracy.
The results show that the 
BLDTT has consistently better performance than the BLTC 
and 
is less sensitive to the particle distribution.

\begin{figure}[htb]
\centering
\begin{tabular}{ccc}
\hspace*{-0.1in}
\includegraphics[trim=4 0 15 10, clip, width=0.34\linewidth] {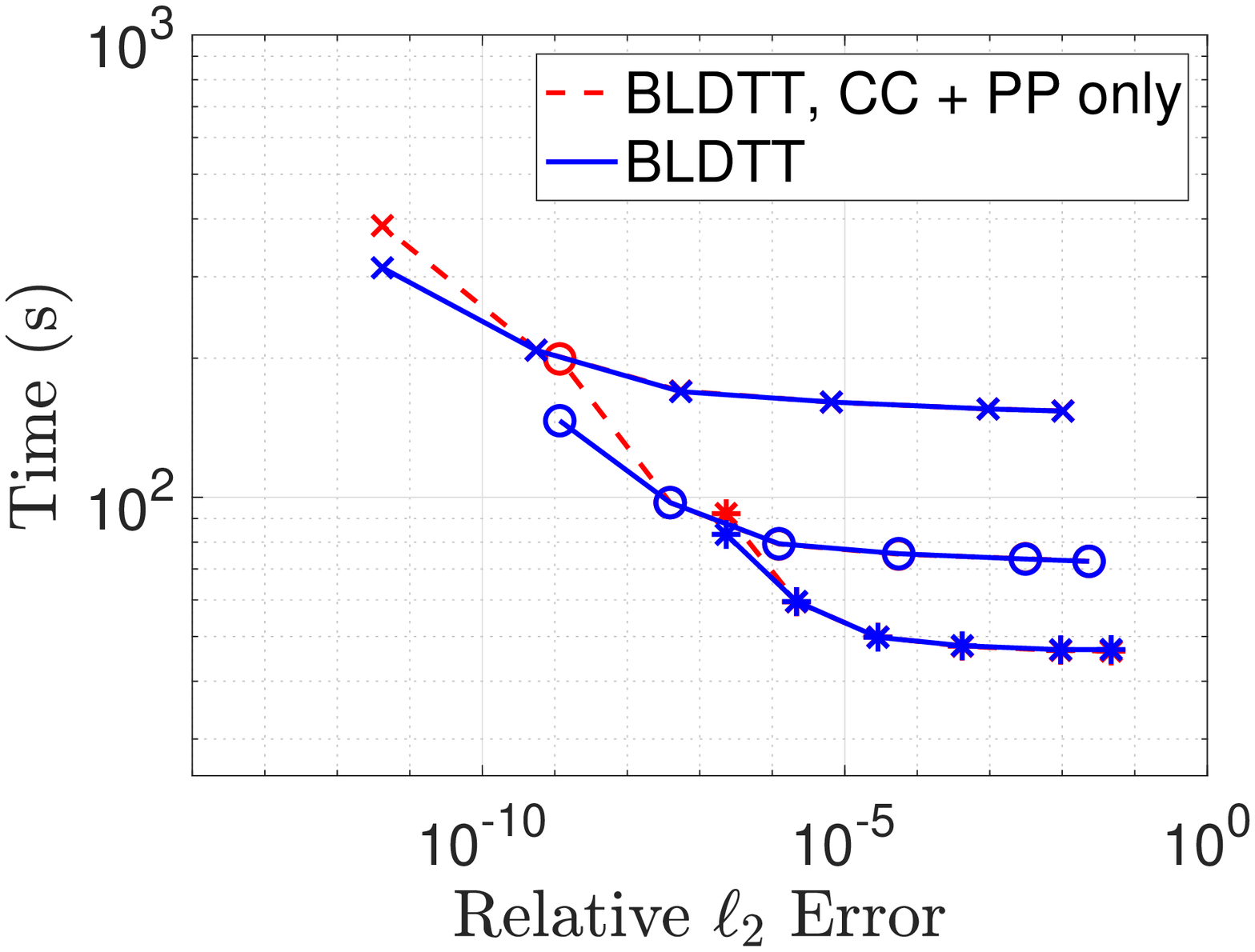} &
\hspace*{-0.2in}
\includegraphics[trim=38 0 15 10, clip, width=0.32\linewidth] {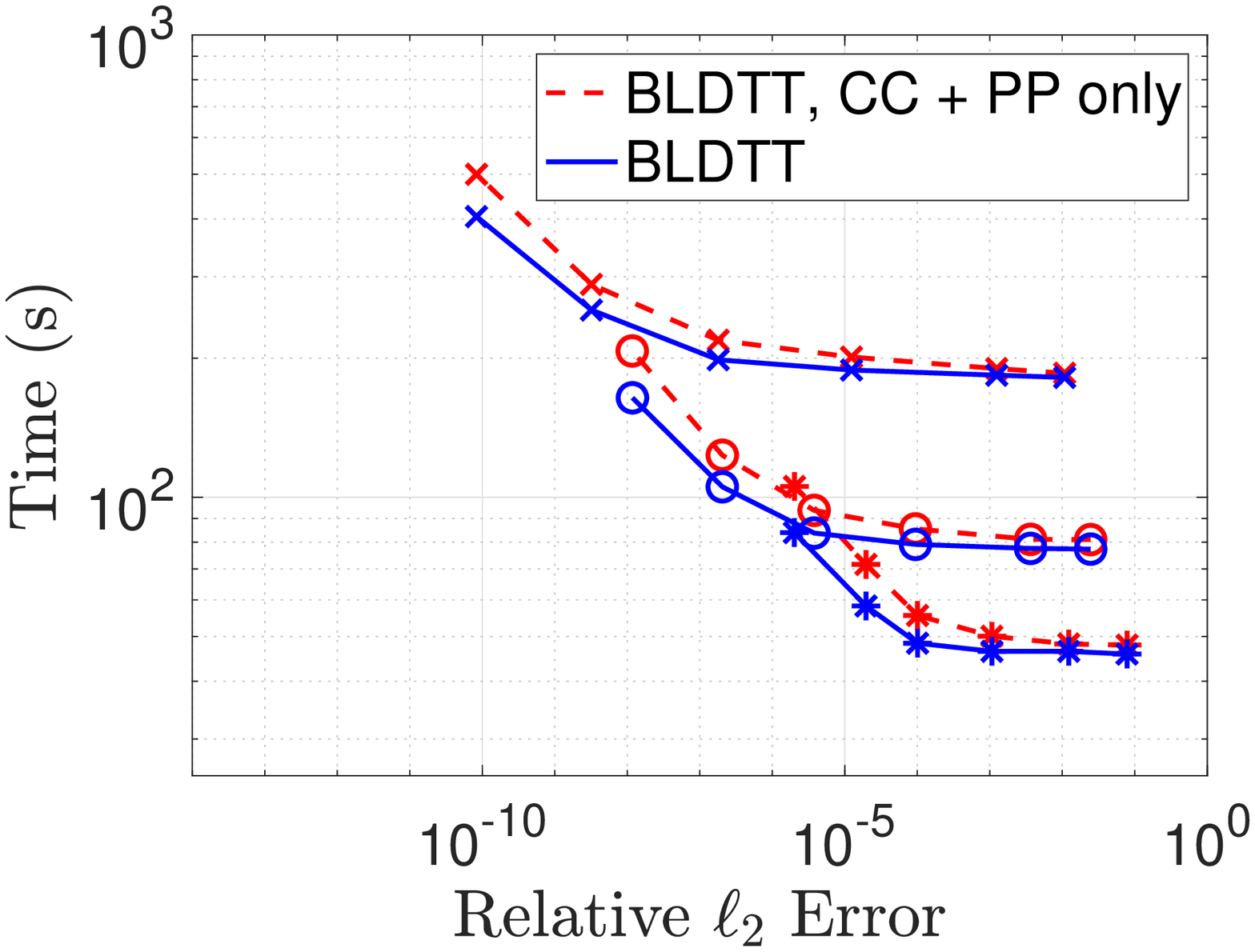} &
\hspace*{-0.2in}
\includegraphics[trim=38 0 15 10, clip, width=0.32\linewidth] {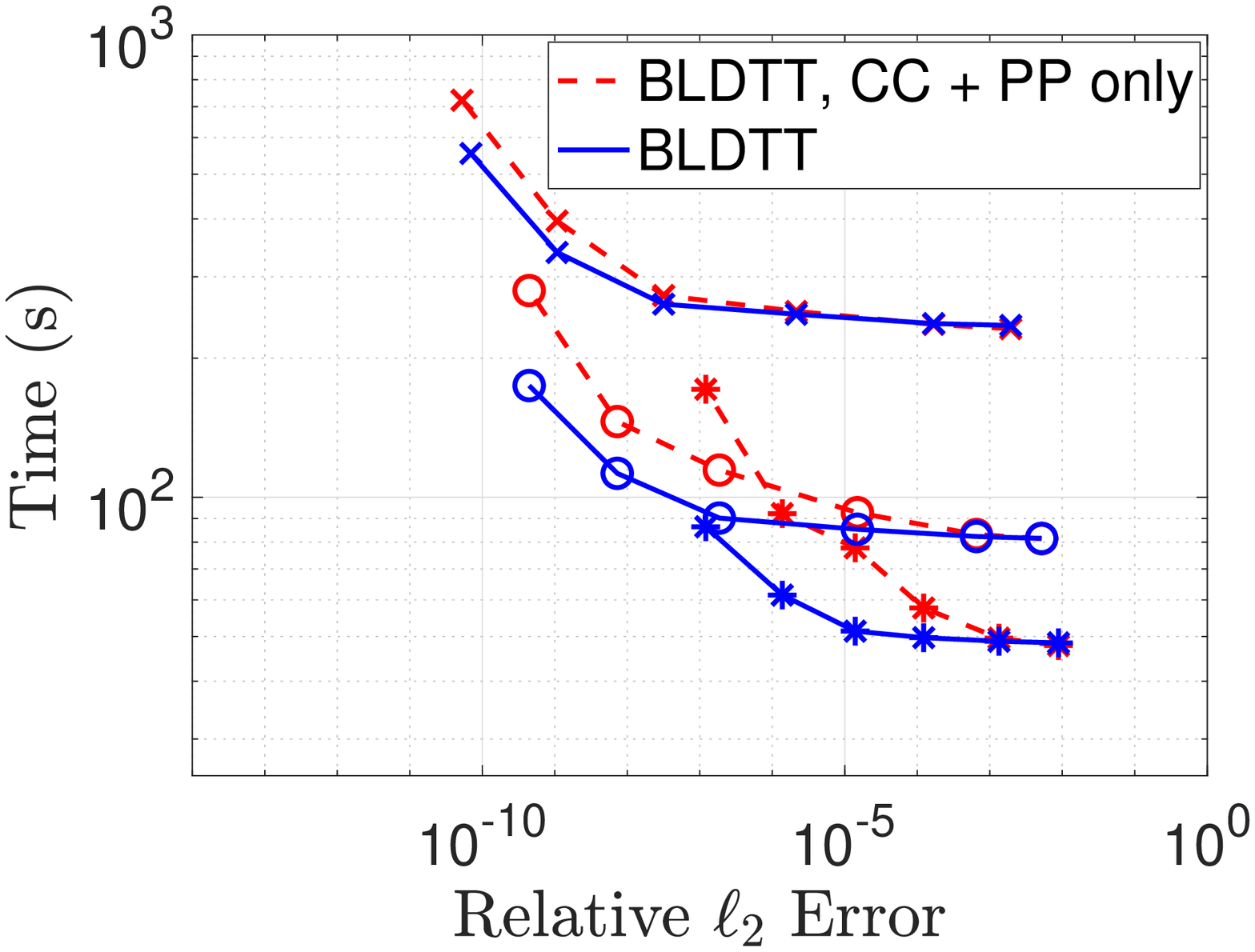} \\
\hspace*{-0.1in}
(a) random uniform & 
\hspace*{-0.2in} 
(b) Gaussian & 
\hspace*{-0.2in}
(c) Plummer \\
\end{tabular}
\caption{
Different particle distributions,
compute time (s) versus relative $\ell_2$ 
$N$=2E7 random particles,
(a) uniform, (b) Gaussian, (c) Plummer, 
BLDTT with only CC and PP interactions ({\color{red}red}, dashed),
BLDTT with PP, PC, CP, and CC interactions ({\color{blue}blue}, solid),
connected curves represent constant MAC 
$\theta$ (0.5 $\times$; 0.7 $\circ$; 0.9 $*$),
interpolation degree $n = 1, 2, 4, 6, 8, 10$ 
increases from right to left on each curve,
simulations ran on one NVIDIA P100 GPU.}
\label{fig:distribution_justcc_parameter_sweep}
\end{figure}

To demonstrate the effect of including PC and CP interactions in the BLDTT, 
Figure~\ref{fig:distribution_justcc_parameter_sweep} 
shows the compute time versus relative $\ell_2$ error 
for the BLDTT  (blue, solid) as presented in this paper 
using PP, PC, CP and CC interactions,
and 
a version of the BLDTT (red, dashed) using only CC and PP interactions.
When only CC and PP interactions are used, 
the interaction between a target cluster and a source cluster
is handled by PP interaction 
if either cluster contains more interpolation points than particles,
whereas the flexibility to choose PC or CP interactions in those cases
improves performance at higher interpolation degree,
especially for the non-uniform particle distributions.

\begin{figure}[htb]
\centering
\begin{tabular}{ccc}
\hspace*{-0.1in}
\includegraphics[trim=4 0 10 4, clip, width=0.35\linewidth] {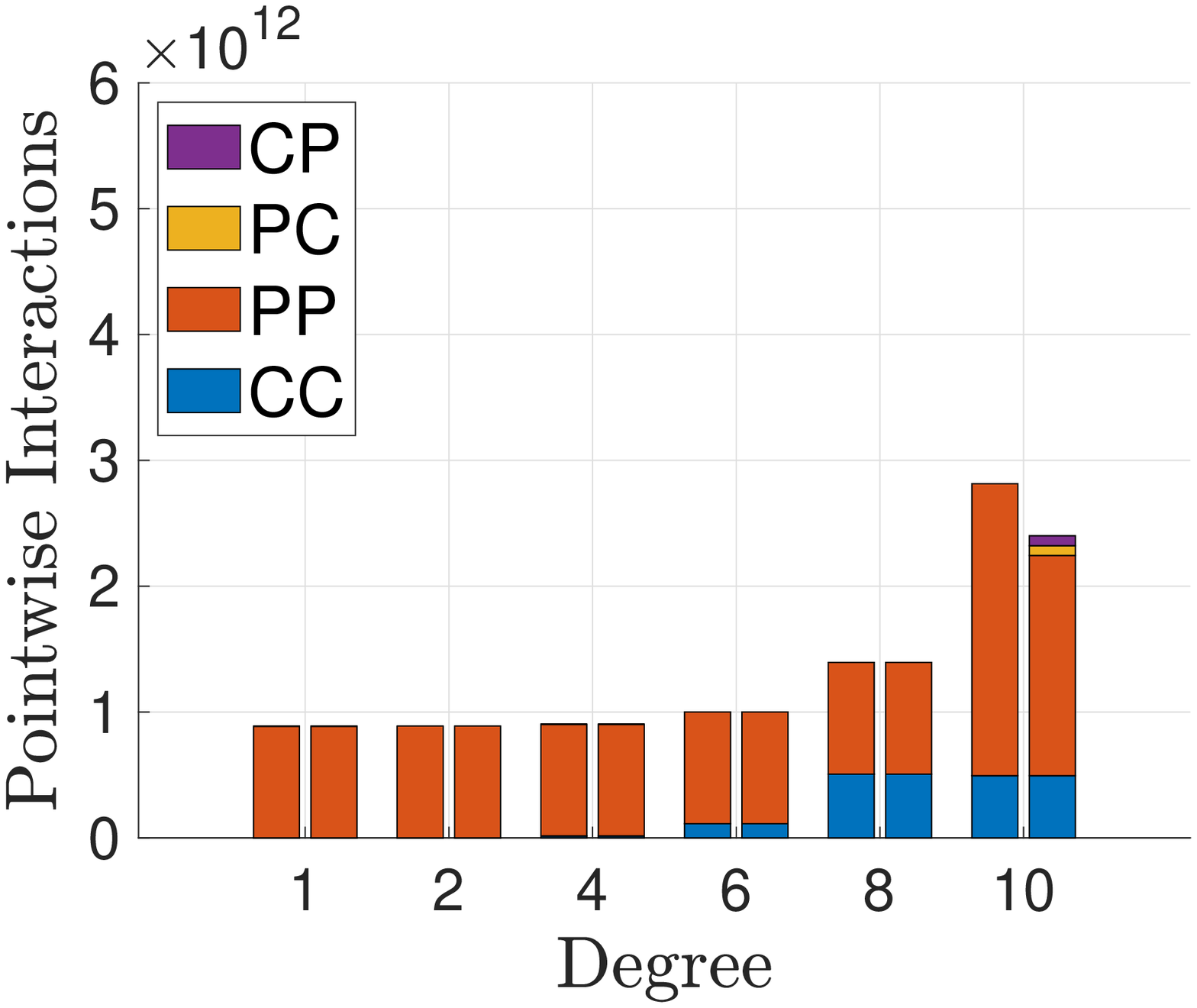} &
\hspace*{-0.2in}
\includegraphics[trim=40 0 10 4, clip, width=0.33\linewidth] {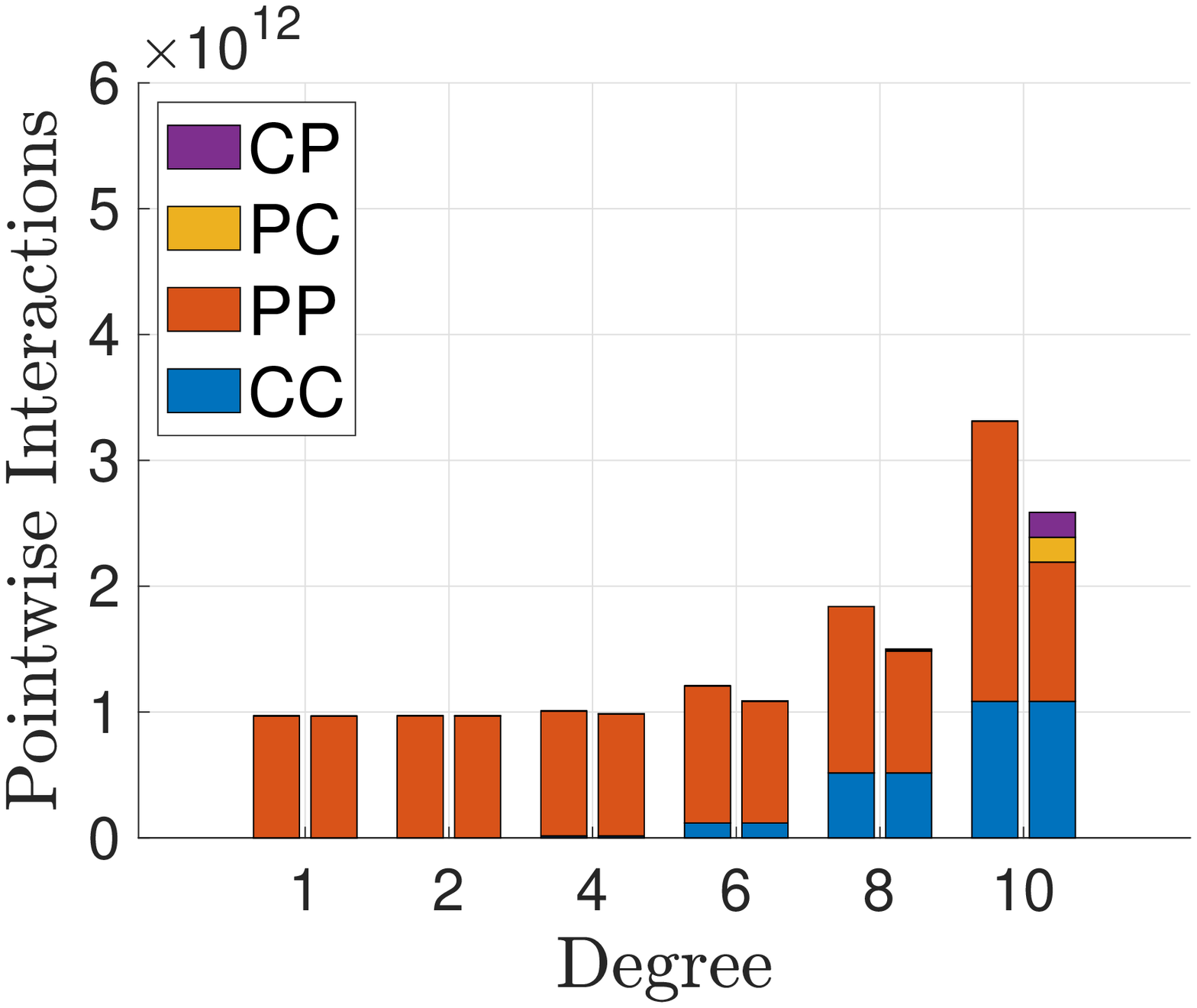} &
\hspace*{-0.2in}
\includegraphics[trim=40 0 10 4, clip, width=0.33\linewidth] {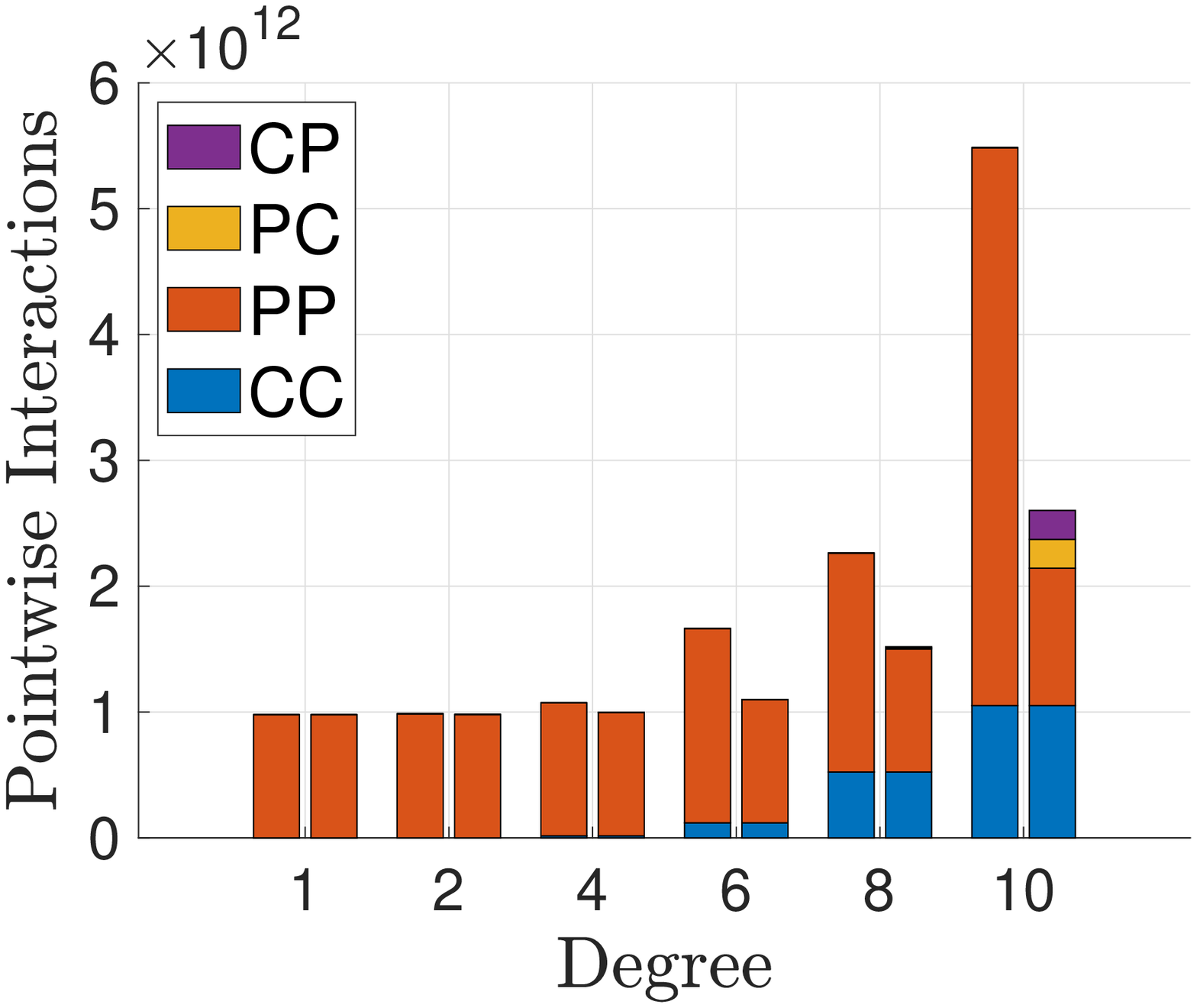} \\
\hspace*{-0.1in}
(a) Random Uniform, $\theta=0.9$ & 
\hspace*{-0.2in}
(b) Gaussian, $\theta=0.9$ & 
\hspace*{-0.2in}
(c) Plummer, $\theta=0.9$ \\
\end{tabular}
\caption{
Different particle distributions,
number of pointwise interactions (kernel evaluations $G(\bx,\by)$),
$N$=2E7 random particles,
(a) uniform, (b) Gaussian, (c) Plummer,
MAC $\theta = 0.9$, interpolation degree $n = 1, 2, 4, 6, 8, 10$,
each pair of bars shows
BLDTT with CC and PP only (left)
and
BLDTT with PP, PC, CP, CC (right), 
direct sum calculation would use 4E14 PP interactions,
simulations ran on one NVIDIA P100 GPU.}
\label{fig:distribution_justcc_interactions_bar}
\end{figure}

To further understand the effect of including PC and CP interactions,
next we compare the number of pointwise interactions 
used by the two versions of the BLDTT,
where by pointwise interaction we mean 
one evaluation of the kernel $G(\bx,\by)$.
Results are shown for MAC $\theta=0.9$ 
and
interpolation degree $n= 1,2,4,6,8,10$,
for the same three random distributions with $N$=2E7 particles as above.
Figure~\ref{fig:distribution_justcc_interactions_bar}
displays results for the four types of interactions in stacked bars,
CC (blue), PP (orange), PC (yellow), CP (purple),
from bottom to top,
where the left bar in each pair is the BLDTT with CC and PP interactions only,
and 
the right bar is the BLDTT with PP, PC, CP, and CC interactions.
In this case a direct sum calculation would use 4E14 PP interactions,
while the BLDTT calculations use less than 6E12 interactions.
The results show that for high degree,
introducing PC and CP interactions into the BLDTT 
significantly reduces the number of PP interactions, 
replacing them with a much smaller number of PC and CP interactions,
and
this effect is more prominent for the nonuniform particle distributions.

\subsection{Non-cubic particle domains}
\begin{figure}[htb]
\centering
\begin{tabular}{ccc}
\hspace{-0.1in}
\includegraphics[trim=100 10 80 20, clip, width=0.33\linewidth] {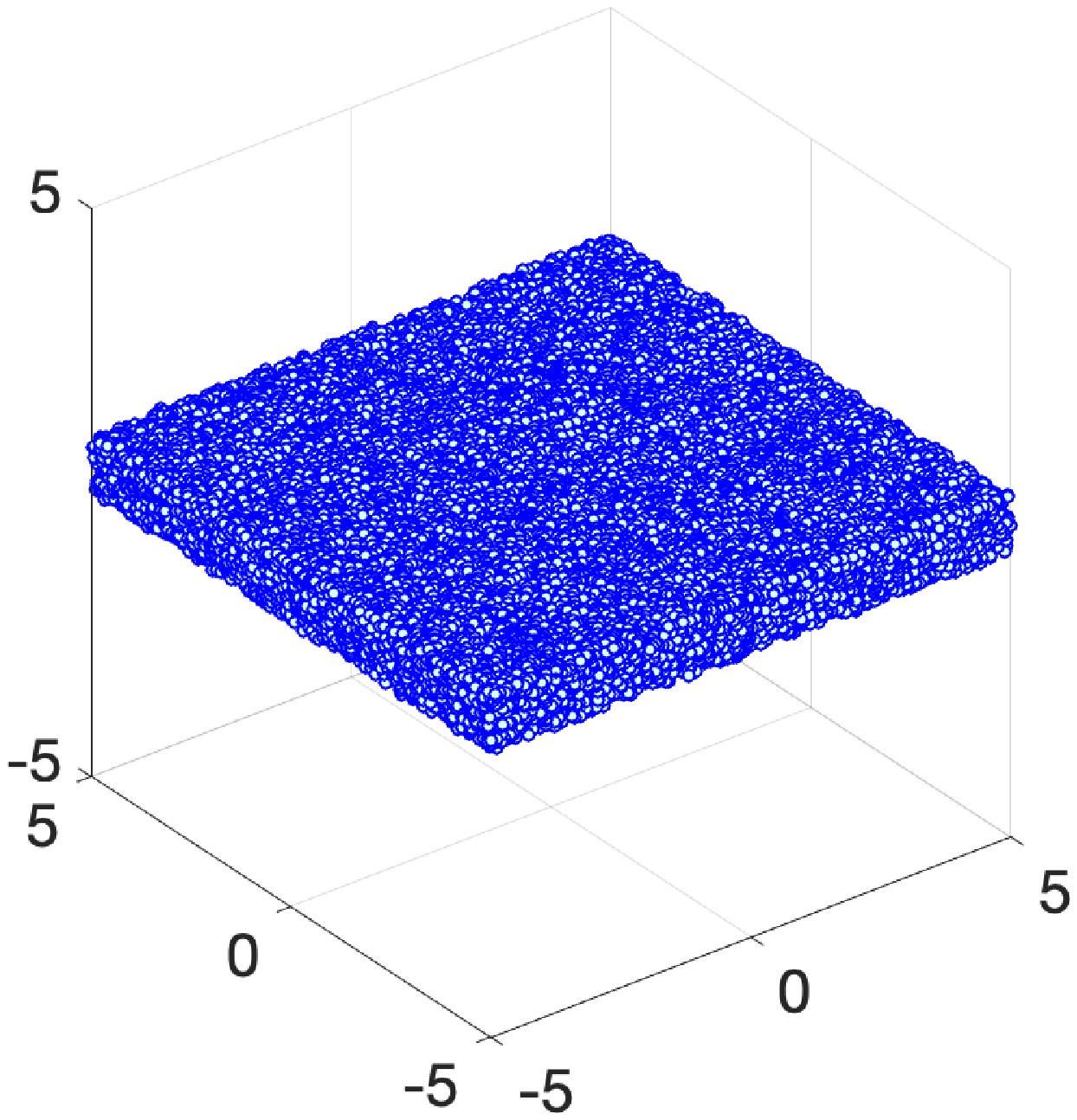}
&
\hspace*{-0.2in}
\includegraphics[trim=100 10 80 20, clip,width=0.33\linewidth] {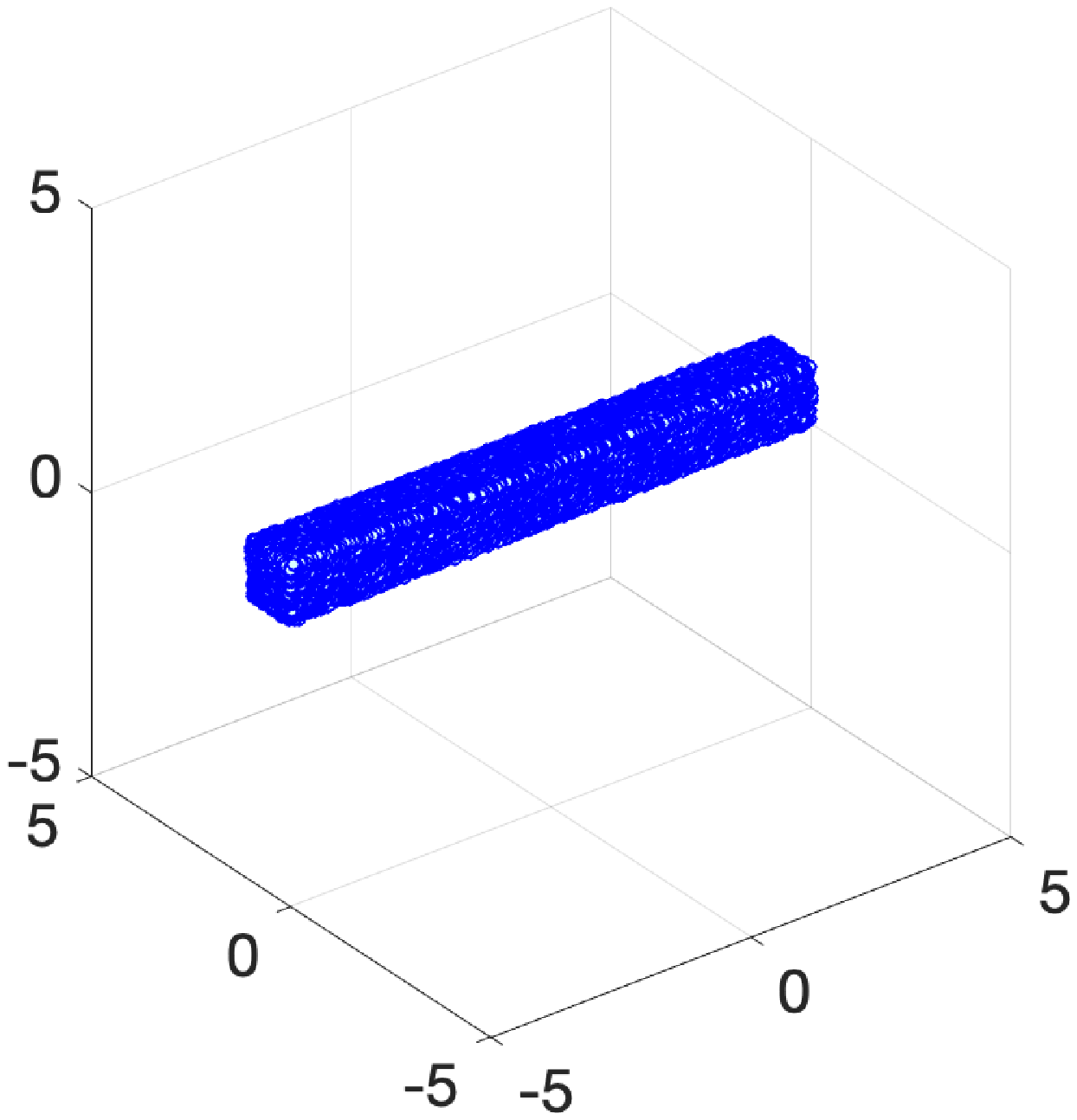} 
&
\hspace*{-0.2in}
\includegraphics[trim=100 10 80 20, clip,width=0.33\linewidth] {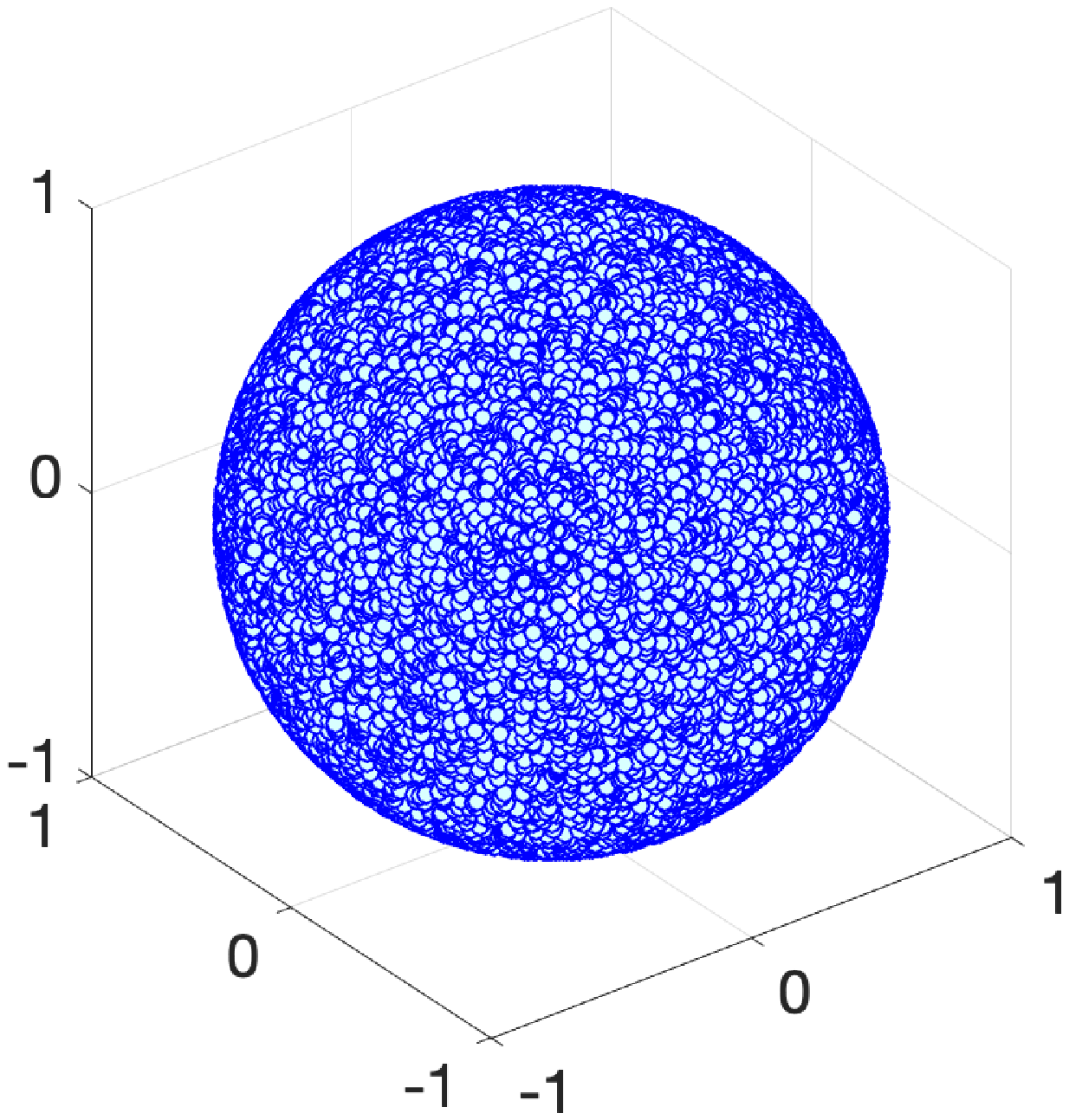}  \\
\hspace{-0.1in}
(a) 2D Slab & 
\hspace*{-0.2in}
(b) 1D Slab  & 
\hspace*{-0.2in}
(c) Spherical Shell \\
\end{tabular}
\caption{
Non-cubic domains, $N$=4E5 random uniformly distributed particles, 
(a) thin slab of dimensions $1 \times 10 \times 10$, 
(b) square rod of dimensions $1 \times 1 \times 10$, 
(c) spherical surface of radius 1.}
\label{fig:domains}
\end{figure}

We demonstrate here the performance of the BLDTT on three examples
with non-cubic particle domains depicted in Figure~\ref{fig:domains}:
(a) thin slab of dimensions $1 \times 10 \times 10$,
(b) square rod of dimensions $1 \times 1 \times 10$,
and (c) spherical surface of radius $1$.
In all cases the particles are random uniformly distributed.

\begin{figure}[htb]
\centering
\begin{tabular}{ccc}
\hspace*{-0.1in}
\includegraphics[trim=4 0 15 10, clip, width=0.34\linewidth] {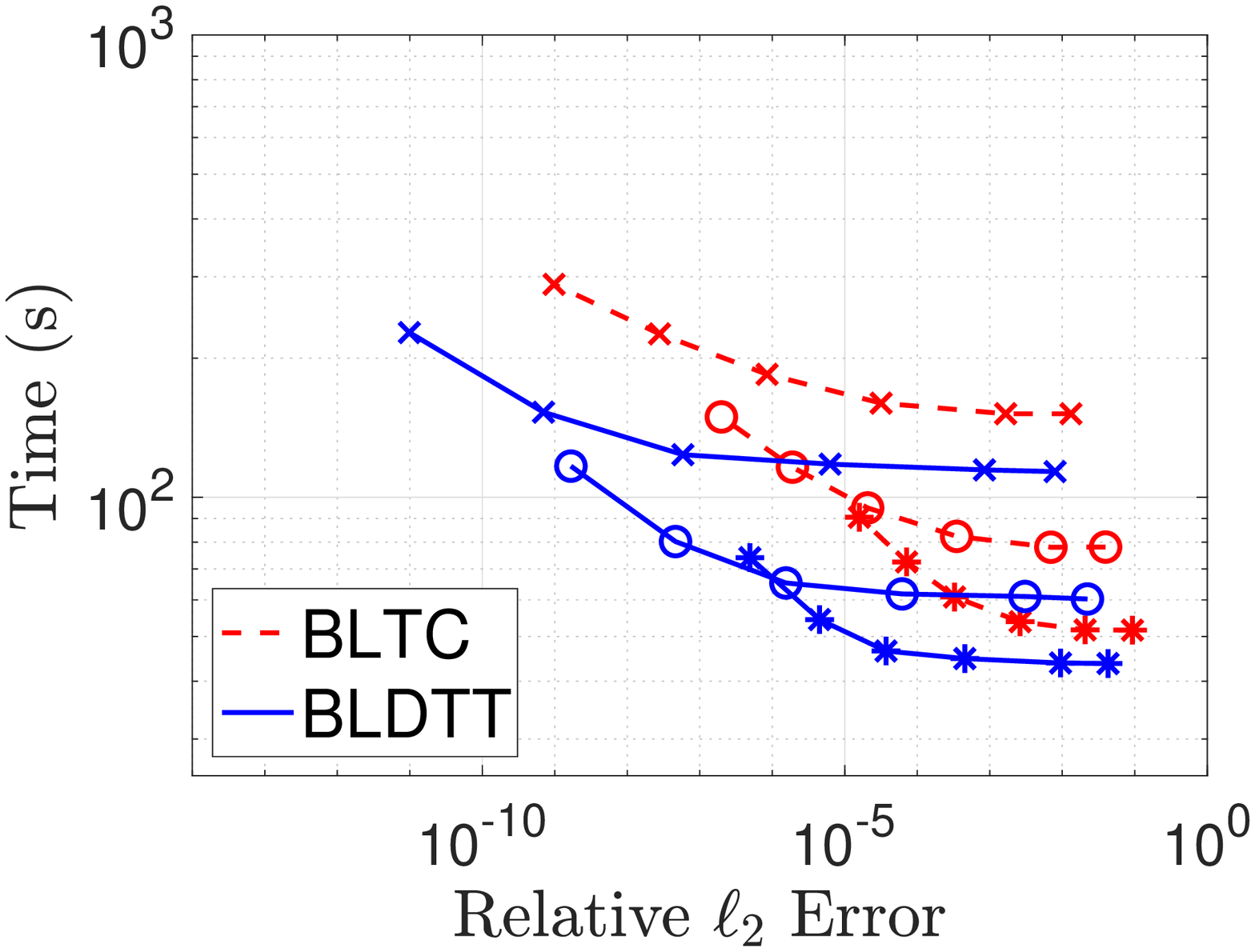} &
\hspace*{-0.2in}
\includegraphics[trim=38 0 15 10, clip, width=0.32\linewidth] {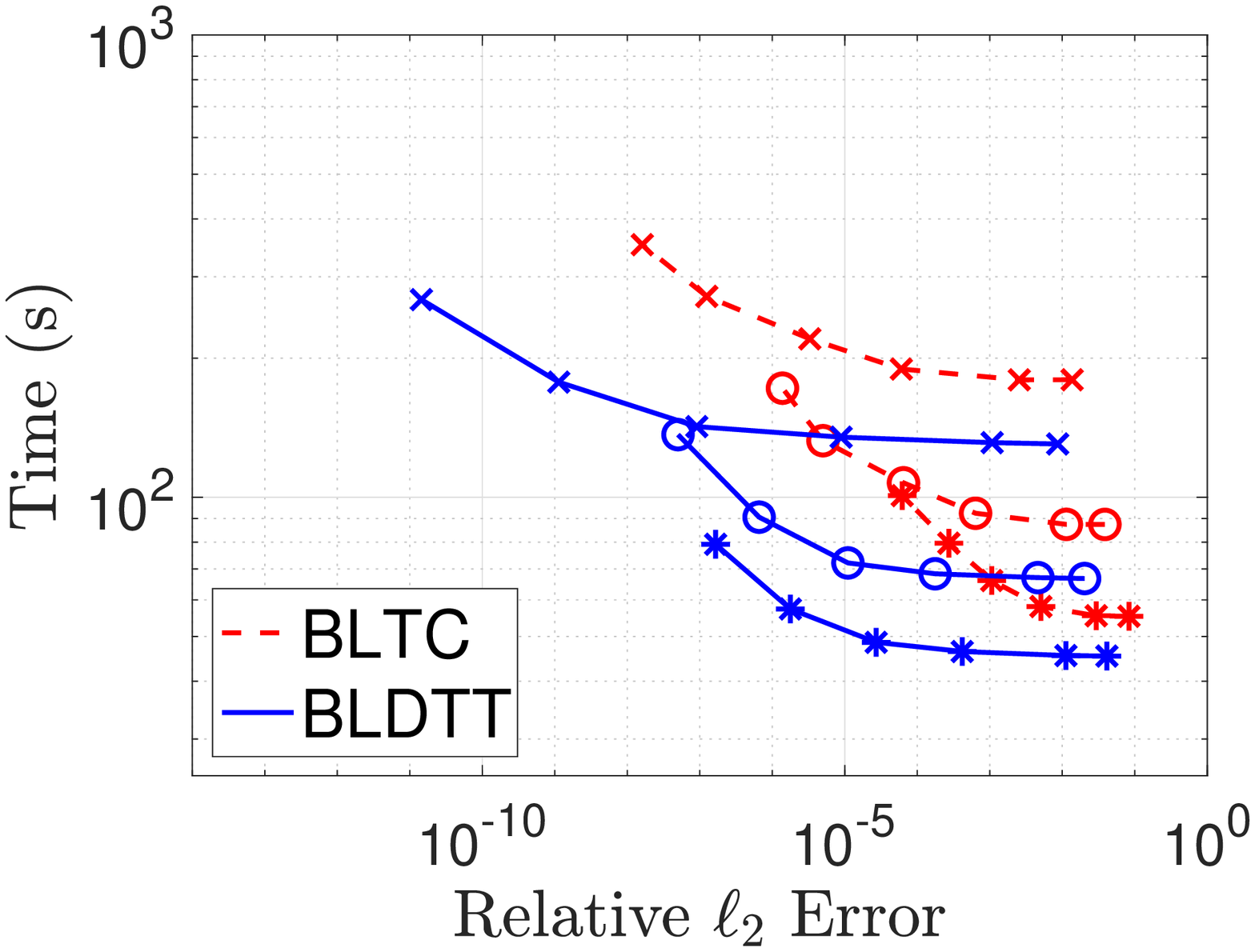} &
\hspace*{-0.2in}
\includegraphics[trim=38 0 15 10, clip, width=0.32\linewidth] {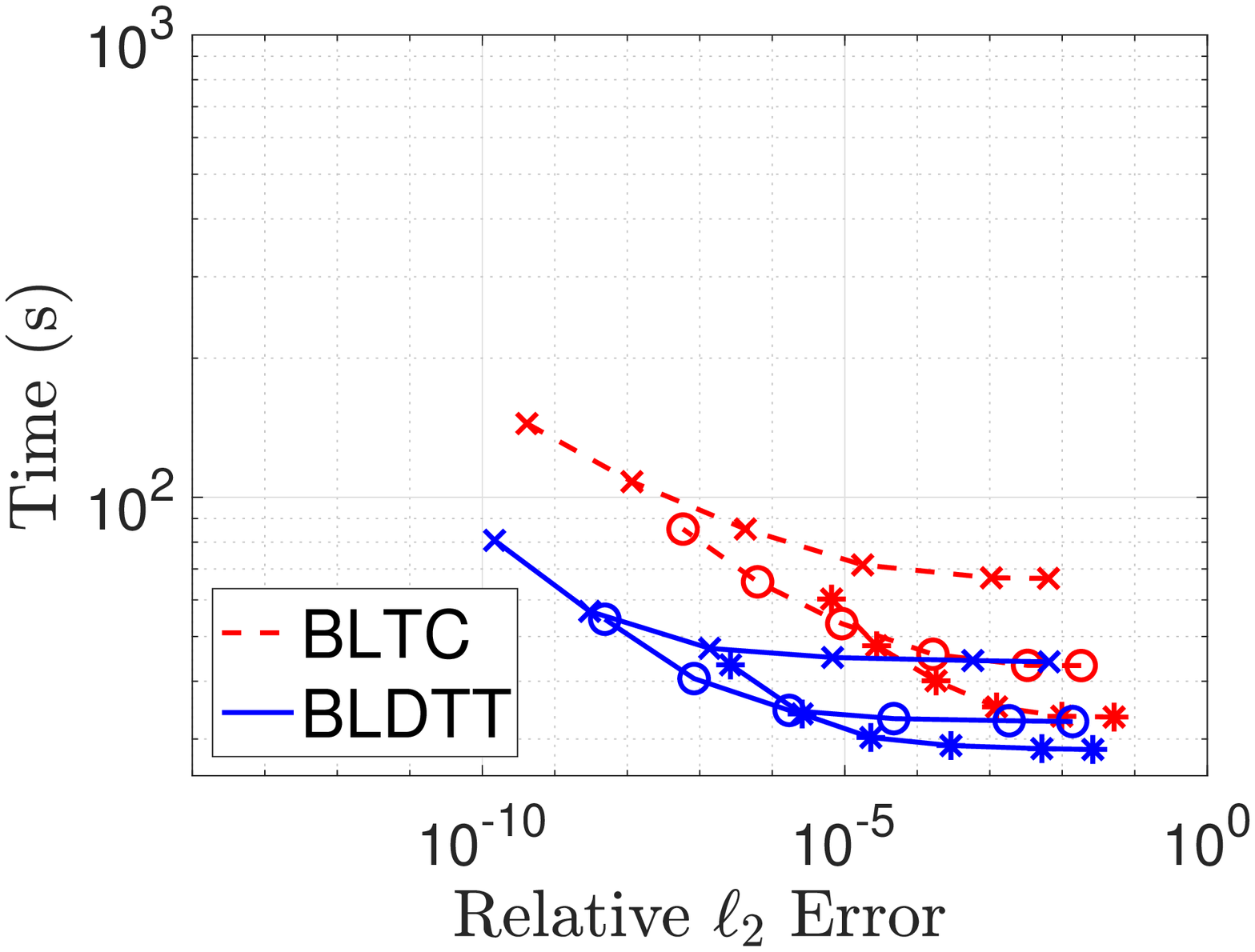} \\
\hspace{-0.1in}
(a) thin slab & 
\hspace*{-0.2in}
(b) square rod  & 
\hspace*{-0.2in}
(c) sphere surface \\
\end{tabular}
\caption{
Non-cubic domains,
$N$=2E7 random uniformly distributed particles,
(a) thin slab of dimensions $1 \times 10 \times 10$, 
(b) square rod of dimensions $1 \times 1 \times 10$, 
(c) sphere surface of radius 1,
compute time (s) versus relative $\ell_2$ error, 
BLTC ({\color{red}red}, dashed),
BLDTT ({\color{blue}blue}, solid),
connected curves represent constant MAC $\theta$ (0.5 $\times$; 0.7 $\circ$; 0.9 $*$),
interpolation degree $n=1,2,4,6,8,10$ increases from right to left on each curve,
simulations ran on one NVIDIA P100 GPU.}
\label{fig:geometry_parameter_sweep}
\end{figure}

Figure~\ref{fig:geometry_parameter_sweep} 
shows the compute time (s) versus the relative $\ell_2$ error  
for the BLDTT (blue, solid) and BLTC (red, dashed) 
on these three examples with $N$=2E7 particles,
using MAC $\theta = 0.5, 0.7, 0.9$ and interpolation degree $n=1,2,4,6,8, 10$.
The results show that the BLDTT has consistently better performance than the BLTC.
Compared to the cubic domain results in Figure~\ref{fig:distribution_parameter_sweep}(a),
the BLDTT achieves similar levels of error 
and
runs somewhat faster for the non-cubic domains.
Heuristically,
the BLDTT run time depends on the complexity of the tree;
in particular,
the tree is an oct-tree for the cubic domain,
close to a quad-tree for the thin slab and sphere surface,
and
close to a binary tree for the square rod.
The results indicate that BLDTT automatically adapts to the complexity
of the tree without requiring explicit reprogramming.

\subsection{Unequal targets and sources}
To demonstrate the performance of the BLDTT with unequal target and source particles, 
we consider a cluster-particle variant of the BLTC 
for comparison~\cite{Boateng:2010aa,Boateng:2013aa}. 
The CP-BLTC builds a tree on the $M$~targets and a set of batches on the $N$~sources,
and
rather than using PP and PC interactions, 
it uses PP and CP interactions.
Instead of an $O(N\log N)$ upward pass to compute proxy charges, 
the CP-BLTC has an $O(M\log M)$ downward pass to interpolate proxy potentials to targets.
While the compute phase of the BLTC is $O(M\log N)$, 
the compute phase of the CP-BLTC is $O(N\log M)$.
Since the compute phase is in general the most expensive part of the algorithm,
we expect the BLTC to perform better than the CP-BLTC when $N > M$, and vice versa.

\begin{figure}[htb]
\centering
\begin{tabular}{cc}
\hspace*{-0.1in}
\includegraphics[trim=4 0 15 10, clip, width=0.50\linewidth] {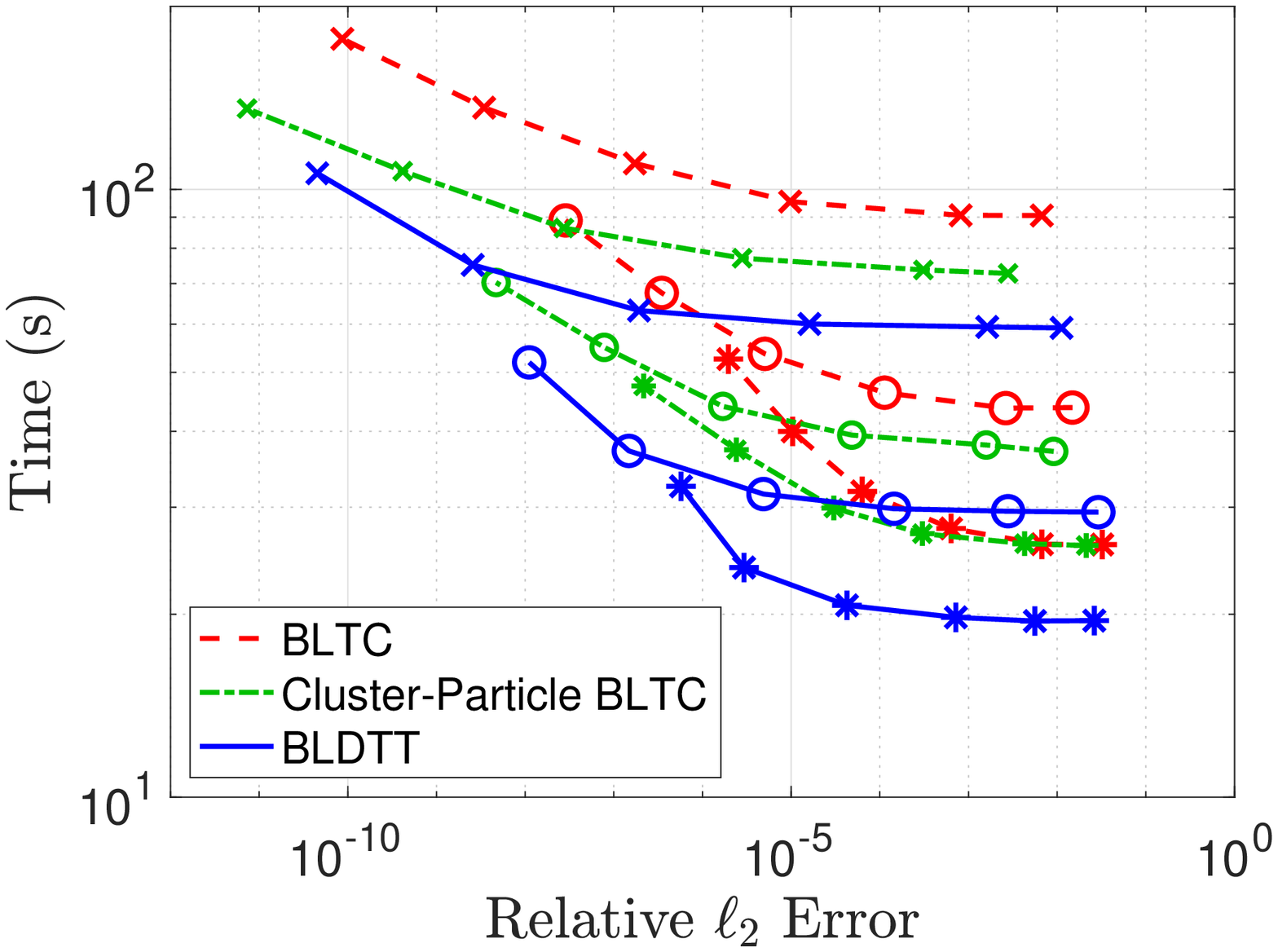} &
\hspace*{-0.20in}
\includegraphics[trim=35 0 15 10, clip, width=0.47\linewidth] {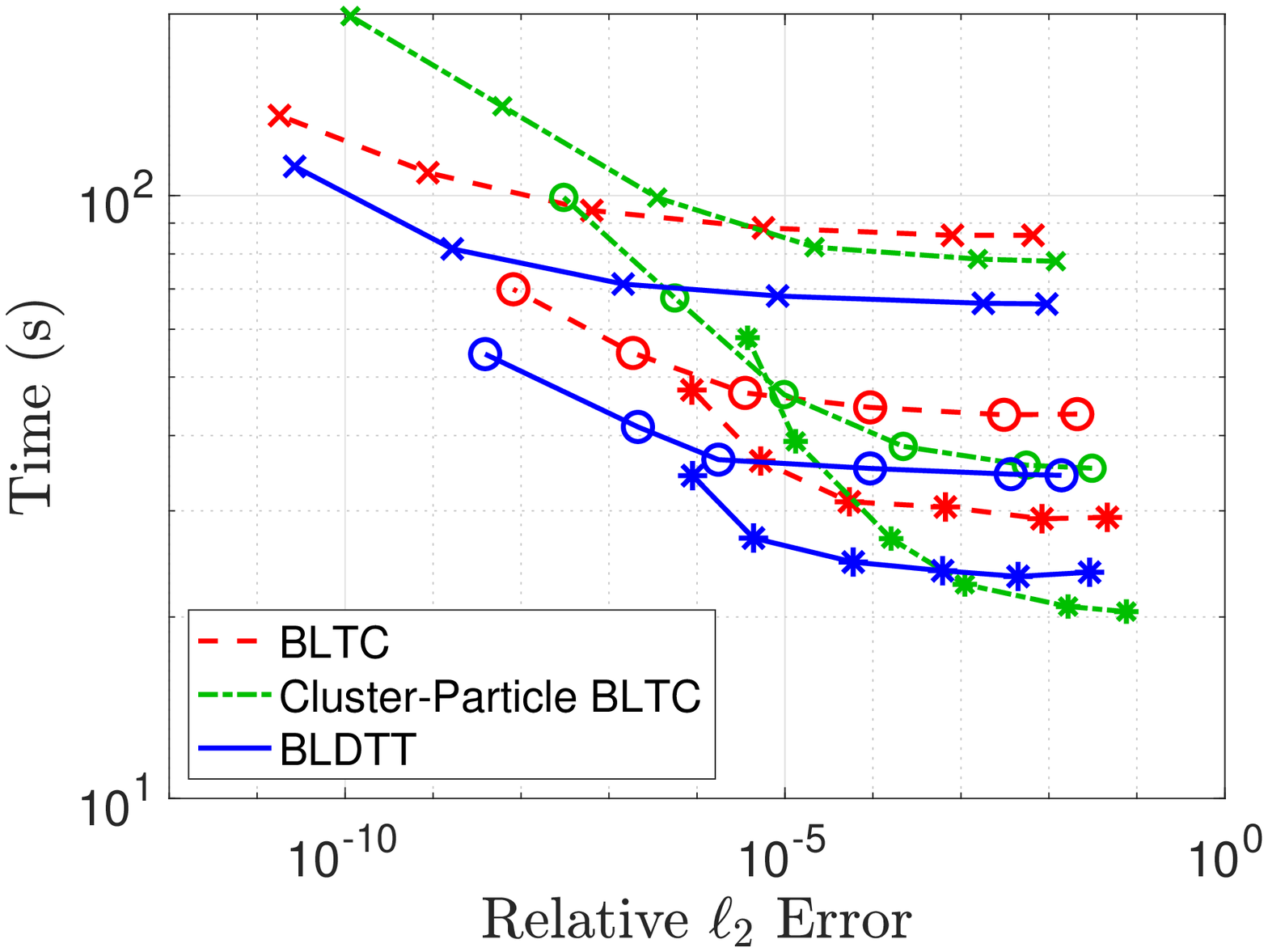} \\
\hspace*{-0.1in}
(a) $M=2$E7, $N=2$E6 & 
\hspace*{-0.38in}
(b) $M=2$E6, $N=2$E7 \\
\end{tabular}
\caption{
Unequal targets and sources,
(a) $M$=2E7 targets, $N$=2E6 sources, 
(b) $M$=2E6 targets, $N$=2E7 sources, 
random uniformly distributed particles,
compute time (s) versus relative $\ell_2$ error,
BLTC ({\color{red}red}, dashed), 
CP-BLTC ({\color{ForestGreen}green}, dash-dotted), 
BLDTT ({\color{blue}blue}, solid), 
connected curves represent constant MAC $\theta$ (0.5 $\times$; 0.7 $\circ$; 0.9 $*$),
interpolation degree $n = 1, 2, 4, 6, 8, 10$,
simulations ran on one NVIDIA P100 GPU.}
\label{fig:uniform_notequal_parameter_sweep}
\end{figure}

Figure~\ref{fig:uniform_notequal_parameter_sweep}
shows the compute time (s) versus relative $\ell_2$ error 
for the BLTC (red, dashed), CP-BLTC (green, dash-dotted), and BLDTT (blue, solid) 
with
(a) $M$=2E7 targets, $N$=2E6 sources, 
(b) $M$=2E6 targets, $N$=2E7 sources,
for MAC $\theta$ and interpolation degree $n$ as above.
For (a) $M > N$, the CP-BLTC outperforms the BLTC,
for (b) $N > M$, the BLTC outperforms the CP-BLTC for errors below 1E-5,
while the BLDTT outperforms the two treecodes in both cases.
Note however in (b) 
that for MAC $\theta = 0.9$ and error larger than 1E-3,
the CP-BLTC runs slightly faster than the BLDTT;
this is due to the cost of the upward pass in the BLDTT,
which is more expensive than the downward pass, 
and with low degree $n$,
makes up a substantial portion of the compute time.
The results demonstrate the ability of the BLDTT to 
efficiently adapt to the case of unequal targets and sources.

\subsection{Other interaction kernels}
\label{section:other-kernels}

\begin{figure}[htb]
\centering
\begin{tabular}{ccc}
\hspace*{-0.1in}
\includegraphics[trim=4 0 15 10, clip, width=0.34\linewidth] {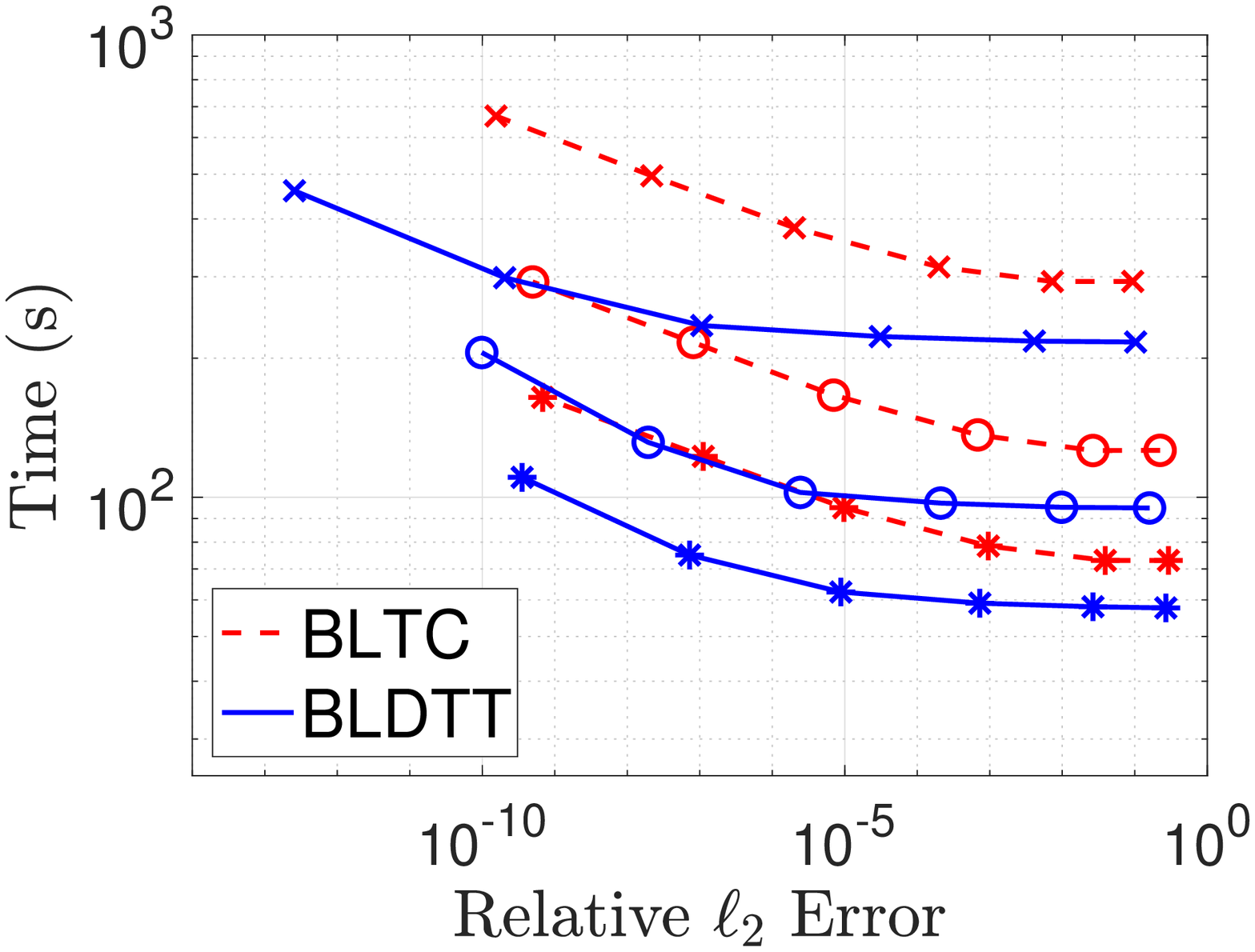} &
\hspace*{-0.2in}
\includegraphics[trim=38 0 15 10, clip, width=0.32\linewidth] {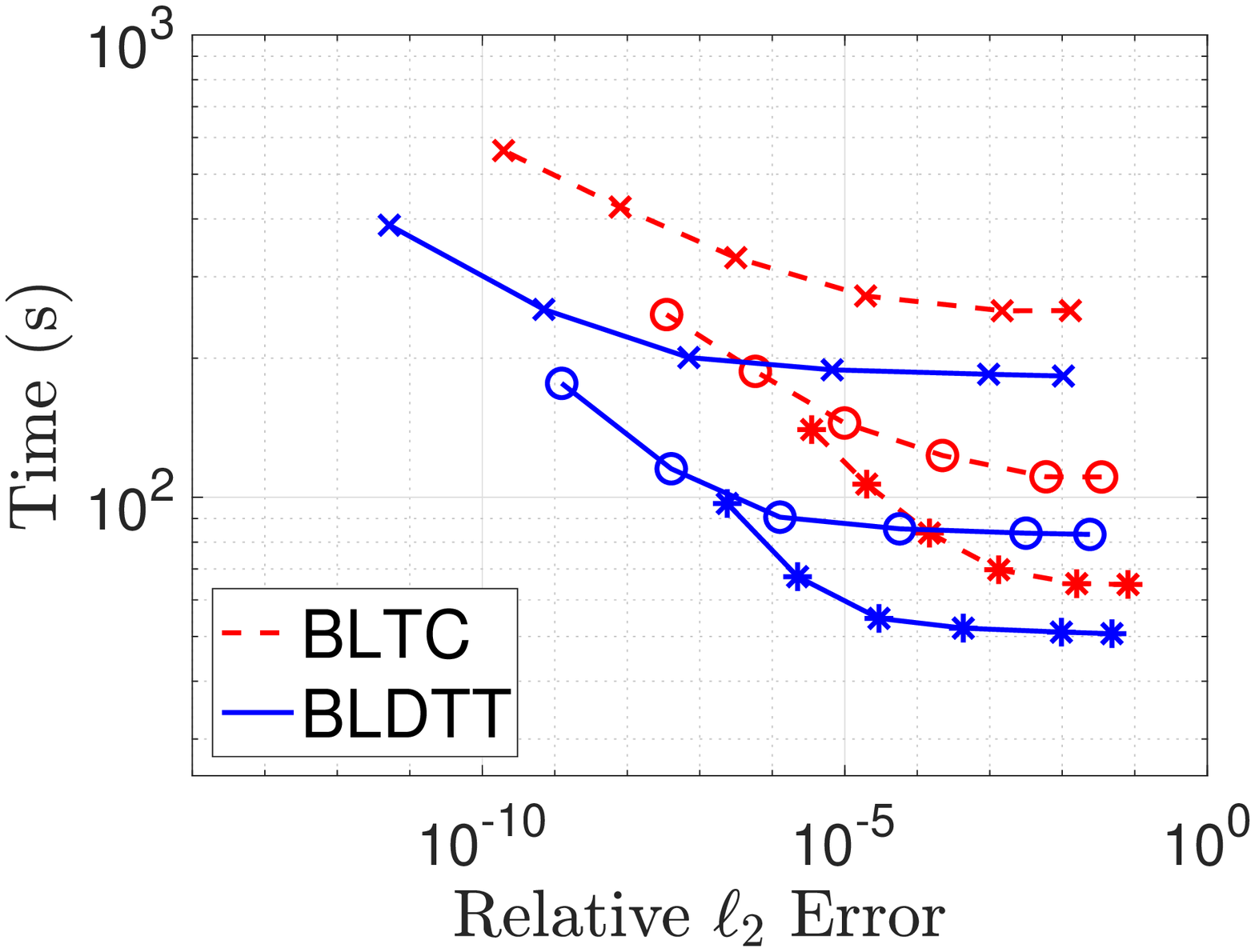} &
\hspace*{-0.2in}
\includegraphics[trim=38 0 15 10, clip, width=0.32\linewidth] {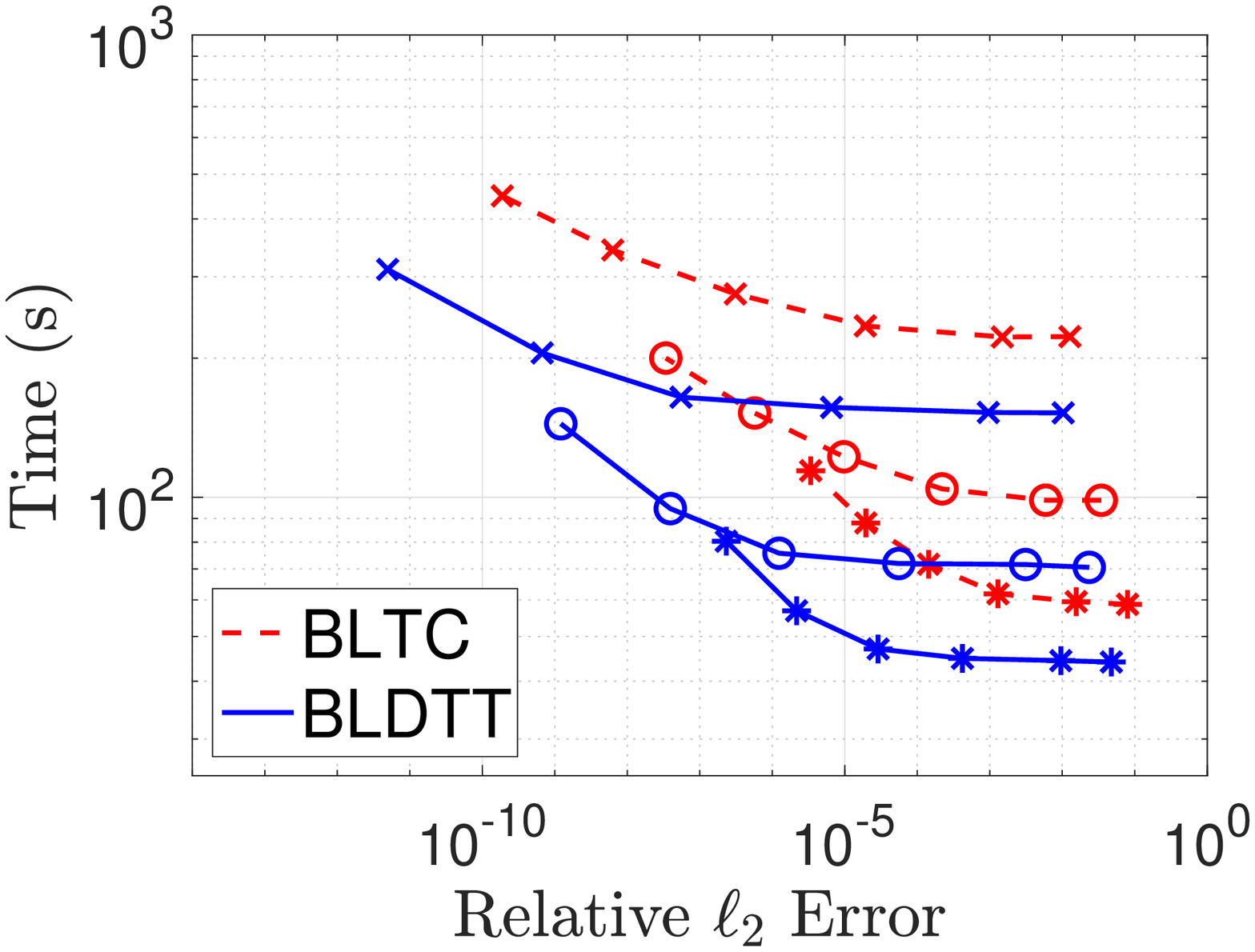} \\
\hspace*{-0.1in}
(a) $\sin(\pi r) / r$ & 
\hspace*{-0.2in}
(b) $\exp(-0.5 r) / r$  & 
\hspace*{-0.2in}
(b) $1/(r^2+0.005^2)^{1/2}$ \\
\end{tabular}
\caption{
Different interaction kernels,
$N$=2E7 random uniformly distributed particles in the cube $[-1,1]^3$,
(a) oscillatory, $\sin(\pi r)/r$, 
(b) Yukawa, $\exp(-0.5 r) / r$, 
(c) regularized Coulomb, $1/(r^2+0.005^2)^{1/2}$, 
compute time (s) versus relative $\ell_2$ error,
BLTC ({\color{red}red}, dashed),
BLDTT ({\color{blue}blue}, solid),
connected curves represent constant MAC $\theta$ (0.5 $\times$; 0.7 $\circ$; 0.9 $*$),
interpolation degree $n = 1, 2, 4, 6, 8, 10$ increases from right to left on each curve,
simulations ran on one NVIDIA P100 GPU.}
\label{fig:uniform_sinoverr_parameter_sweep}
\end{figure}

In previous sections we considered particles interacting through the Coulomb potential.
Here we demomstrate the performance of the BLDTT on three other interaction kernels:
(a) an oscillatory kernel, $\sin(\pi r)/r$, 
(b) a Yukawa kernel, $\exp(-0.5r)/r$, 
and (c) a regularized Coulomb kernel, $1/(r^2+\epsilon^2)^{1/2}$, with $\epsilon = 0.005$.
Figure~\ref{fig:uniform_sinoverr_parameter_sweep} 
shows the compute time (s) versus relative $\ell_2$ error  
for the BLDTT (blue, solid) and BLTC (red, dashed) 
for these three kernels on $N$=2E7 random uniformly distributed particles 
in the cube $[-1,1]^3$
for various values of the MAC $\theta$ and interpolation degree $n$. 
The results show that the BLDTT has consistently better performance than the BLTC.
Comparing with the Coulomb potential resuts in Figure~\ref{fig:distribution_parameter_sweep}(a), 
we see that the BLDTT has similar performance for the various interaction kernels,
reflecting the kernel-independent nature of the method.

\subsection{MPI strong scaling}\label{sec:mpi}

\begin{figure}[htb]
\centering
\begin{tabular}{cc}
\includegraphics[trim=50 20 50 20, clip, width=0.38\linewidth] {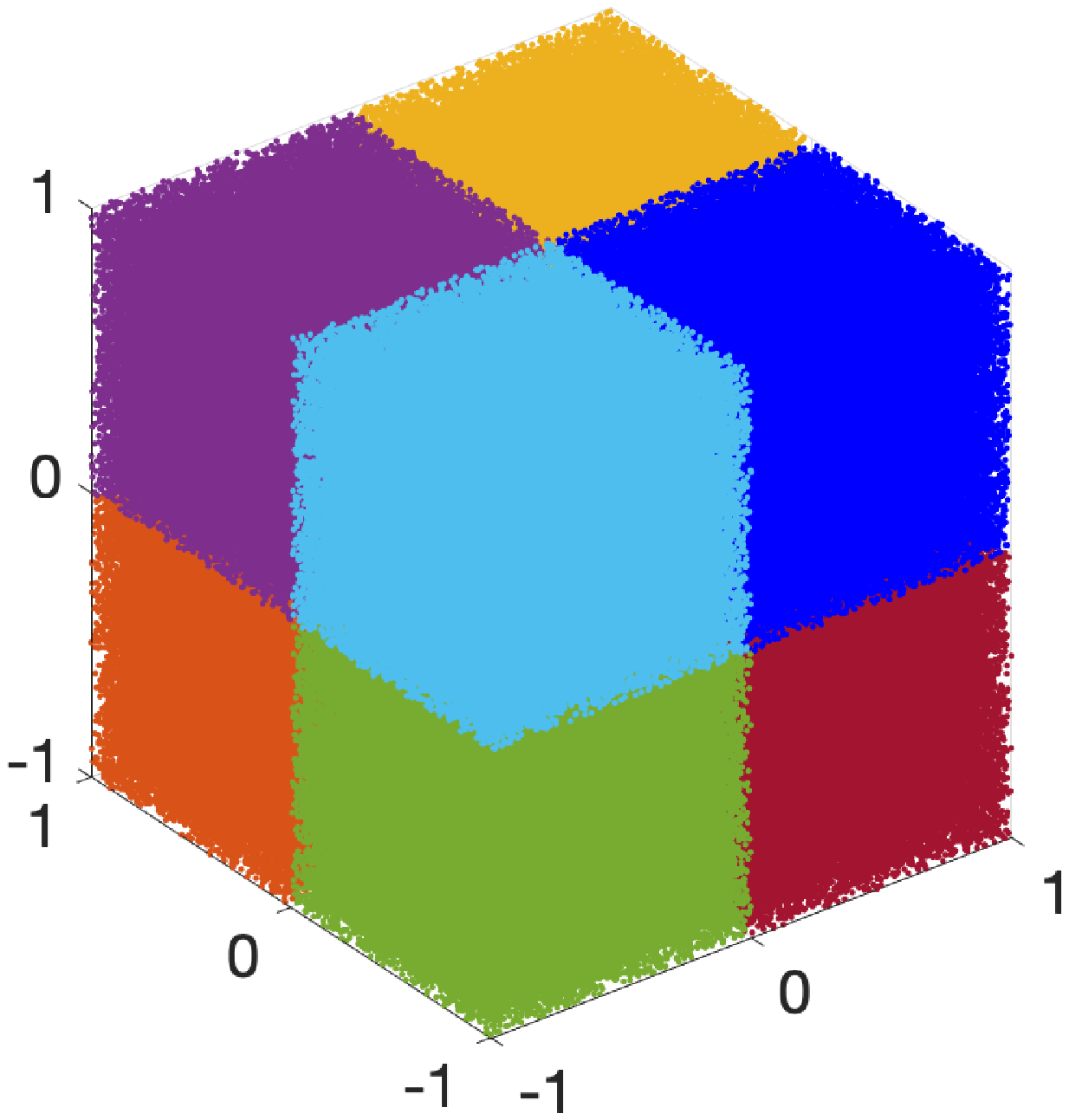} &
\includegraphics[trim=50 20 50 20, clip, width=0.38\linewidth] {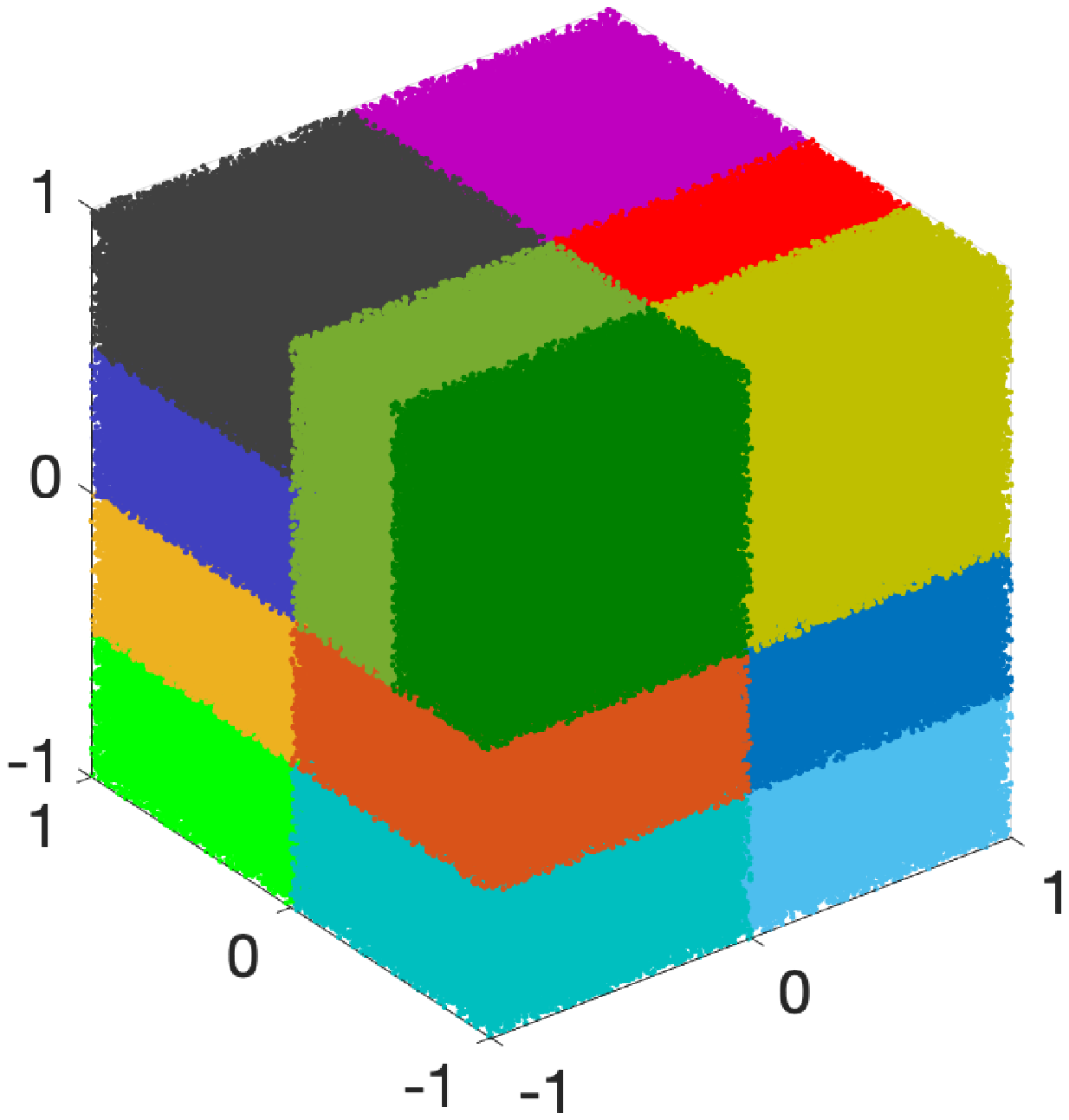} \\
(a) & (b)
\end{tabular}
\caption{
Examples of domain decomposition,
$N$=1.6E6 random uniformly distributed particles in the cube $[-1,1]^3$,
(a) 8 ranks with 2E5 particles per rank,
(b) 16 ranks with 1E5 particles per rank, 
colors represent particles residing on different ranks, 
partitioning by Trilinos Zoltan~\cite{zoltan-website,Boman:2012aa}.}
\label{fig:dist_16ranks}
\end{figure}

Finally, we demonstrate the MPI strong scaling of the BLDTT up to 32 NVIDIA P100 GPUs
with one MPI rank per GPU.
The particles are partitioned into geometrically localized domains by
Trilinos Zoltan~\cite{zoltan-website,Boman:2012aa}.
Figure~\ref{fig:dist_16ranks} depicts a sample domain decomposition 
for $N$=1.6E6 random uniformly distributed particles in the cube $[-1,1]^3$, (a) across 8 ranks with 2E5 particles per rank,
and (b) across 16 ranks with 1E5 particles per rank.
Colors represent particles residing on different ranks.

\begin{figure}[htb]
\centering
\begin{tabular}{ccc}
\hspace*{-0.1in}
\includegraphics[trim=4 0 10 10, clip, width=0.34\linewidth] {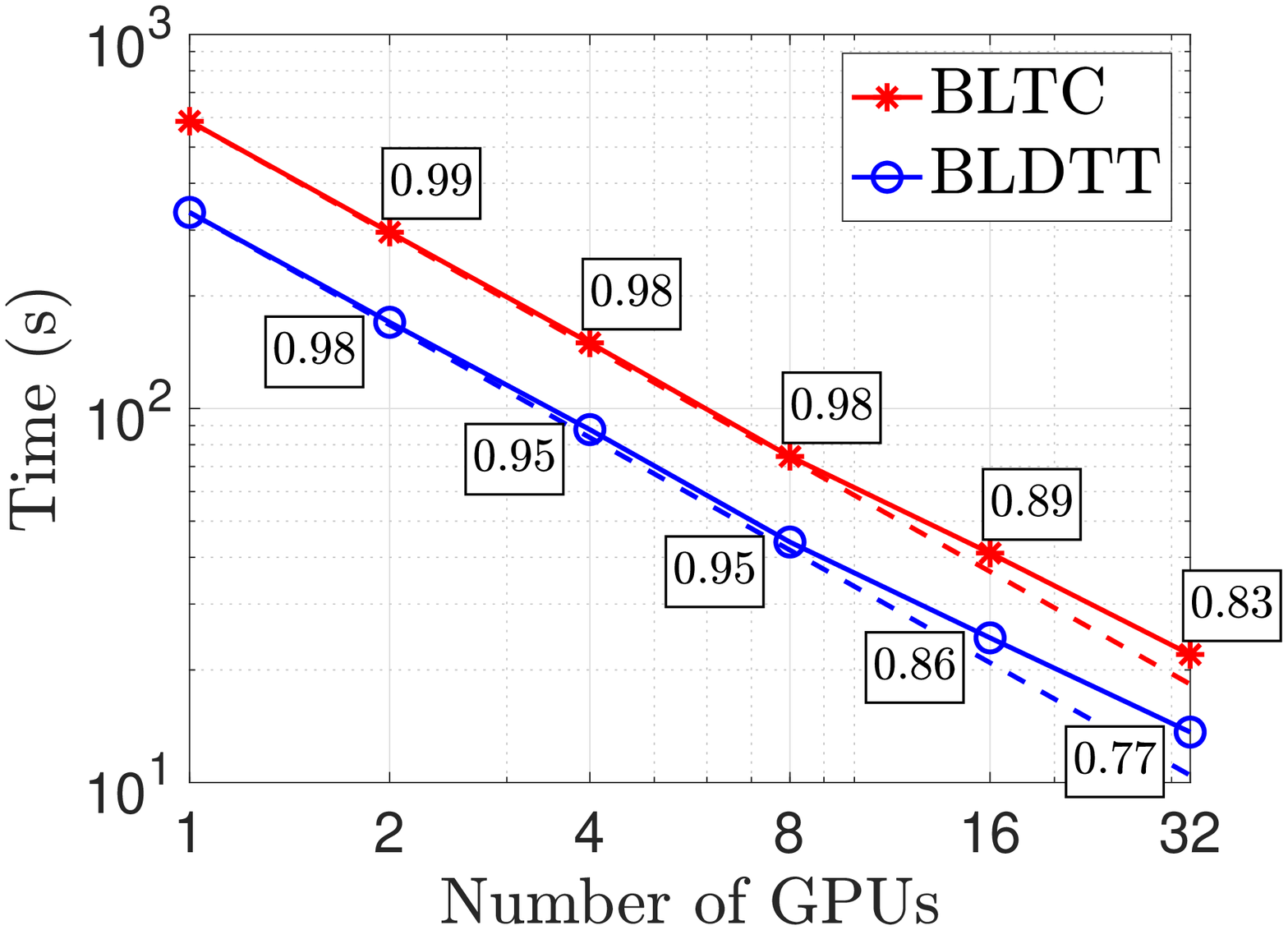} &
\hspace*{-0.2in}
\includegraphics[trim=38 0 10 10, clip, width=0.32\linewidth] {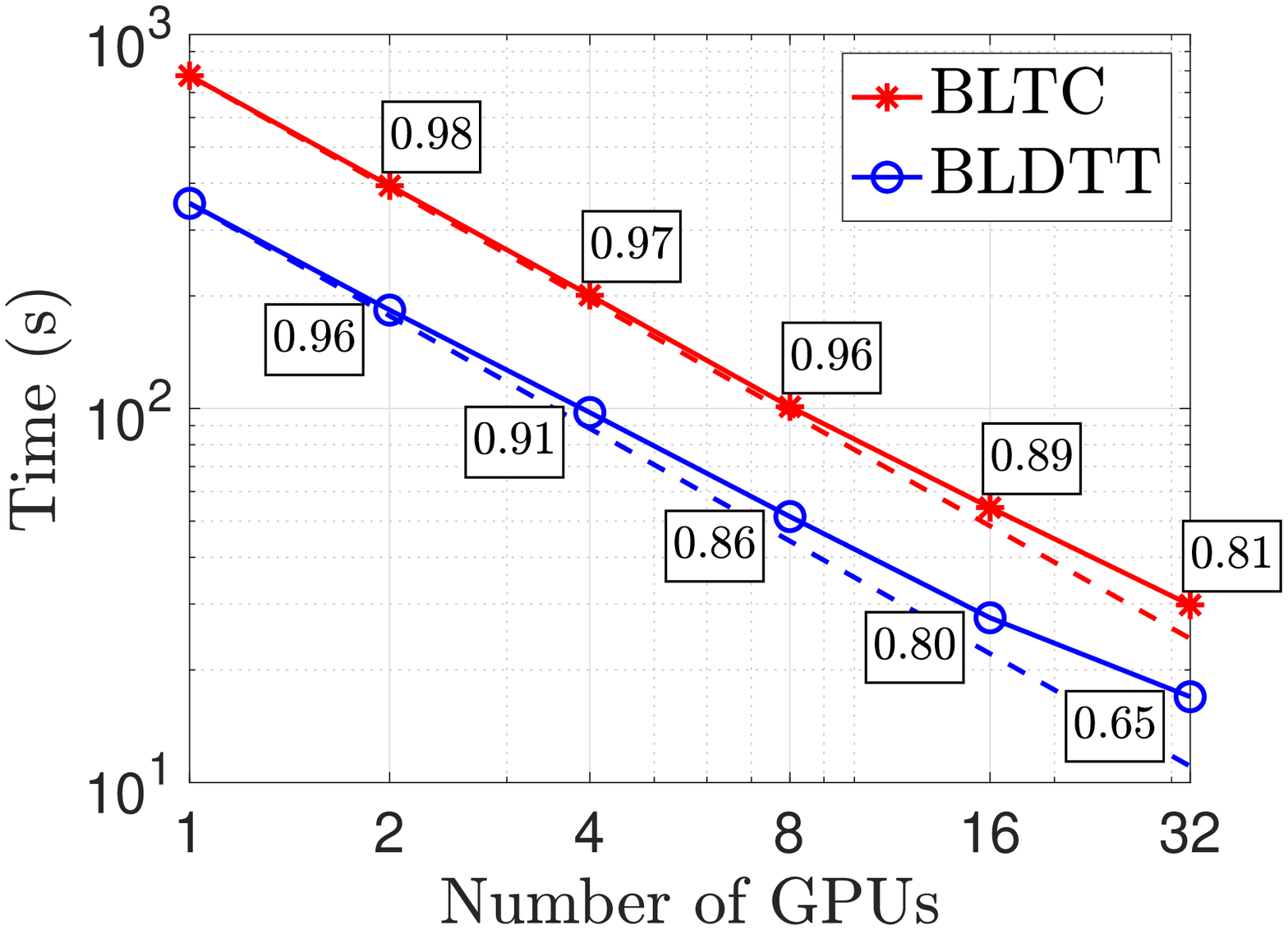} &
\hspace*{-0.2in}
\includegraphics[trim=38 0 10 10, clip, width=0.32\linewidth] {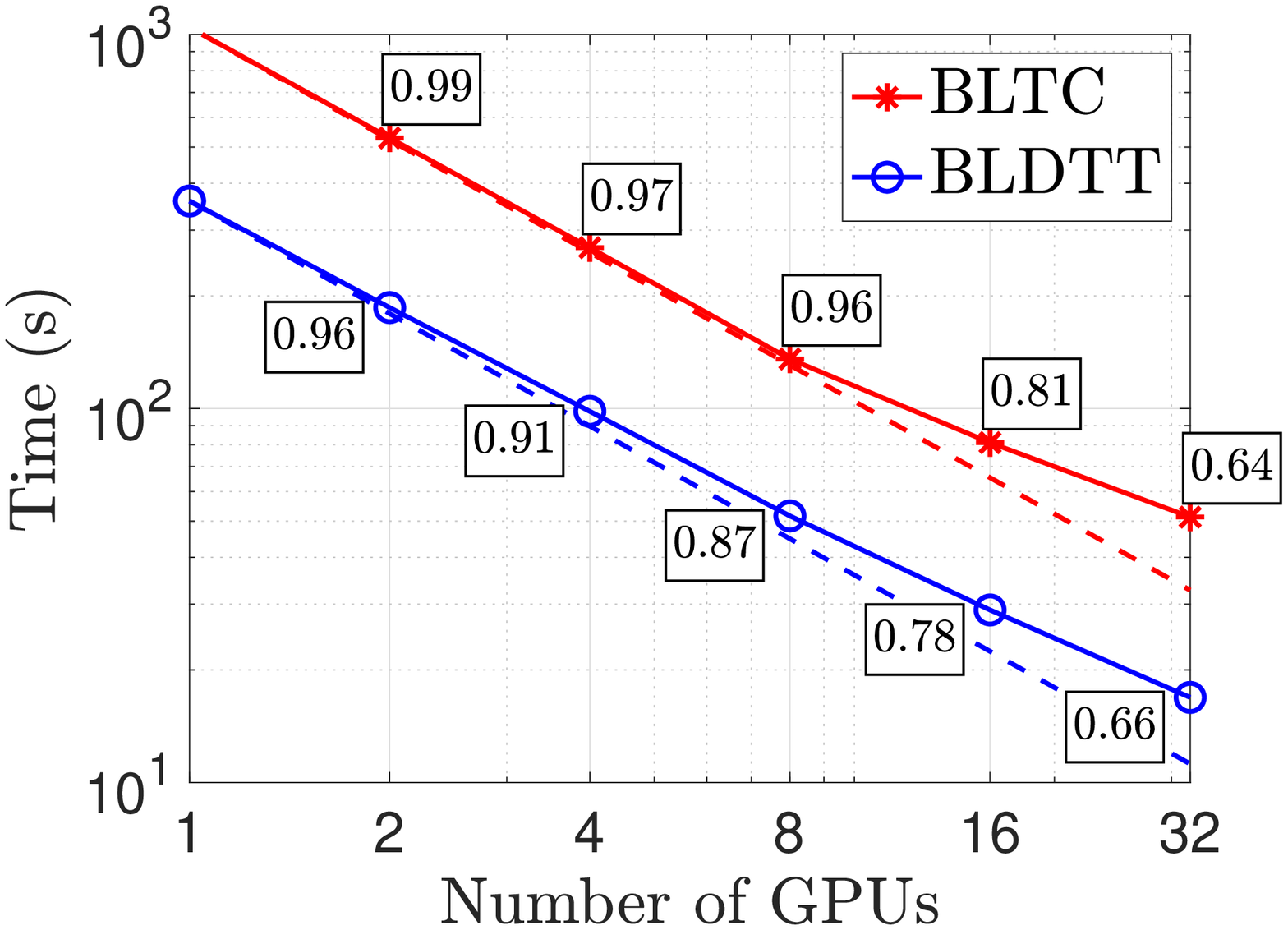}  \\
\hspace*{-0.1in}
(a) Uniform & 
\hspace*{-0.2in}
(b) Gaussian  & 
\hspace*{-0.2in}
(c) Plummer \\
\end{tabular}
\caption{
MPI strong scaling,
$N$=64E6 particles, 
(a) random uniform, (b) Gaussian, (c) Plummer distributions,
MAC $\theta = 0.7$, interpolation degree $n = 8$
yielding 7-8 digit accuracy, 
compute time (s) versus number of GPUs,
BLTC (red), BLDTT (blue),
ideal scaling (dashed lines), 
parallel efficiency (boxed numbers).}
\label{fig:strong_scaling}
\end{figure}

We consider a problem with $N$=64E6 particles 
using MAC $\theta=0.7$ and interpolation degree $n=8$ yielding error $\approx$~1E-8.
Figure~\ref{fig:strong_scaling} shows the compute time versus the number of GPUs
for (a) random uniform, (b) Gaussian, and (c) Plummer distributions
for the BLDTT (blue) and BLTC (red),
where dashed lines indicate ideal scaling 
and 
the boxed numbers show the parallel efficiency.
As was shown earlier for one GPU in Figure~\ref{fig:distribution_parameter_sweep},
the BLDTT is consistently faster than the BLTC up to 32~GPUs,
and
the speedup improves for the nonuniform particle distributions.
The BLDTT and BLTC have generally similar parallel efficiency for all three distributions;
for example on 32 GPUs, 
the BLDTT has parallel efficiency 77\%, 65\%, and 66\% for the uniform, Gaussian, and Plummer distributions, 
compared to 83\%, 81\%, and 64\% for the BLTC.

\begin{figure}[htb]
\centering
\begin{tabular}{cc}
\hspace*{-0.1in}
\includegraphics[trim=0 0 0 0, clip, width=0.51\linewidth] {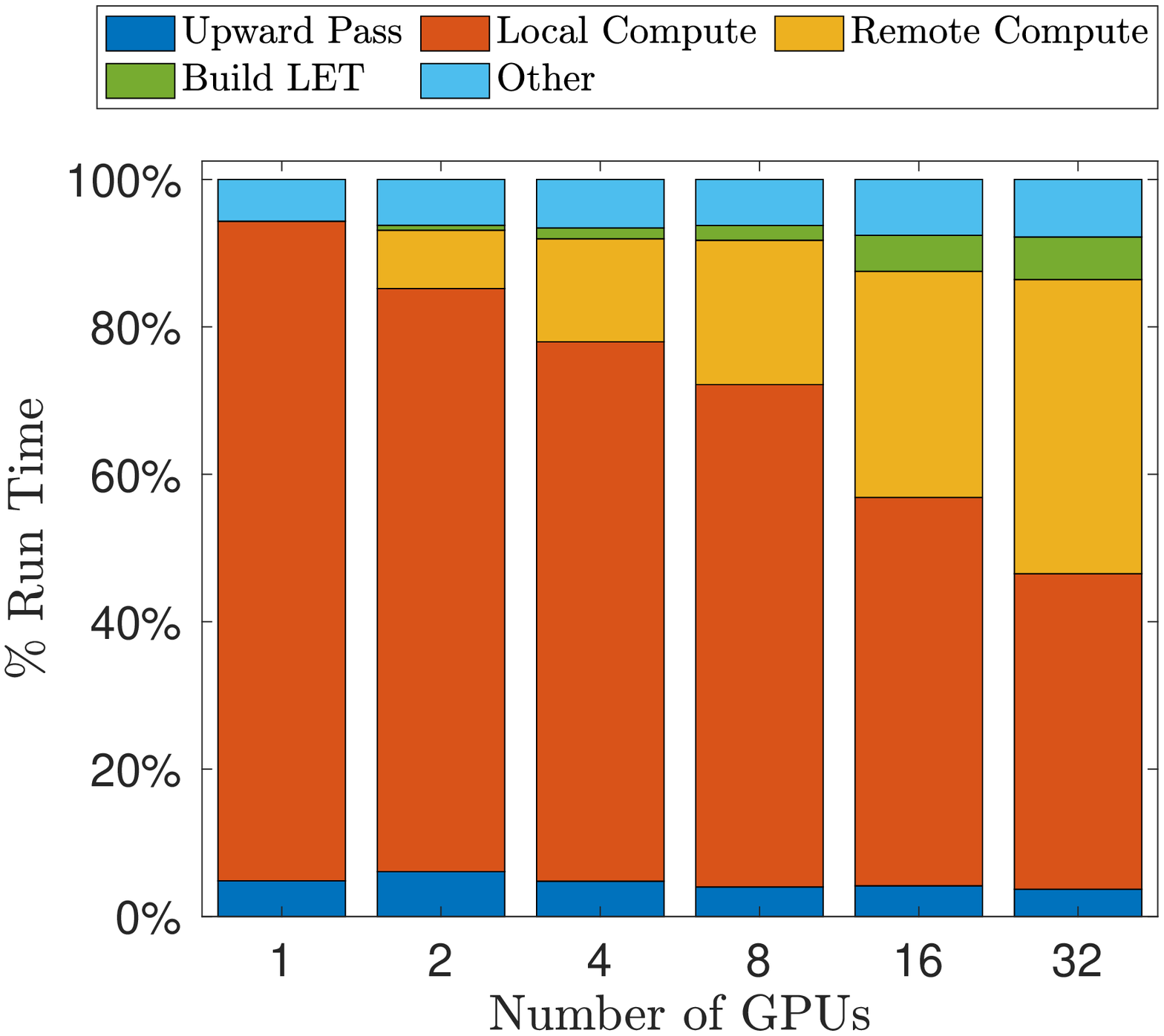} &
\hspace*{-0.2in}
\includegraphics[trim=20 0 0 0, clip, width=0.49\linewidth] {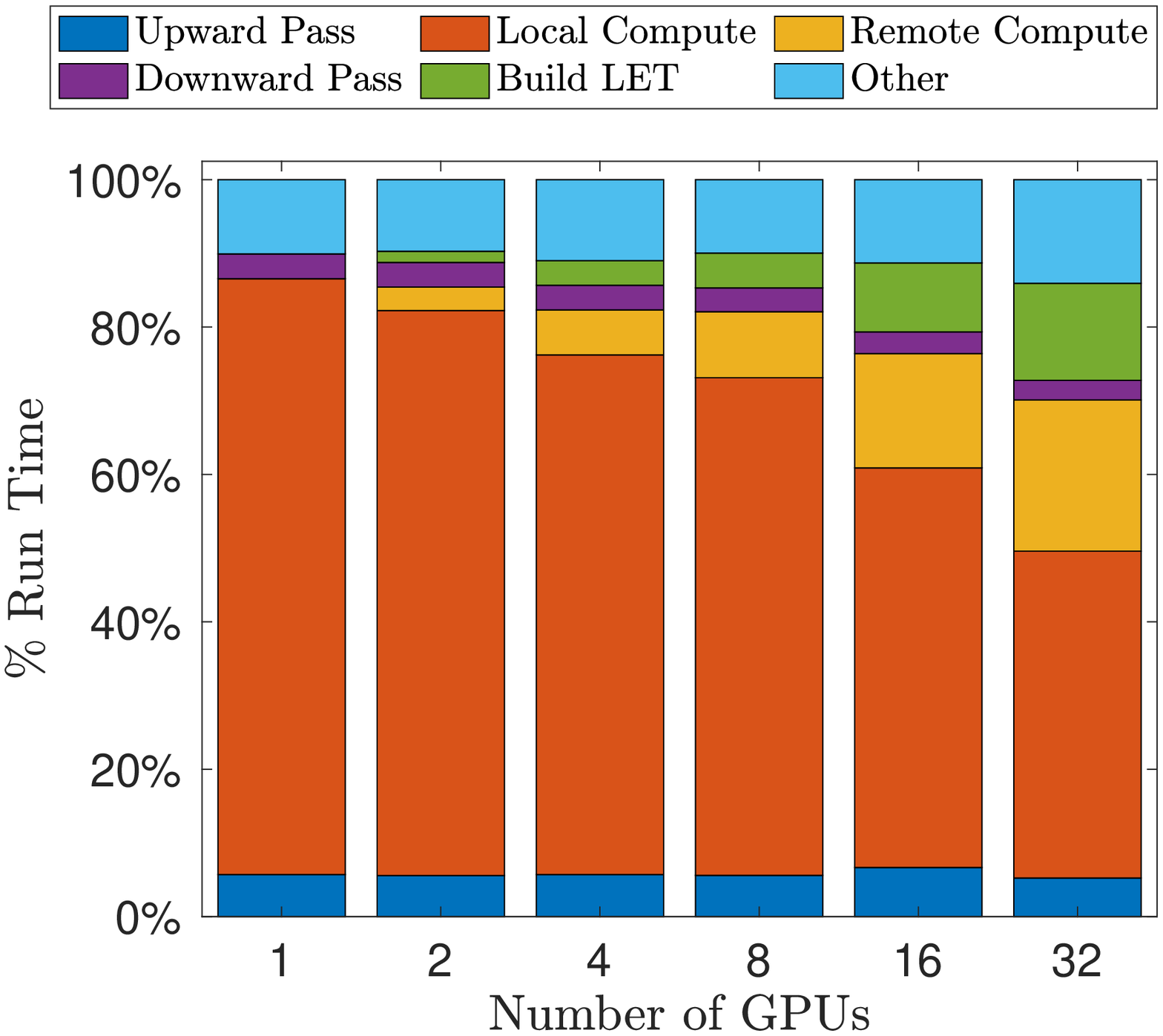} \\
\hspace*{-0.1in}
(a) BLTC & 
\hspace*{-0.2in}
(b) BLDTT \\
\end{tabular}
\caption{
Component breakdown of run time across 1 to 32 NVIDIA P100 GPUs,
64E6 random uniformly distributed particles in the cube $[-1,1]^3$,
MAC $\theta = 0.7$, interpolation degree $n = 8$, error $\approx$~1E-8,
(a) BLTC,
(b) BLDTT,
upward pass (blue), 
compute due to local sources and source clusters (orange),
compute due to remote sources and source clusters (yellow), 
downward pass (purple),
LET construction and communication (green), 
and other (light blue), which includes tree building and interaction list building,
breakdown is based on timing results for most expensive MPI rank in each case.}
\label{fig:strong_scaling_cc}
\end{figure}

The slightly lower parallel efficiency for the BLDTT compared to the BLTC 
can be explained by the greater efficiency of the BLDTT algorithm itself.
Figure~\ref{fig:strong_scaling_cc} shows the component breakdown as a percentage of 
run time (total wall clock time) of the (a) BLTC and (b) BLDTT
for the uniform distribution results in Figure~\ref{fig:strong_scaling}(a).
The components shown are the upward pass (blue), 
compute due to local sources and source clusters (orange), 
compute due to remote sources and source clusters (yellow), 
downward pass (purple),
LET construction and communication (green), 
and other (light blue), which includes tree building and interaction list building.
The breakdown is based on timing results for the most expensive MPI rank in each case.
Note that the LET construction accounts for an increasing percentage of the 
run time as the number of ranks increases.
This is to be expected as more particles reside on remote ranks and must be communicated, 
and this is the primary factor that impedes ideal parallel scaling.
The LET construction time is nearly identical for the BLTC and BLDTT,
however since the BLDTT computations are more efficient,
the LET construction accounts for a larger percentage of the run time 
and 
this results in the lower parallel efficiency seen in Figure~\ref{fig:strong_scaling}
as the number of ranks increases.

\section{Conclusion}
\label{section:conclusion}

We presented the barycentric Lagrange dual tree traversal (BLDTT)
fast summation method for particle interactions,
and its OpenACC implementation
with MPI remote memory access for distributed memory parallelization
running on multiple GPUs.
The BLDTT builds adaptive trees of clusters on the target particles and source particles,
where each cluster is the minimal bounding box of its particles,
and
a parent cluster may have 8, 4, or 2 children.
The BLDTT uses a dual tree traversal strategy~\cite{Appel:1985aa,Dehnen:2002aa}
to determine well-separated target and source clusters
that interact through
particle-cluster (PC), cluster-particle (CP), or 
cluster-cluster (CC) approximations based on 
barycentric Lagrange interpolation at proxy particles
defined by tensor product Chebyshev points of the 2nd kind
in each cluster~\cite{Berrut:2004aa,Wang:2020aa}.
The BLDTT has an upward pass and downward pass 
similar those in the FMM~\cite{Greengard:1987aa},
although it relies on interlevel polynomial interpolation 
rather than translation of expansion coefficients.
The BLDTT is kernel-independent, requiring only kernel evaluations.
The distributed memory parallelization of the BLDTT
uses recursive coordinate bisection for domain decomposition 
and MPI remote memory access to build 
locally essential trees on each rank~\cite{Warren:1992aa}.
The PP, PC, CP, and CC interactions all have a direct sum form
that efficiently maps onto the GPU, 
where target particles provide an outer level of parallelism, 
and source particles provide an inner level of parallelism.
The BLDTT code is part of the BaryTree library 
for fast summation of particle interactions available on 
GitHub~\cite{barytree-website}.
 
The performance of the BLDTT was compared with that of the BLTC,
an earlier particle-cluster barycentric Lagrange treecode~\cite{Wang:2020aa}.
For the systems of size $N$=1E5 to $N$=1E8 studied here
running on a single GPU,
the BLTC scales like $O(N\log N)$,
while the BLDTT scales like $O(N)$. 
The BLDTT was applied to
different random particle distributions (uniform, Gaussian, Plummer),
different particle domains (thin slab, square rod, spherical surface),
unequal sets of target and source particles,
and
different interaction kernels (oscillatory, Yukawa, 
singular and regularized Coulomb).
%
The BLDTT had consistently better performance than the BLTC,
showing its ability to adapt to different situations.
Finally,
the MPI strong scaling of the BLDTT and BLTC was demonstrated on up to 32 GPUs 
for $N$=64E6 particles with 7-8 digit accuracy.
Across these simulations 
the parallel efficiency of both codes is better than 64\%,
while the BLDTT is 1.5-2.5$\times$ faster than the BLTC.

Future work to further improve the efficiency of the BLDTT
could investigate 
overlapping communication and computation, 
building tree nodes that span multiple ranks, 
using mixed-precision arithmetic,
and
employing barycentric Hermite interpolation~\cite{Krasny:2019aa}.
We also anticipate applying the BLDTT to speed up
integral equation based
Poisson--Boltzmann implicit solvent computations~\cite{Geng:2013aa} 
and
density functional theory calculations~\cite{Vaughn:2020aa}. 

\section{Acknowledgements}
This work was supported by National Science Foundation grant DMS-1819094, 
Extreme Science and Engineering Discovery Environment (XSEDE) 
grants ACI-1548562, ASC-190062, 
and 
the Mcubed program 
and 
Michigan Institute for Computational Discovery and Engineering (MICDE) 
at the University of Michigan.

\bibliographystyle{elsarticle-num}
\bibliography{./library}

\clearpage

\appendix
\section{Preliminaries}
The derivation of the upward pass and downward pass rely on a property of interpolating polynomials.
For a general function $f(x)$,
the degree $n$ interpolation polynomial $p_n(x)$ is given by
\begin{equation}
    p_n(x) = \sum_{k=0}^n f(s_k) L_k(x)
\end{equation}
where $s_k$ denote a set of interpolation points.
Now consider the special case where $f(x) = L_{\ell}(x)$, a degree $n$ polynomial.  
Then
\begin{equation}
    p_n(x) = \sum_{k=0}^n L_{\ell}(s_k) L_k(x).
\end{equation}
Notice that $p_n(x)$ and $L_{\ell}(x)$ are both degree $n$ polynomials, and by construction, they take on the same values at the set of $n+1$ interpolation $s_k$. 
Therefore, $p_n(x)=L_{\ell}(x)$, and we have the following relation
\begin{equation}
    L_{\ell}(x) = \sum_{k=0}^n L_{\ell}(s_k) L_k(x),
    \label{eqn:interpolation-of-polynomial}
\end{equation}
that is,
the degree $n$ interpolating polynomial of a degree $n$ polynomial is itself.
Equation~\eqref{eqn:interpolation-of-polynomial} will be used in the appendices to follow, where $L_\ell(x)$ and $L_k(x)$ refer to interpolating polynomials in parent and child clusters.

\section{Details of upward pass}
\label{section:appendix-upward pass}



Recall Eq.~\eqref{eqn:proxy-charges} 
for the definition of the proxy charges $\widehat{q}_k$ 
of a parent source cluster $C_s$ in the 1D case,
\begin{equation}
    \widehat{q}_k =  \sum_{y_j\in C_s} L_{k}(y_{j})q_j,
    \label{eqn:appendix-proxy-charges}
\end{equation}
where $y_j, q_j$ are the source cluster particles and charges,
and
$L_{k}(y)$ are the Lagrange polynomials associated with the cluster.
Also consider the parent's eight child clusters $C_s^{i}$, $i=1:8$ 
with Lagrange polynomials $L^i_{k_i}(y)$,
interpolation points $s_{k_i}$,
and proxy charges $\widehat{q}_{k_i}$.
The parent proxy charge $\widehat{q}_k$ in Eq.~\eqref{eqn:appendix-proxy-charges} can be 
expressed in terms of the child proxy charges $\widehat{q}_{k_i}$ as follows,
\begin{subequations}
\begin{align}
\widehat{q}_k 
&= \sum_{y_j\in C_s} L_{k}(y_{j})q_j \label{eqn:1d-upward pass-derivation-a}\\
&= \sum_{i=1}^8 \sum_{y_j \in C_s^{i}} L_{k}(y_{j})q_j \label{eqn:1d-upward pass-derivation-b}\\
&= \sum_{i=1}^8\sum_{y_j\in C_s^{i}}
\left( \sum_{k_i=0}^n L_{k}(s_{k_i})L_{k_i}(y_{j}) \right)q_j \label{eqn:1d-upward pass-derivation-c}\\
&= \sum_{i=1}^8 \sum_{k_i=0}^n
L_{k}(s_{k_i})
\sum_{y_j\in C_s^{i}} 
L_{k_i}(y_{j})q_j \label{eqn:1d-upward pass-derivation-d}\\
&= \sum_{i=1}^8\sum_{k_i=0}^n L_{k}(s_{k_i})\widehat{q}_{k_i}.\label{eqn:1d-upward pass-derivation-e}
\end{align}
\end{subequations}

Equation~\eqref{eqn:1d-upward pass-derivation-a} 
is the definition of the parent proxy charges,
Eq.~\eqref{eqn:1d-upward pass-derivation-b} 
splits this into the sum over the eight child clusters,
Eq.~\eqref{eqn:1d-upward pass-derivation-c} 
uses the relation in Eq.~\eqref{eqn:interpolation-of-polynomial},
Eq.~\eqref{eqn:1d-upward pass-derivation-d} rearranges the sums,
and 
Eq.~\eqref{eqn:1d-upward pass-derivation-e} 
applies the definition of the child proxy charges.
This result extends in a straightforward way to 3D.
In summary,
the upward pass ascends the source tree from the leaves to the root, 
computing the parent proxy charges of each source cluster
from the child proxy charges as described here.

\section{Details of downward pass}
\label{section:appendix-downward pass}

The downward pass adds the CP and CC interactions to the potentials $\phi(x_i)$.
For simplicity of notation,
the formulas are written in 1d with straightforward extension to 3D,
and
we consider the case in which the tree has two levels ($L=2$).
Recall Eq.~\eqref{eqn:1d-direct-downpass}, 
which interpolates the proxy potentials $\phi(t^m_{k_m})$
at each level in the tree
directly to the target particles $x_i$,
\begin{equation}
    \phi(x_i) \mathrel{+}= 
    \sum_{m = 1}^2 \sum_{k_m=0}^n L^m_{k_m}(x_{i})\phi(t^m_{k_m}),
    \label{eqn:appendix-interpolation}
\end{equation}
where $L^m_{k_m}(x_{i})$ is a Lagrange polynomial associated with
the cluster $C_t^m$ at level $m$ containing the target~$x_i$,
and
$\mathrel{+}=$ indicates that the right side is aggregated 
with the PP and PC interactions already computed in the DTT.
Then the right side of Eq.~\eqref{eqn:appendix-interpolation}
can be rewritten as follows,
\begin{subequations}
\begin{align}
    \sum_{m = 1}^2 \sum_{k_m=0}^n L^m_{k_m}(x_{i})\phi(t^m_{k_m})
    &= \sum_{k_1=0}^n L^1_{k_1}(x_{i})\phi(t^1_{k_1}) +
       \sum_{k_2=0}^n L^2_{k_2}(x_{i})\phi(t^2_{k_2}) \label{eqn:downpass-derivation-a} \\
    &= \sum_{k_1=0}^n L^1_{k_1}(x_{i})\phi(t^1_{k_1}) +
       \sum_{k_2=0}^n \left( \sum_{k_1=0}^n L^2_{k_2}(t^1_{k_1})L^1_{k_1}(x_i) \right) \phi(t^2_{k_2})
       \label{eqn:downpass-derivation-b} \\
    &= \sum_{k_1=0}^n L^1_{k_1}(x_{i})\phi(t^1_{k_1}) +
       \sum_{k_1=0}^n L^1_{k_1}(x_i) \left( \sum_{k_2=0}^n   L^2_{k_2}(t^1_{k_1}) \phi(t^2_{k_2}) \right) \label{eqn:downpass-derivation-c} \\
    &= \sum_{k_1=0}^n L^1_{k_1}(x_{i}) \left( \phi(t^1_{k_1}) + \sum_{k_2=0}^n   L^2_{k_2}(t^1_{k_1}) \phi(t^2_{k_2}) \right). \label{eqn:downpass-derivation-d}
\end{align}
\label{eqn:downpass-derivation}
\end{subequations}
Steps a,c,d are straightforward definitions and algebra,
while step b
relies on the interpolation relation in 
Eq.~\eqref{eqn:interpolation-of-polynomial}. 
Then the alternative version of Eq.~\eqref{eqn:appendix-interpolation} is
\begin{equation}
    \phi(x_i) \mathrel{+}=
    \sum_{k_1=0}^n L^1_{k_1}(x_{i}) \left( \phi(t^1_{k_1}) + \sum_{k_2=0}^n   L^2_{k_2}(t^1_{k_1}) \phi(t^2_{k_2}) \right), 
    \label{eqn:appendix-interpolation-2}
    \end{equation}
which corresponds to Eq.~\eqref{eqn:1d-downpass-leaves},
where the terms in parentheses
correspond to Eq.~\eqref{eqn:1d-parent-child-downpass};
the second term is the parent-to-child interpolation
and
the first term is aggregation with proxy potentials in the leaves
previously computed in the DTT.
In summary,
instead of interpolating from $t^1_{k_1}$ to $x_i$
\textit{and} 
from $t^2_{k_2}$ to $x_i$ (step a),
one interpolates from $t^2_{k_2}$ to $t^1_{k_1}$,
aggregates with previously computed results at $t^1_{k_1}$,
and
finally interpolates from $t^1_{k_1}$ to $x_i$ (step d).  
This procedure generalizes to accommodate trees of any depth.


\end{document}